\documentclass[preprint2]{aastex}
\usepackage{lineno}
\usepackage{amsmath}
\usepackage{url}
\shorttitle{Timing of Fermi Pulsars}
\shortauthors{Ray et al.}

\begin{document}

\title{Precise $\gamma$-Ray Timing and Radio Observations of 17 Fermi $\gamma$-Ray Pulsars}

\author{
P.~S.~Ray\altaffilmark{1,2},
M.~Kerr\altaffilmark{3},
D.~Parent\altaffilmark{4},
A.~A.~Abdo\altaffilmark{5},
L.~Guillemot\altaffilmark{6},
S.~M.~Ransom\altaffilmark{7},
N.~Rea\altaffilmark{8},
M.~T.~Wolff\altaffilmark{1},
A.~Makeev\altaffilmark{4},
M.~S.~E.~Roberts\altaffilmark{9},
F.~Camilo\altaffilmark{10},
M.~Dormody\altaffilmark{11},
P.~C.~C.~Freire\altaffilmark{6},
J.~E.~Grove\altaffilmark{1},
C.~Gwon\altaffilmark{1},
A.~K.~Harding\altaffilmark{12},
S.~Johnston\altaffilmark{13},
M.~Keith\altaffilmark{13},
M.~Kramer\altaffilmark{14,6},
P.~F.~Michelson\altaffilmark{3},
R.~W.~Romani\altaffilmark{3},
P.~M.~Saz~Parkinson\altaffilmark{11},
D.~J.~Thompson\altaffilmark{12},
P.~Weltevrede\altaffilmark{14},
K.~S.~Wood\altaffilmark{1},
M.~Ziegler\altaffilmark{11},
}
\altaffiltext{1}{Space Science Division, Naval Research Laboratory, Washington, DC 20375, USA}
\altaffiltext{2}{email: Paul.Ray@nrl.navy.mil}
\altaffiltext{3}{W. W. Hansen Experimental Physics Laboratory, Kavli Institute for Particle Astrophysics and Cosmology, Department of Physics and SLAC National Accelerator Laboratory, Stanford University, Stanford, CA 94305, USA}
\altaffiltext{4}{Center of Earth Observation and Space Research, College of Science, George Mason University, Fairfax, VA 22030, resident at Naval Research Laboratory, Washington, DC 20375}
\altaffiltext{5}{National Research Council Research Associate, National Academy of Sciences, Washington, DC 20001, resident at Naval Research Laboratory, Washington, DC 20375}
\altaffiltext{6}{Max-Planck-Institut f\"ur Radioastronomie, Auf dem H\"ugel 69, 53121 Bonn, Germany}
\altaffiltext{7}{National Radio Astronomy Observatory (NRAO), Charlottesville, VA 22903, USA}
\altaffiltext{8}{Institut de Ciencies de l'Espai (IEEC-CSIC), Campus UAB, 08193 Barcelona, Spain}
\altaffiltext{9}{Eureka Scientific, Oakland, CA 94602, USA}
\altaffiltext{10}{Columbia Astrophysics Laboratory, Columbia University, New York, NY 10027, USA}
\altaffiltext{11}{Santa Cruz Institute for Particle Physics, Department of Physics and Department of Astronomy and Astrophysics, University of California at Santa Cruz, Santa Cruz, CA 95064, USA}
\altaffiltext{12}{NASA Goddard Space Flight Center, Greenbelt, MD 20771, USA}
\altaffiltext{13}{CSIRO Astronomy and Space Science, Australia Telescope National Facility, Epping NSW 1710, Australia}
\altaffiltext{14}{Jodrell Bank Centre for Astrophysics, School of Physics and Astronomy, The University of Manchester, M13 9PL, UK}

\begin{abstract}
We present precise phase-connected pulse timing solutions for 16 $\gamma$-ray-selected pulsars recently discovered using the Large Area Telescope (LAT) on the \textit{Fermi Gamma-ray Space Telescope} plus one very faint radio pulsar (PSR J1124$-$5916) that is more effectively timed with the LAT. We describe the analysis techniques including a maximum likelihood method for determining pulse times of arrival from unbinned photon data. A major result of this work is improved position determinations, which are crucial for multi-wavelength follow up. For most of the pulsars, we overlay the timing localizations on X-ray images from \textit{Swift} and describe the status of X-ray counterpart associations. We report glitches measured in PSRs J0007+7303, J1124$-$5916, and J1813$-$1246. We analyze a new 20~ks \textit{Chandra} ACIS observation of PSR J0633+0632 that reveals an arcminute-scale X-ray nebula extending to the south of the pulsar. We were also able to precisely localize the X-ray point source counterpart to the pulsar and find a spectrum that can be described by an absorbed blackbody or neutron star atmosphere with a hard powerlaw component. Another \textit{Chandra} ACIS image of PSR J1732$-$3131 reveals a faint X-ray point source at a location consistent with the timing position of the pulsar. Finally, we present a compilation of new and archival searches for radio pulsations from each of the $\gamma$-ray-selected pulsars as well as a new Parkes radio observation of PSR J1124$-$5916 to establish the $\gamma$-ray to radio phase offset.
\end{abstract}


\section{Introduction}

Pulsar timing involves making precise measurements of pulse times of arrival (TOAs) at an observatory (or spacecraft) and then fitting the parameters of a `timing model' to those measurements.
This powerful technique enables extremely high precision measurements that probe numerous topics in fundamental physics and astrophysics. This is due to the ability to construct a coherent timing model that accounts for every rotation of the neutron star over periods of years. Precise timing measurements on radio pulsars have yielded many fundamental advances including the first indirect detection of energy loss due to gravitational radiation \citep{tfm79} and confirmation of many effects predicted by General Relativity \citep{s03,kw09}. 

Until recently, pulsar timing was only practical in the radio and, in some cases, soft X-ray bands \citep[for example]{jh05,lrc+09}. For radio and X-ray quiet/faint pulsars discovered with the Large Area Telescope (LAT) on \textit{Fermi}, the only option is to time them directly using the $\gamma$-ray data. Earlier instruments, such as EGRET on the \textit{Compton Gamma-Ray Observatory}, required very long exposures to even detect a handful of $\gamma$-ray pulsars and it only observed them occasionally, typically during a few 2-week viewing periods spread over the 9-year mission.  With the LAT, we have a vastly more powerful instrument for long-term pulsar studies. First, its effective area ($\sim 8000$ cm$^2$ at 1 GeV), energy coverage (20 MeV to $>300$ GeV) and point spread function ($\sim 0.8^\circ$ at 1 GeV) are greatly improved, providing a large increase in instantaneous sensitivity over EGRET \citep{LATpaper}. Second, because \textit{Fermi} operates in a continuous all-sky survey mode with a very large field of view ($\sim 2.4$ sr), it accumulates data on all pulsars in the sky roughly uniformly at all times. This allows long evenly-sampled timing observations of all pulsars detectable with \textit{Fermi}.

In this paper, we describe the techniques developed for precise timing of pulsars using the $\gamma$-ray photon data provided by the LAT. We then apply this method to the first 16 $\gamma$-ray-selected pulsars discovered in blind searches of LAT data \citep{LATBlindSearch} plus one additional radio pulsar (PSR J1124$-$5916), which is too faint for routine radio timing \citep{cmg+02}. The timing models presented here are updated versions of those used for these pulsars in the First LAT Catalog of Gamma-ray Pulsars \citep{PulsarCatalog}, and this paper documents the methods used to create those models.  In the case of the bright Vela and Geminga pulsars, these methods were used to provide high-precision timing models used for phase-resolved analysis \citep{Vela2,LATGeminga}. 

Timing observations provide a wealth of important information critical to the understanding of these newly-identified pulsars. First, one gets a measurement of the period and period derivative of the source. Having these two numbers allows us to derive estimates of several key parameters including the characteristic age, the inferred dipole magnetic field strength, and the spindown energy loss rate. These parameters are fundamental to understanding the astrophysics of the system. For example, the characteristic age is useful (though certainly not definitive) in the context of arguments for or against associations with supernovae and pulsar wind nebulae (PWNe).

The next critical parameter in the timing  model is the pulsar position. Estimating the source position from the reconstructed photon arrival directions can yield localizations that are good to a few arcmin, but to do better than this requires timing.  For young or middle-aged pulsars, the LAT can measure pulse arrival times with accuracies of order a millisecond\footnote{The accuracy of a pulse time of arrival measurement is determined by the photon statistics and the sharpness of the features in the pulse profile. It is always considerably larger than the $\sim 1$ $\mu$s accuracy on individual photon event times recorded by the LAT.}, which can be fit to determine positions to arcsecond accuracy. Accurate positions then allow deep counterpart searches in the X-ray, optical, and radio bands, and remove the effects of position error on the remaining timing parameters, most notably the spin-down rate.

Once the basic spin and position parameters are well determined, timing allows us to investigate the rotational irregularities that are common in young pulsars. The primary phenomena are timing noise and glitches. Glitches are sudden increases in pulse frequency with a magnitude in the range $\Delta \nu/\nu \sim 10^{-10} - 10^{-5}$, which provide valuable information about the superfluid interior of neutron stars \citep[for example]{acp03,lel99}. Timing noise is unmodeled low-frequency (often quasi-periodic) noise observed in the residuals of many pulsars after all the deterministic spin-down effects have been removed. The magnitude of the timing noise has been shown to correlate with frequency derivative (i.e. torque) \citep{cd85,ant+94,hlk10}, but its nature remains poorly understood.

Using the timing positions for these pulsars, we have also undertaken deep radio observations of the $\gamma$-ray-selected pulsars to search for radio pulsations. These searches have resulted in three discoveries of radio pulsations, which have been published elsewhere \citep{crr+09,MGROPaper}. Here, we compile the upper limits from our observations, and from the literature, for the remaining pulsars. These are important inputs to population statistics and modeling of these apparently radio quiet pulsars \citep[for example]{yr95,sgh07}, as well as for guiding future deeper searches.

\section{Methods}

\subsection{Data Selection}


For the current analysis we use LAT data from 2008 August 4 through at least 2010 February 4, the first 18 months of LAT survey operations. 
We select LAT events from the most restrictive ``diffuse'' class of the ``Pass 6'' event reconstructions \citep{LATpaper} with a zenith angle of $<105^\circ$ to reduce contamination from atmospheric secondary $\gamma$ rays from near the Earth's limb.  For each pulsar, we find an optimal radius and low energy cut to maximize the pulse detection significance. The radius cuts ranged from 0.5--1.6$^\circ$, while the low energy cuts ranged from 50--900 MeV. Only photons that pass these cuts are included in the timing analysis.

The number of photons surviving these cuts ranged from 1,174 (PSR J0633+0632) to 14,875 (PSR J1836+5925) in our 18 months of observing, a span in which the pulsars completed several hundred million rotations.  This emphasizes the unique nature of timing pulsars using extremely sparse $\gamma$-ray data.  Typically only of order 100 photons go into each TOA determination. In addition, unlike with radio pulsar timing, the integration time per TOA is equal to the spacing between TOAs, requiring the model to maintain phase accuracy over a much longer time than is required for radio pulsar timing where the integration time for a TOA is only minutes or hours.  We constructed initial models  using the \texttt{prepfold} tool from the PRESTO pulsar analysis software package\footnote{Available from \url{http://www.cv.nrao.edu/~sransom/presto/}}.  This tool performs epoch folding searches over narrow ranges of frequency and frequency first and second derivatives around the $\nu$ and $\dot\nu$ values found from the blind search to maximize the signal-to-noise ratio.  Combined with searching over a grid of possible pulsar positions, we are able to arrive at an initial model that maintains coherence well enough for TOAs to be determined and the pulsar timing to proceed as described below.

\subsection{Geocentering}

Pulsar timing software generally expects pulse TOAs to be measured at an observatory that is at a fixed geographic location on the Earth.  Observations from a spacecraft in orbit about the Earth obviously do not satisfy this condition and this must be accounted for before computing a pulse arrival time. One could go directly to a time scale (such as TDB) at the solar system barycenter, but this requires a precise knowledge of the pulsar location before the correction can be done and removes the possibility of fitting for the pulsar position as part of the timing model. Instead, in order to remove the effects of the spacecraft motion on the photon arrival times while maintaining the ability to fit for astrometric parameters in the timing model, we correct the measured times to a fictitious observatory located at the Earth's geocenter.

LAT photon times are recorded in Mission Elapsed Time (MET), which is referenced to Terrestrial Time (TT) via the MJDREF keyword in the FITS file header\footnote{See OGIP Memo OGIP/93-003 \url{http://heasarc.gsfc.nasa.gov/docs/heasarc/ofwg/docs/rates/ogip_93_003/ogip_93_003.html}}.  Time is maintained onboard the spacecraft to an accuracy of better than 1 $\mu$s using a GPS receiver \citep{DAS08}. 

The geocentric time is the satellite time corrected for geometric light travel time to the geocenter. It does not include relativistic terms in the correction. The geocentric photon time $t_\mathrm{geo}$ is defined as
\begin{equation}
t_\mathrm{geo} = t_\mathrm{obs} + \frac{\mathbf{r}_\mathrm{sat}}{c} \cdot \mathbf{\hat{n}}_\mathrm{psr},
\end{equation}
where $\mathbf{r}_\mathrm{sat}$ is the vector pointing from the geocenter to the spacecraft, $\mathbf{\hat{n}}_\mathrm{psr}$ is a unit vector pointing in the direction of the pulsar (here assumed to be at an infinite distance), and $c$ is the speed of light.

This correction is applied using the \textit{Fermi} science tool\footnote{\url{http://fermi.gsfc.nasa.gov/ssc/data/analysis/documentation/}} \texttt{gtbary} with the \texttt{tcorrect=geo} option. After this correction, the time system for the events is still TT, but all times are referenced to the geocenter.  This correction has a maximum amplitude of 23.2 ms. Therefore, an error in the assumed pulsar direction as large as 1$^\circ$ causes a maximum error in the corrected time of only 0.4 ms.

\subsection{TOA Determination}

\begin{figure}
\includegraphics[width=3.0in]{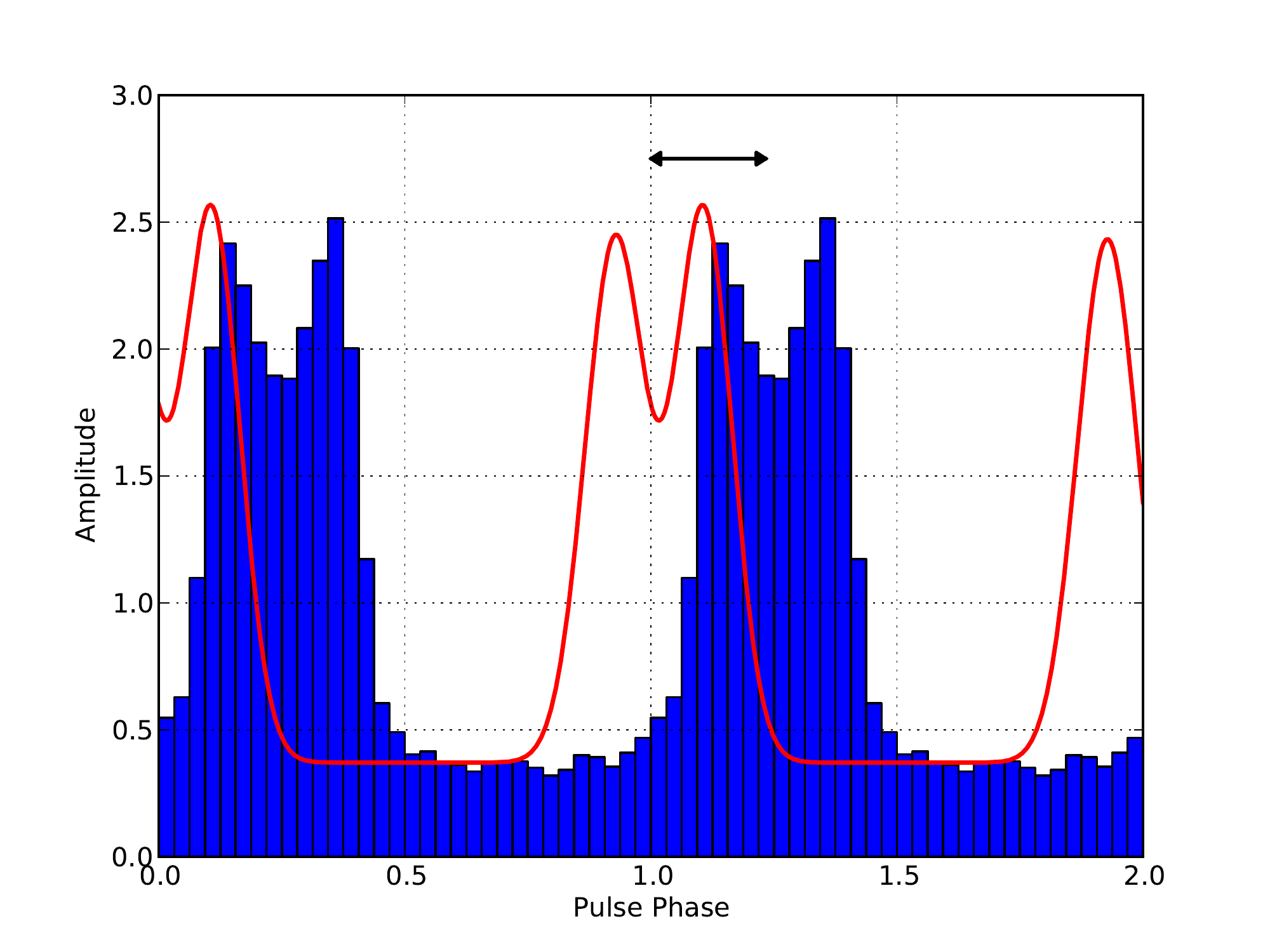}
\caption{Example of a TOA measurement. The blue histogram is a binned pulse profile generated from the observed photons (two cycles are shown for clarity). The red curve is a two Gaussian template profile, where the point at phase 0.0 (or equivalently 1.0) defines the fiducial point. The black arrow represents the measured phase offset ($\Delta$) required to align the profile with the template.  \label{fig:lctemplate}}
\end{figure}

A TOA is determined from the photon times in a segment of data by first assigning pulse phases to each photon based on an initial model, then measuring the phase offset ($\Delta$) required to align a standard template profile with the measured pulse profile (see Figure~\ref{fig:lctemplate}). This offset is then converted to a time using the pulse period ($P$) and added to the observation start time, $T_0$, to become the measured TOA.
\begin{equation}
\mathrm{TOA} = T_0 + \Delta \times P
\end{equation}
This TOA is the time when the fiducial point on the pulse profile arrived at the observatory, for a representative pulse during the observation interval. In the case of the geocentered LAT events the TOA is for a fictitious observatory at the geocenter (observatory code \texttt{coe} in \textsc{Tempo2}; \citet{hem06}). The measurement can be made using binned pulse profiles (as in Figure~\ref{fig:lctemplate}) or directly from the unbinned photon phases, as described below.

For this work, we divide the full 18-month observation interval into segments of equal duration and determine a pulse time of arrival from each segment.  The length of each segment is a balance between signal-to-noise ratio and time resolution.  Longer integrations result in better signal-to-noise ratio and smaller statistical measurement errors on each TOA.  On the other hand, shorter integrations provide finer time resolution that better samples the annual sinusoidal signal caused by the Earth's motion around the Sun and the timing noise in some very noisy young pulsars. Therefore we try to achieve at least 1 TOA per month.  Fainter pulsars that require a substantial fraction of a year, or longer, per TOA measurement will be difficult to time with the LAT.

For most radio pulsar timing, the TOAs are determined from binned data. The start time of the observation is precisely known from the observatory clock. During the observation, data are folded using \textit{predicted} phases for the pulsar based on a provisional ephemeris (e.g. using the \texttt{-polyco} option to \textsc{Tempo2}), and a binned profile for that observation is computed.  The arrival time is computed by cross-correlating the observed profile with a high signal-to-noise template profile with the same binning.  The accuracy of this measurement is improved if the cross-correlation is implemented as a fit to a linear phase gradient in the Fourier domain (an application of the Fourier shift theorem), rather than as a simple time-domain cross correlation \citep{tay92}. Finally the TOA is determined as the observation start time plus the measured phase offset (converted back into time units). 

The binned TOA determination method can also be applied to photon data, such as that from the LAT, by computing the predicted phase for each photon and building a binned pulse profile from the events.  However, since we must make TOA measurements based on a small number of detected photons (often $<100$ photons go into each TOA), we can improve the TOA determinations by using an unbinned likelihood analysis to compute the TOA directly from the set of photon phases. This has been discussed before \citep{lrc+09}, but we have developed and generalized the technique and describe it in detail here.

We use an unbinned maximum likelihood method to estimate both the light curve template and the TOAs with associated errors.  In the likelihood formulation, the template is interpreted as a periodic probability density function to observe a photon at a given phase, $f(\phi;\/ \lambda,\/ \Delta)$, with $\lambda$ some set of parameters describing the light curve morphology and $\Delta$ accounting for the phase shift between the template and the given data set. The TOA is determined by $\Delta$.  The template is normalized such that $\int{d\phi}\ f(\phi;\/ \lambda,\/ \Delta) = 1$.

We first start with a description of the template, $f(\phi;\/ \lambda,\/ \Delta)$, which must be a continuous function that can be evaluated at any value of $\phi$.  In many cases, the statistics are sufficiently limited that the pulse profile can be described as a sum of a constant background component and a small number of gaussian peaks.  That is,
\begin{equation}
f(\phi) = (1 - \sum_{i = 1}^{N_p} p_i) + \sum_{i = 1}^{N_p} p_i\ g(\phi,\/x_i,\/\sigma_i),
\end{equation}
with $N_p$ the number of gaussian peaks, $p_i$ the fraction of the total emission belonging to each peak, and $g(x,\sigma)$ a gaussian with mean $x$ and standard deviation $\sigma$.  The domain of the gaussian functions is assumed to be wrapped to $[0,1)$.  Here, $\Delta$ can be associated with $x_1$, the location of the first peak, while the remaining parameters are subsumed in $\lambda$.

With increasing statistics, the complexity of GeV light curves is no longer well-represented by a simple sum of components.  Bridge emission and peak asymmetry appears, and in general, no simple functional form is sufficient to describe the profile (e.g. the Vela pulsar \citep{Vela2}).  In this case, we prefer kernel density estimation (KDE) methods \citep{dJRS86}.  These methods result in a faithful, non-parametric representation of the light curve.  However, even for bright pulsars, the available statistics are such that KDE methods produce a template with broadened peaks and ``noisy'' valleys, neither of which is desirable for the calculation of a TOA.  A good estimator for the template should provide a smooth template (ignore fluctuations) while simultaneously preserving the structure and sharpness of the peaks, which is important because the template sharpness is a factor in the accuracy of the TOA measurements.  We outline two approaches we have found effective below.

The first forms the basis for the H-test statistic often used to assess pulsation significance in the absence of a template \citep{dJRS89}.  Coefficients of a Fourier expansion are estimated directly from the unbinned phases.  For $n$ photons, the coefficients for the $k$th harmonic are
\begin{equation}
\alpha_k = \frac{1}{n} \sum_{i=1}^{n} \cos(k\phi_i), \beta_k = \frac{1}{n} \sum_{i=1}^{n} \sin(k\phi_i)
\end{equation}
and the light curve is given by
\begin{equation}
f(\phi) = 1 + 2 \sum_{k = 1}^{m} \alpha_k \cos(k\phi) + \beta_k \sin(k\phi).
\end{equation}
The only free parameter is the overall phase of the light curve; variation can be implemented with the Fourier shift theorem or simply by adding a constant phase to the data.  The number of harmonics retained should offer an optimum balance between peak ``sharpness'' and noise in the remainder of the profile. We call this the `empirical Fourier' (EF) method.

The second method is a gaussian KDE \emph{with a phase-dependent bandwidth}, the idea being to use smaller bandwidth for the peaks while smoothing the valleys with a broader kernel.  Here, $f(\phi) = \sum_{i=1}^n g(\phi,\/\phi_i,\/\sigma_i)$, 
with $g$ again the standard gaussian.  The bandwidth is determined by $\sigma_i = (f_{\max} - f_{\min})/f(\phi_i)\sqrt{n}$.
Lest the reader worry about this circular definition, in practice we begin with a phase-independent bandwidth $\sigma=\sqrt{n}$ and iterate. As with the Fourier expansion, the only free parameter is the overall template offset.

\begin{figure}
\includegraphics[width=3.0in]{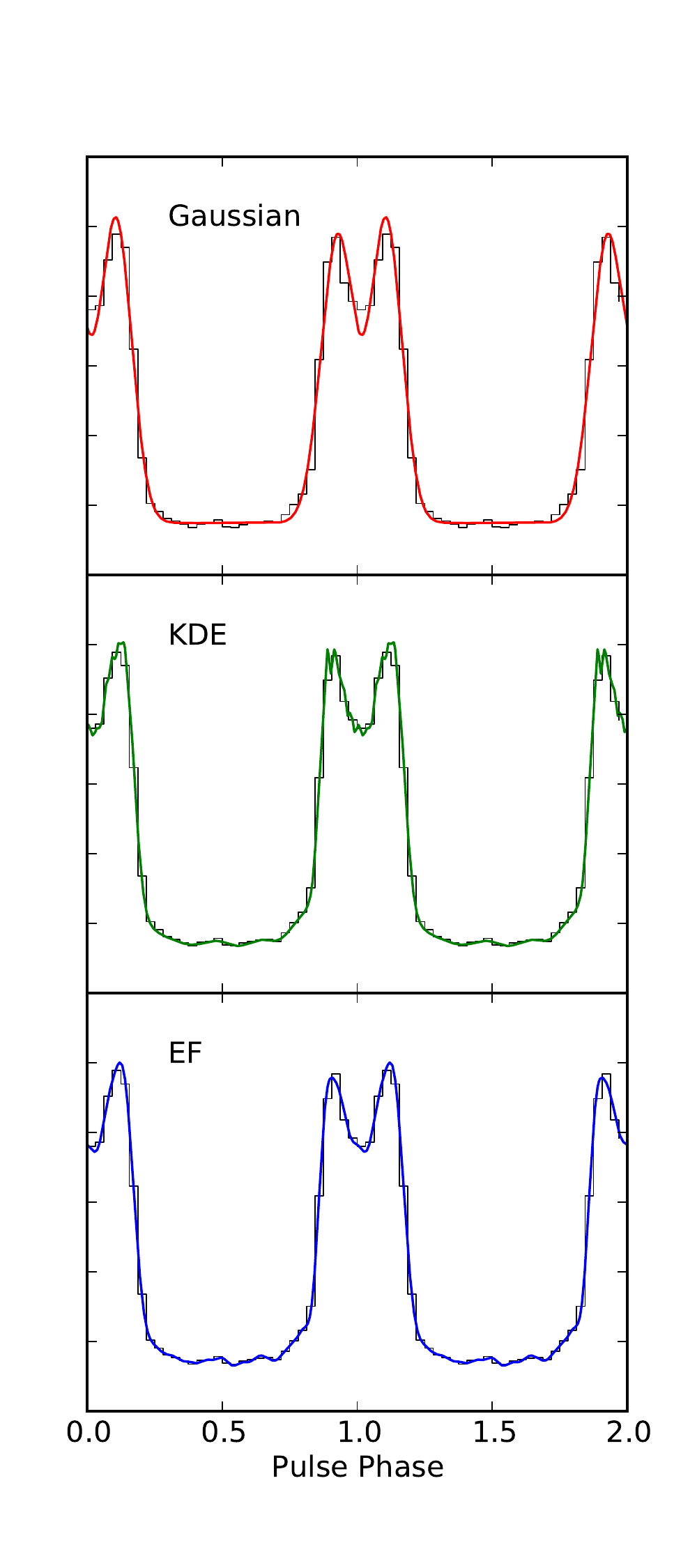}
\caption{The pulse profile of PSR J0007+7303 fitted with a 2 Gaussian, Kernel Density Estimator (KDE), and Empirical Fourier (EF) template with 16 harmonics. The black histogram shows the measured pulse profile with 32 bins, but the templates are fitted to the unbinned photon phases, as described in the text \label{fig:templates}}
\end{figure}

In summary, for our template, we choose one of the three above methods (Gaussian, EF, KDE) that produces the best results, as evidenced by the smallest RMS residuals. A comparison of a pulse profile fitted with the three different templates is shown in Figure~\ref{fig:templates}. The template choice is documented for each pulsar.

With the template defined, the next step is to fit for the TOA from each segment of data, always using the chosen template to define the pulse profile and fiducial point. For the fitting, we start with an approximate timing solution (say from a $\nu$-$\dot{\nu}$ search) and fold the photon arrival times to obtain a set of phases $\{\phi_1, \phi_2, ... , \phi_n\}$.  The probability to observe these data, given the light curve model, is formally inverted to form the log likelihood for the parameters, $\sum_{i=1}^n \log f(\lambda, \Delta; \phi_i)$.  The parameters are varied to maximize the log likelihood.  In the case of a multi-gaussian template, the full dataset is used to determine $\lambda$, while in determining TOAs, only $\Delta$ is fit.  The likelihood surface generally has a gaussian shape near the best-fit value for $\Delta$, and we estimate the error on $\Delta$ by measuring and inverting the curvature of the log likelihood function at the best-fit value.  Thus, to determine a TOA, the above fit is carried out for each subset to estimate for $\Delta$ and its error $\sigma_{\Delta}$.  


We mention here an additional challenge brought on by morphology of GeV light curves.  An appreciable fraction of pulsars observed so far present light curves with two peaks of similar height with a separation close to 0.5 periods.  These light curves are approximately invariant under a half-period translation, and statistical fluctuations may then lead to a likelihood maximum associated with the ``wrong'' peak.  When this happens, a blind search for the maximum likelihood will result in a TOA off by $0.5/\nu$ seconds which must be excluded from the timing solution fit.  To avoid loss of data, rather than employing a blind search, we ``track'' the solution.  That is, provided the trial solution is sufficiently good (and this is always the case with iteration), the drift of the actual arrival time from the predicted arrival time is much less than half of a period.  We then simply restrict the search for the likelihood maximum to within a range that excludes the ``wrong'' peak.

\subsection{Fitting Timing Models}

The measured sets of TOAs are then fitted to a timing model using the pulsar timing software \textsc{Tempo2} \citep{hem06,ehm06}. There are many parameters that can be used in the timing models.  For all pulsars we fit for pulse frequency ($\nu$) and frequency first derivative ($\dot{\nu}$) and frequency second derivative ($\ddot{\nu}$).  We fit for $\ddot{\nu}$ as a measure of the timing noise present in each pulsar.  In most cases, a significant $\ddot{\nu}$ is not detected and we report a 2 $\sigma$ upper limit on the magnitude $|\ddot{\nu}|$.  In the cases where $\ddot{\nu}$ is measured, we attribute this solely to timing noise as the $\ddot{\nu}$ expected from any reasonable braking index would be immeasurable over our 18 month data span. If this is still insufficient to whiten the residuals, we add a third frequency derivative, or harmonically related sinusoids (\texttt{WAVE} parameters in \textsc{Tempo2}, see \citet{hlk+04}) to the fit until a satisfactory model is achieved. Note that because of large covariances between the parameters, one should avoid fitting the position and WAVE parameters at the same time.  


Finally, in three cases (PSRs J0007+7303, J1124$-$5916, J1813$-$2332), a glitch was observed and several glitch parameters were added to the fit, as described in \S\ref{sec:results}.

In our models, the absolute phase 0.0 is arbitrary. In the case of pulsars with both radio and $\gamma$-ray emission the convention is usually to assign phase 0 as the peak of the radio pulse (as a proxy for the more physically meaningful point of closest approach of the magnetic axis to the line of sight). However, since we don't observe radio pulsations from most of these pulsars we have not attempted to define a particular phase 0. However, we do report the parameter TZRMJD for our models, which is the reference for phase 0.0. Phase 0.0 is the pulse phase at the time TZRMJD at the geocenter at infinite frequency.

As a final note, we want to emphasize that different timing models are appropriate for different purposes.  One of the primary goals of this work is to use the capability of pulsar timing with the LAT to make accurate localizations of $\gamma$-ray-selected pulsars, thus enabling multiwavelength studies of potential counterparts.  A secondary goal is characterizing the timing noise in this set of pulsars.  For other purposes, different models are appropriate. In particular, for many studies it would be preferable to freeze the position using an accurately known counterpart position (from \textit{Chandra} X-ray observations, for example) to reduce the number of free parameters in the model.

The \textsc{Tempo2} timing models described here will all be made available electronically at the Fermi Science Support Center (FSSC) web site\footnote{http://fermi.gsfc.nasa.gov/ssc/data/access/lat/ephems/}.

\subsection{A \textsc{Tempo2} Plugin For Assigning Photon Pulse Phases}

One important use of the timing models presented here is to be able to assign an accurate pulse phase to each photon in a LAT observation of a particular pulsar.  This is needed for studies of the $\gamma$-ray light curve, phase-resolved spectroscopy, or ``gating'' the data on the off-pulse region to blank out a pulsar to enable studies of faint sources nearby (e.g. Cyg X-3 \citep{CygX3}).  The standard \textit{Fermi} Science Tool \texttt{gtpphase} tool was developed for this application. However it suffers from the limitation that it cannot represent the full complexity of pulsar timing models that include frequency derivatives above $\ddot{\nu}$, glitches, parallax, proper motion, or \texttt{WAVE} parameters. \textsc{Tempo2}, on the other hand, allows all of these as well as several additional orbital models for pulsars in binary systems.

For these reasons, we have implemented a graphical plugin for calculating pulsar phases for \emph{Fermi}-LAT data with \textsc{Tempo2}, called \texttt{fermi\_plug.C}. This plugin takes LAT event (``FT1'') files with the photon arrival dates, spacecraft (``FT2'') files with the satellite position as a function of time, and \textsc{Tempo2} timing solutions and writes photon pulse phases in the FT1 event file. It uses the same spacecraft position interpolation algorithm as implemented in the \textit{Fermi} science tool \texttt{gtbary}\footnote{See \url{http://fermi.gsfc.nasa.gov/ssc/library/support/psr_tools_anatomy/}} and derives barycentric photon times with analogous methods. The barycentric times are then treated as TOAs to find the pulsar phases relative to the absolute phase reference given by the TZRMJD parameter in the input ephemeris. The plugin thus allows \emph{Fermi}-LAT data analysis with an ephemeris built from radio, X-ray or $\gamma$-ray TOAs, with virtually unlimited complexity in the timing model. The plugin has been shown to
reproduce the results from the Science Tools when working in \texttt{tempo1}
emulation mode, but using this mode is not required. It is available in the \textsc{Tempo2} sourceforge distribution\footnote{http://tempo2.sourceforge.net/} and from the Fermi Science Support Center\footnote{\url{http://fermi.gsfc.nasa.gov/ssc/data/analysis/user/}}. This plugin is suitable for use with any of the timing models presented here.

\section{Results}
\label{sec:results}

In this section, we present details of the timing models for each of the 17 pulsars listed in Table \ref{tab:names}. The models are determined from the data set as described above. The statistical errors on the parameters are the (single parameter 1-$\sigma$) uncertainties reported by \textsc{Tempo2} from the fits.  For the pulsars with no $\ddot{\nu}$ required in the model, we estimate the statistical error in the position fit from a fit with $\ddot{\nu}$ free, because this results in a more conservative error estimate that better accounts for the correlations between the astrometric and spin parameters.  

The errors on the position in the tables and shown in the figures are statistical only (though in the figures they are 95\% confidence, rather than 1-$\sigma$ since that is the standard practice for LAT error ellipses), and thus underestimate the true error on the position determinations.  This is predominately because our span include only 1.5 periods of the annual sinusoid induced by an error in the position.  Timing instabilities present on a similar timescale can thus perturb the fitted position.  In addition, there are strong covariances between the astrometric and spin parameters that mean that the parameters are not actually determined to as high a precision as indicated by the 1-parameter statistical errors.  Because knowing the true positional uncertainties is very important for counterpart searches at other wavelengths, we have tried to quantify the magnitude of this effect by Monte Carlo simulation.  We make the assumption that over the short span of data we have, the measured values of $\ddot{\nu}$ and $\dddot{\nu}$ are dominated by timing noise, not the secular spindown of the pulsar.  To estimate the magnitude of the systematic error, we generate many simulated sets of TOAs, each one using the measured timing parameters for the pulsar, with the exception of $\ddot{\nu}$ and $\dddot{\nu}$.  For those two parameters, we replace them with normally-distributed random values with mean zero and standard deviation equal to the measured value, or the upper limit in the case where $\ddot{\nu}$ is not significantly detected.  For $\nu$ and $\dot{\nu}$ we use random values distributed around the measured value with the measured uncertainty.  Each trial set of TOAs is then fit with a 1-year sine wave plus a polynomial up to order $\ddot{\nu}$ (see Figure~\ref{fig:simsys}) and the magnitude of the sine wave is converted to a position offset (see Appendix \ref{app:pos}).  We then compute the position uncertainty by finding the position offset that 68\% of the trials are lower than.  It is important to note that our simulations include random, uncorrelated, measurement errors as appropriate for the particular pulsar, so these position error estimates are of the \textit{total} uncertainty, including both systematic and statistical components.  Also, the fidelity of the estimates depends on how well our random polynomial model of the timing noise describes the actual situation, which is not well understood and may vary from pulsar to pulsar.  Therefore, these estimates should be considered indicative of the magnitude of the position error, but not precise bounds on the systematic errors. Finally, in this analysis, we just consider the total position offset, so it yields an intermediate value in the cases where the position error region from the timing is highly elliptical.

\begin{figure}
\includegraphics[width=3.0in]{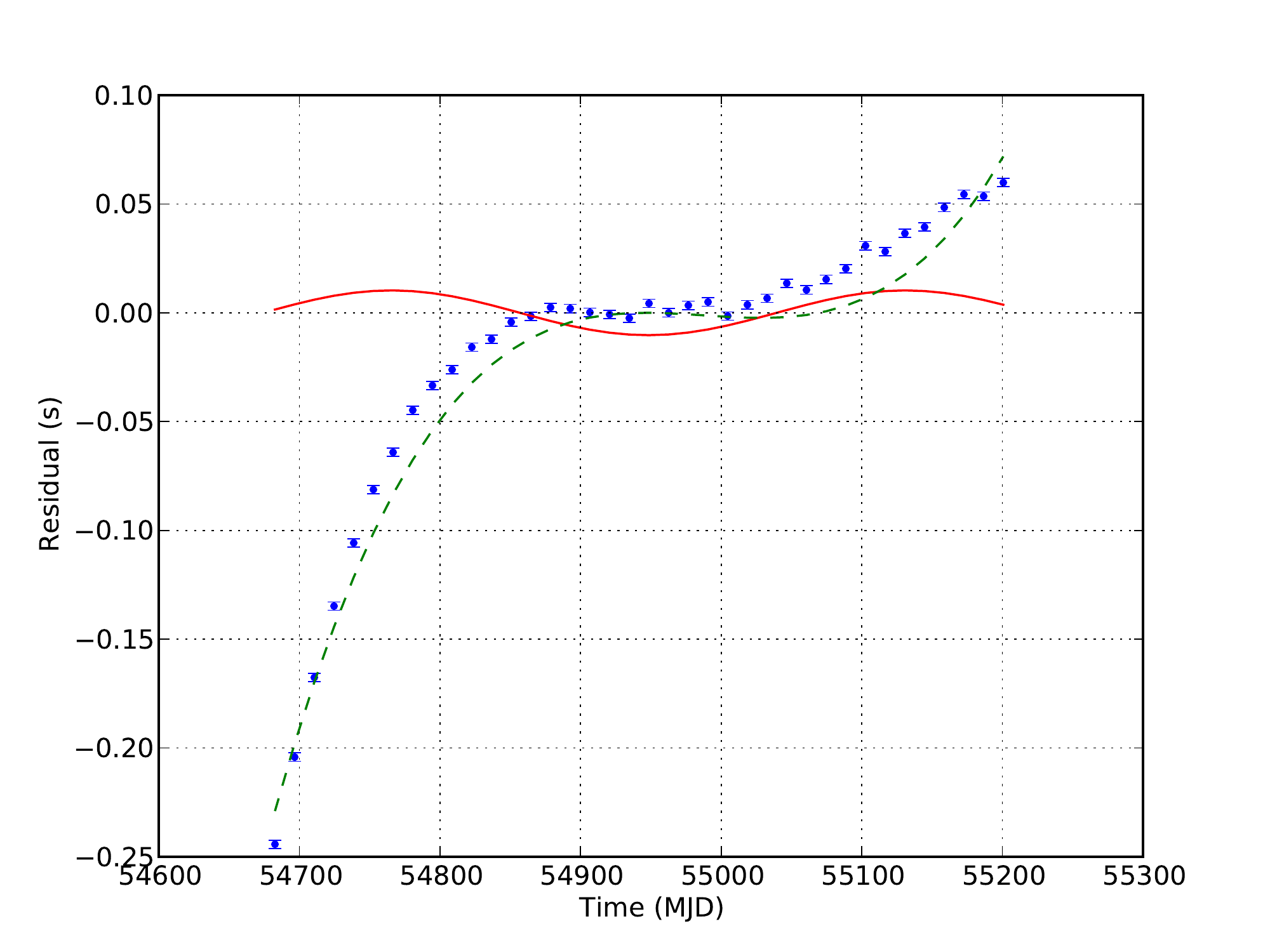}
\caption{Example of a Monte Carlo trial to estimate the systematic error on the timing position. The blue points with error bars are the simulated TOAs, which are fitted to the sum of a 1-year sinusoid (solid red line) and a third order polynomial (dashed green line), as described in the text.\label{fig:simsys}}
\end{figure}
For all timing models we use the JPL DE405 planetary ephemeris \citep{de405}. All reported frequencies and epochs are referenced to the TDB time system \citep{s92} as has been the standard for pulsar work\footnote{Note that the default time system for \textsc{Tempo2} is TCB, but we override this default and use TDB as the time units for our models.}. The clock correction procedure is TT(TAI) and all fits are made with weighting by the TOA error estimates enabled (\texttt{MODE 1}). The reference time (TZRMJD) is for the geocenter at infinite frequency. The validity range for each model is included in the tables. Care should be taken when attempting to use these models outside of that range.  In particular, models that include significant $\ddot{\nu}$ or higher order derivatives, or \texttt{WAVE} parameters, will extrapolate very poorly outside the fit range, since timing noise is a stochastic process and neither of those parameterizations reflect a physical model of the process. 

In the following subsections, we discuss some of the main results from the timing of the individual pulsars. 
For each pulsar, we present the timing model in a table, the post-fit timing residuals, a 2-D phaseogram, and a pulse profile.  In all cases, the optimized data selections used for the pulsar timing are also used to construct the 2-D phaseogram and pulse profile figures.
The phaseograms are raw photon counts and are not exposure-corrected, so the apparent variations in brightness that can be seen are from exposure variations resulting from the $\sim 55$-day precession period of the spacecraft orbit, the change in rocking angle during the mission, spacecraft reboots, or automatic repoints in response to $\gamma$-ray bursts.  The fluxes of $\gamma$-ray pulsars is expected to be constant on time scales of days to months.

In the discovery paper \citep{LATBlindSearch}, five of the pulsars were given names of the form JHHMM+DD because the declinations were not known with sufficient precision to justify a name of the form JHHMM+DDMM.  In all five cases, we now know the position well enough to confidently add the additional precision to the names, as shown in Table~\ref{tab:names}.  Also, in five cases (PSRs J1418$-$6058, J1741$-$2054, J1809$-$2332, J1813$-$1246, and J1958+2846) the current best-fit timing position would result in a different last two digits of the declination than given in the discovery paper, although in several cases we know the name to be correct based on the X-ray counterpart position. In all cases, we follow the IAU preference for not changing a source name once it is given and we continue to use the original names, except where we have only added precision, as described above. See the sections on each individual source below for a discussion of the confidence in the previously proposed counterpart associations.

\begin{deluxetable}{llcr}
\tablewidth{0pt}
\tablecaption{Pulsars Timed with the Fermi LAT\label{tab:names}}
\tablehead{
\colhead{Name} & \colhead{Prev. Name} & \colhead{Period} & \colhead{$\dot{E}$} \\
\colhead{} & \colhead{} & \colhead{(ms)} & \colhead{($10^{34}$ erg s$^{-1}$)}
}
\startdata
J0007+7303 &	\nodata  & 315.9 & 45.2 \\
J0357+3205   & J0357+32  & 444.1 & 0.6 \\
J0633+0632 &	 \nodata & 297.4 & 11.9 \\
J1124$-$5916 & \nodata & 135.5 & 1195.0 \\
J1418$-$6058 & \nodata & 110.6 & 494.8 \\
J1459$-$6053 & J1459$-$60 & 103.2 & 90.9 \\
J1732$-$3131 & J1732$-$31 & 196.5 & 14.5 \\
J1741$-$2054 & \nodata & 413.7 & 0.9 \\
J1809$-$2332 & \nodata & 146.8 & 42.9 \\
J1813$-$1246 & \nodata & \phn48.1 & 624.1 \\
J1826$-$1256 & \nodata & 110.2 & 358.0 \\
J1836+5925 & \nodata & 173.3 & 1.1 \\
J1907+0602 & J1907+06 & 106.6 & 282.7 \\
J1958+2846 & \nodata & 290.0 & 34.2 \\
J2021+4026 & \nodata & 265.3 & 11.6 \\
J2032+4127 & \nodata & 143.2 & 27.3 \\
J2238+5903 & J2238+59 & 162.7 & 88.9 \\
\enddata
\end{deluxetable}

\subsection{PSR J0007+7303}

\label{sec:cta1}

\begin{figure}
\includegraphics[width=3.5in]{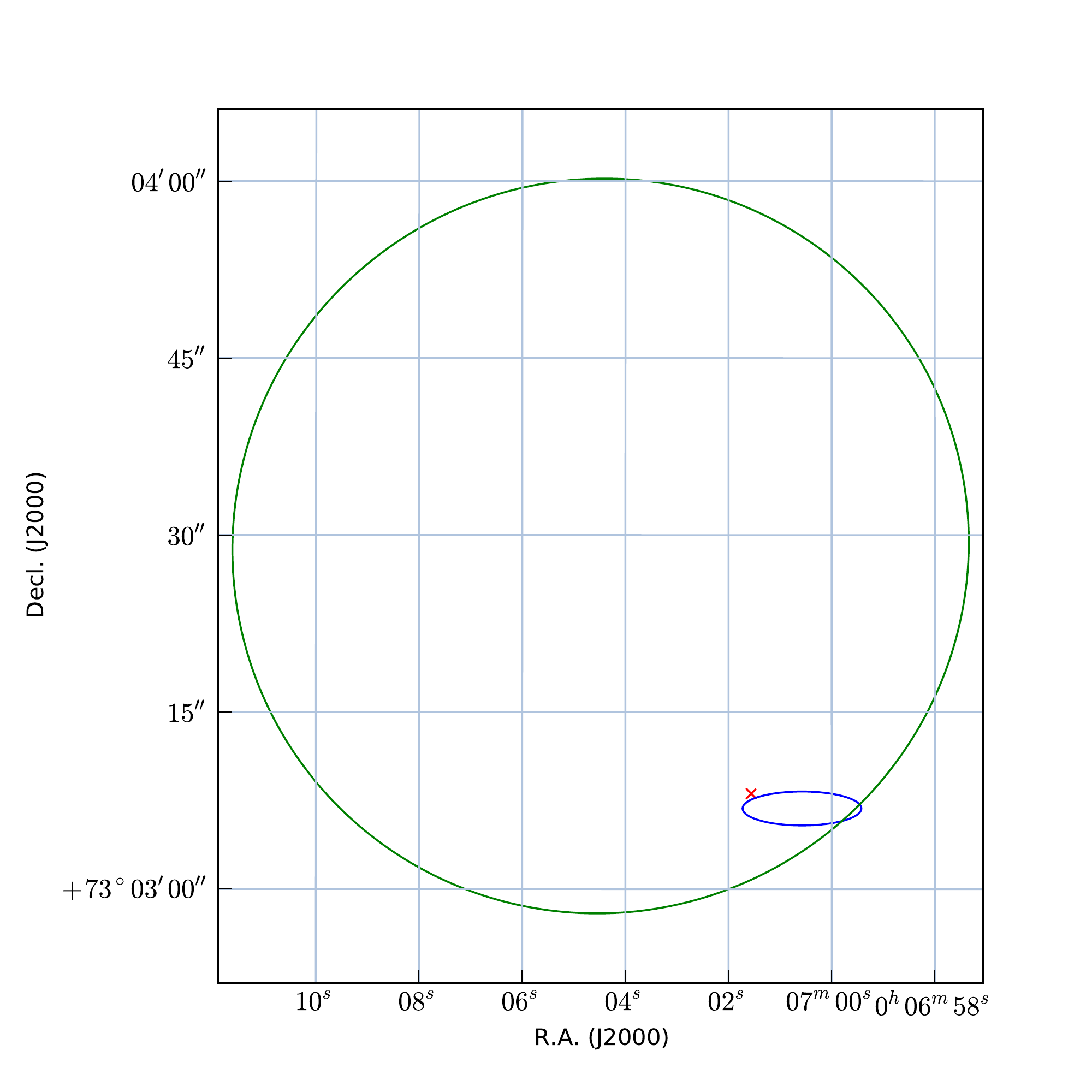} 
\caption{Timing position for PSR J0007+7303 (blue ellipse). The large green ellipse is the LAT 95\% confidence localization of 1FGL J0007.0+7303, based on 18 months of data. The red X is the \textit{Chandra} position of RX J0007.0+7303, which is 4.4\arcsec\ from the best timing position \citep{hgc+04}. \label{pos:0007} } 
\end{figure}

\begin{figure}
\includegraphics[width=3.0in]{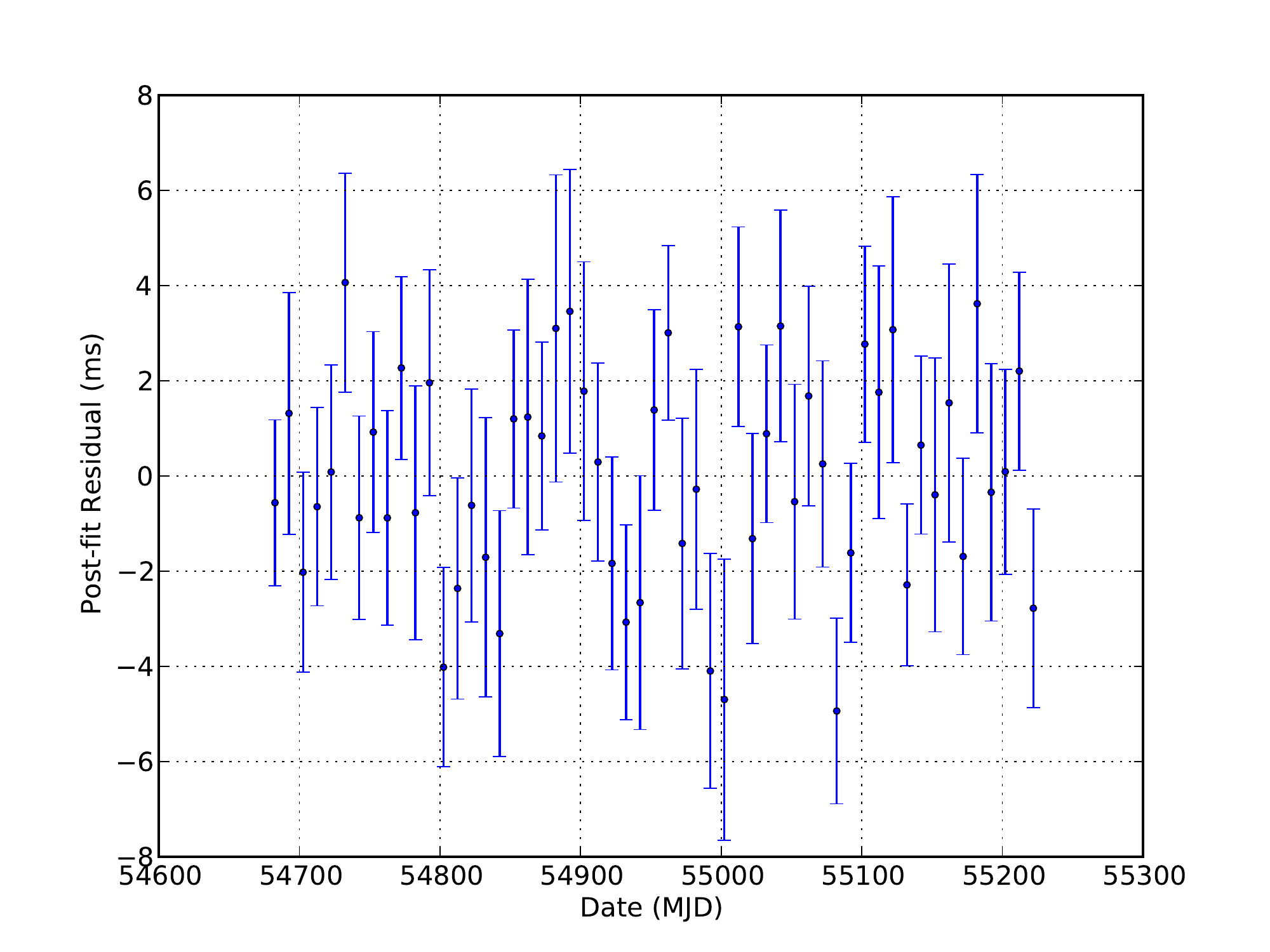}
\caption{Post-fit timing residuals for PSR J0007+7303.\label{resid:0007}}
\end{figure}

\begin{figure}
\includegraphics[width=3.0in]{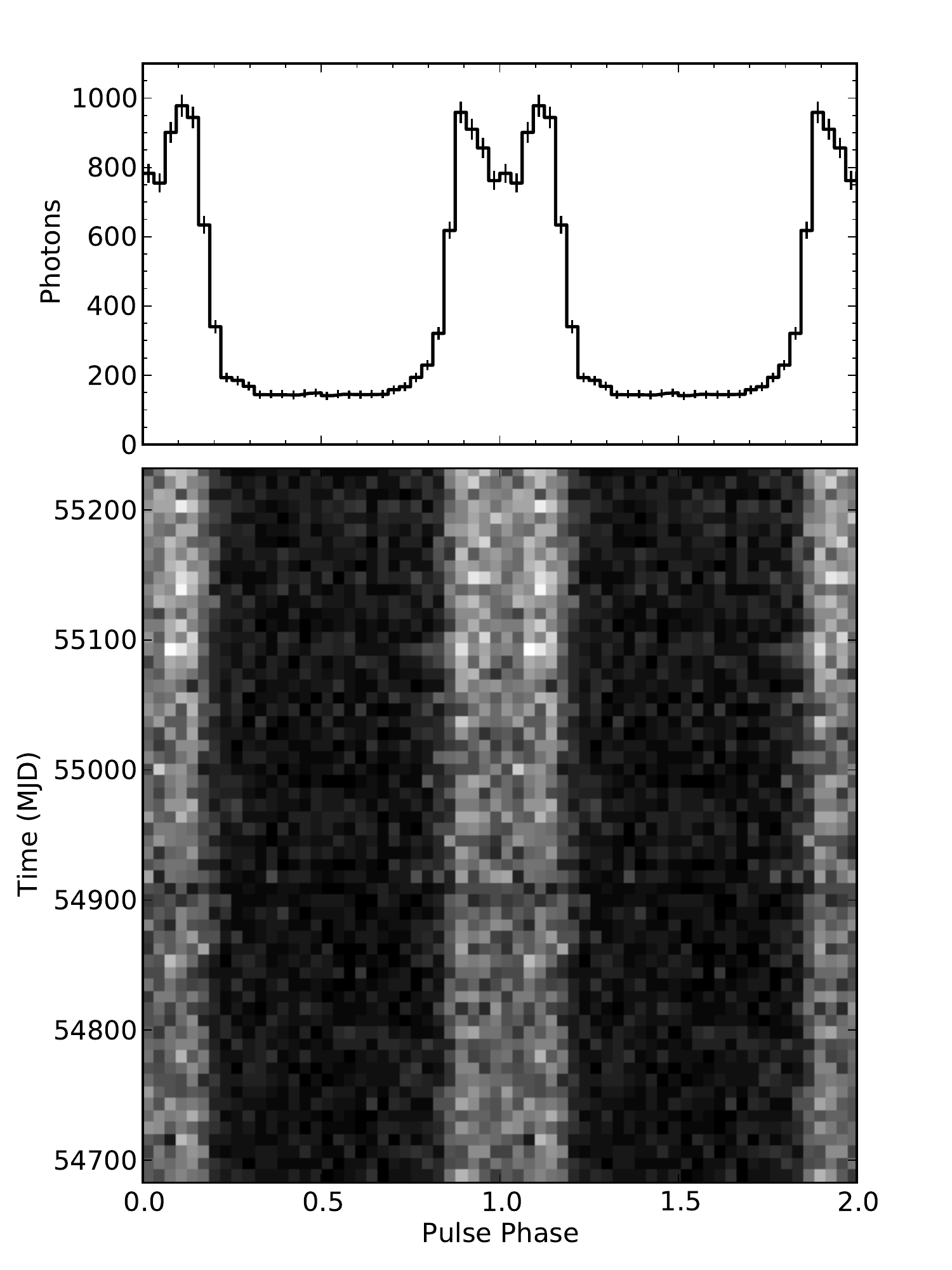}
\caption{2-D phaseogram and pulse profile of PSR J0007+7303.  Two rotations are shown on the X-axis. The photons were selected according to the ROI and $E_\mathrm{min}$ in Table~\ref{tab:0007}. In this and the following phaseogram plots, the grey scale is the number of photons in each phase/time bin, without any correction for exposure, so apparent brightness changes are caused by the precession period of the \textit{Fermi} satellite, interruptions in science operations, or from operational changes in the rocking pattern in sky survey mode. The fiducial point corresponding to TZRMJD is phase 0.0.  This and all following pulse profiles are constructed with 32 bins across the pulse period.\label{phaseogram:0007}}
\end{figure}

The timing model parameters for this pulsar are displayed in Table~\ref{tab:0007} and the timing position determination, post-fit residuals, 2-D phaseogram, and folded pulse profile for this pulsar are shown in Figures \ref{pos:0007}, \ref{resid:0007}, and \ref{phaseogram:0007}, respectively.
 
This was the first pulsar discovered in a blind search of $\gamma$-ray data \citep{CTA1} and is believed to be the pulsar powering the compact PWN RX J0007.0+7303 near the center of the shell-type supernova remnant CTA1. As seen in Figure \ref{pos:0007}, our timing position provides independent confirmation of that conclusion.  In addition, we have detected a glitch in this pulsar on 2009 May 1 with a magnitude $\Delta \nu/\nu = 5.53(1) \times 10^{-7}$, a typical glitch magnitude for a pulsar of this age. When we fit for position in the timing model, the glitch can be fully accounted for by a simple $\Delta \nu$ at the time of the glitch.  However, when we hold the position fixed at the \textit{Chandra} position of the point source (00:07:01.56, 73:03:08.3; see \citet{hgc+04}) we find that an additional parameter is required.  This can be modeled as a change in the frequency first derivative at the glitch of $\Delta \dot{\nu}/\dot{\nu}$ of $0.0010(2)$.  It is important to note that $\ddot{\nu}$ and the glitch $\Delta\dot{\nu}$ are highly covariant and additional data will likely be required to determine whether timing noise or a frequency derivative change at the glitch are the correct model for the observed behavior.   The properties of this source and the glitch will be discussed in more detail in a future paper (Abdo et al. 2011, in prep).

\subsection{PSR J0357+3205}

\begin{figure}
\includegraphics[width=3.5in]{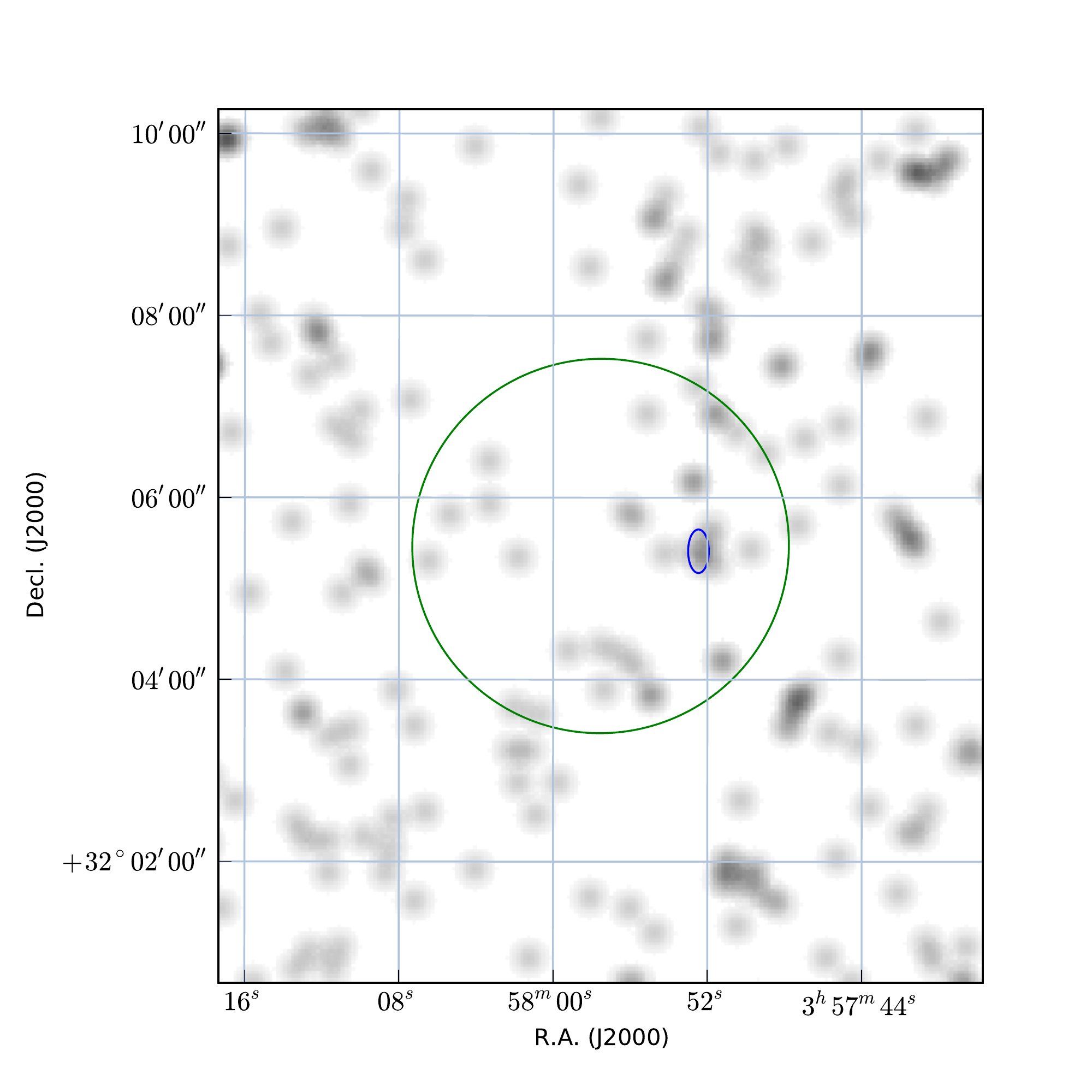} 
\caption{Timing position for PSR J0357+3205 (blue ellipse). The large green ellipse is the LAT position of 1FGL J0357.8+3205, based on 18 months of data. The background 0.2--10 keV X-ray image is from a 2.6 ks \textit{Swift} observation (ObsID 00031299001), smoothed with a gaussian with $\sigma = 7$\arcsec.  \label{pos:0357} }
\end{figure}

\begin{figure}
\includegraphics[width=3.0in]{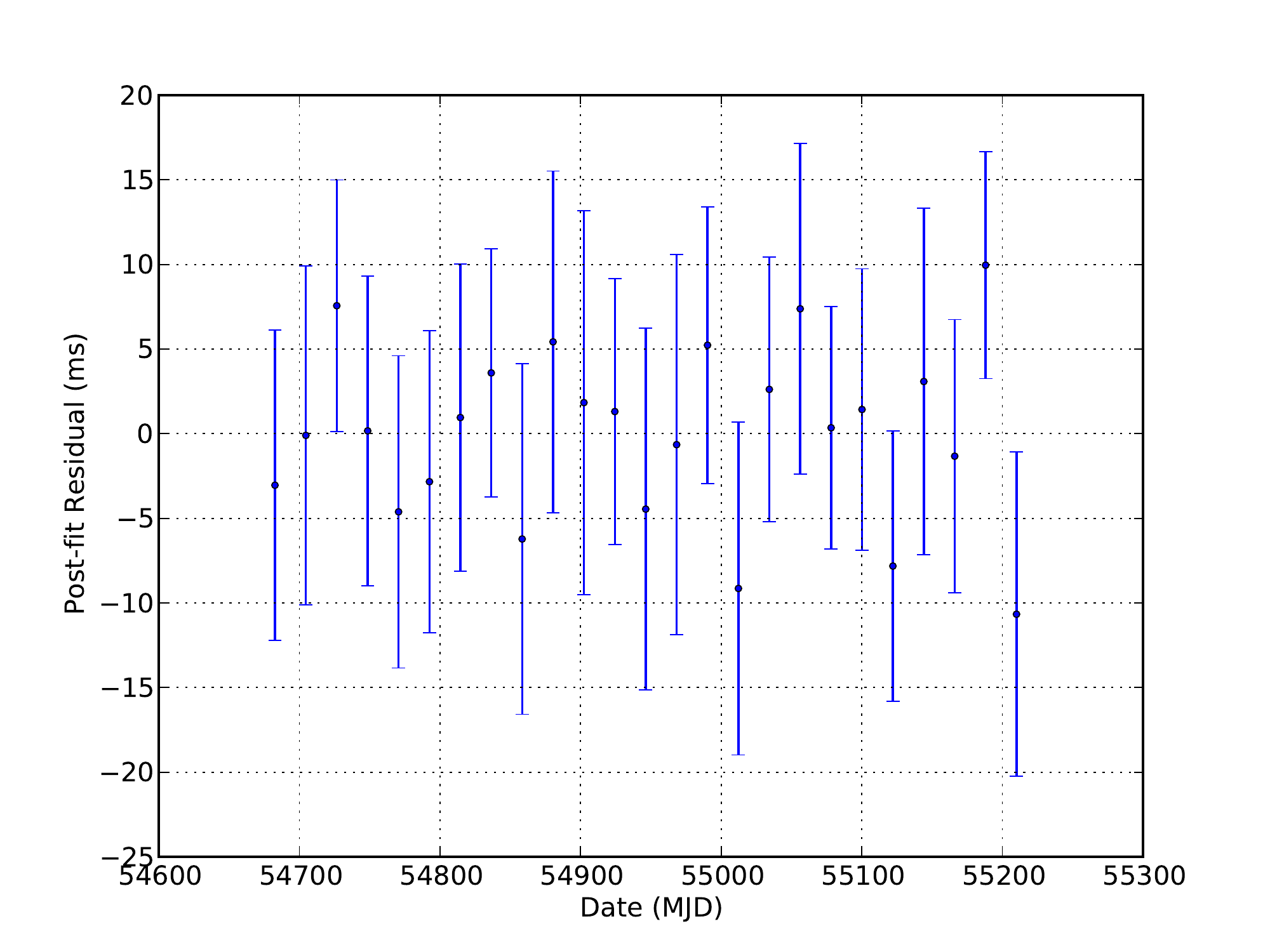}
\caption{Post-fit timing residuals for PSR J0357+3205.\label{resid:0357}}
\end{figure}

\begin{figure}
\includegraphics[width=3.0in]{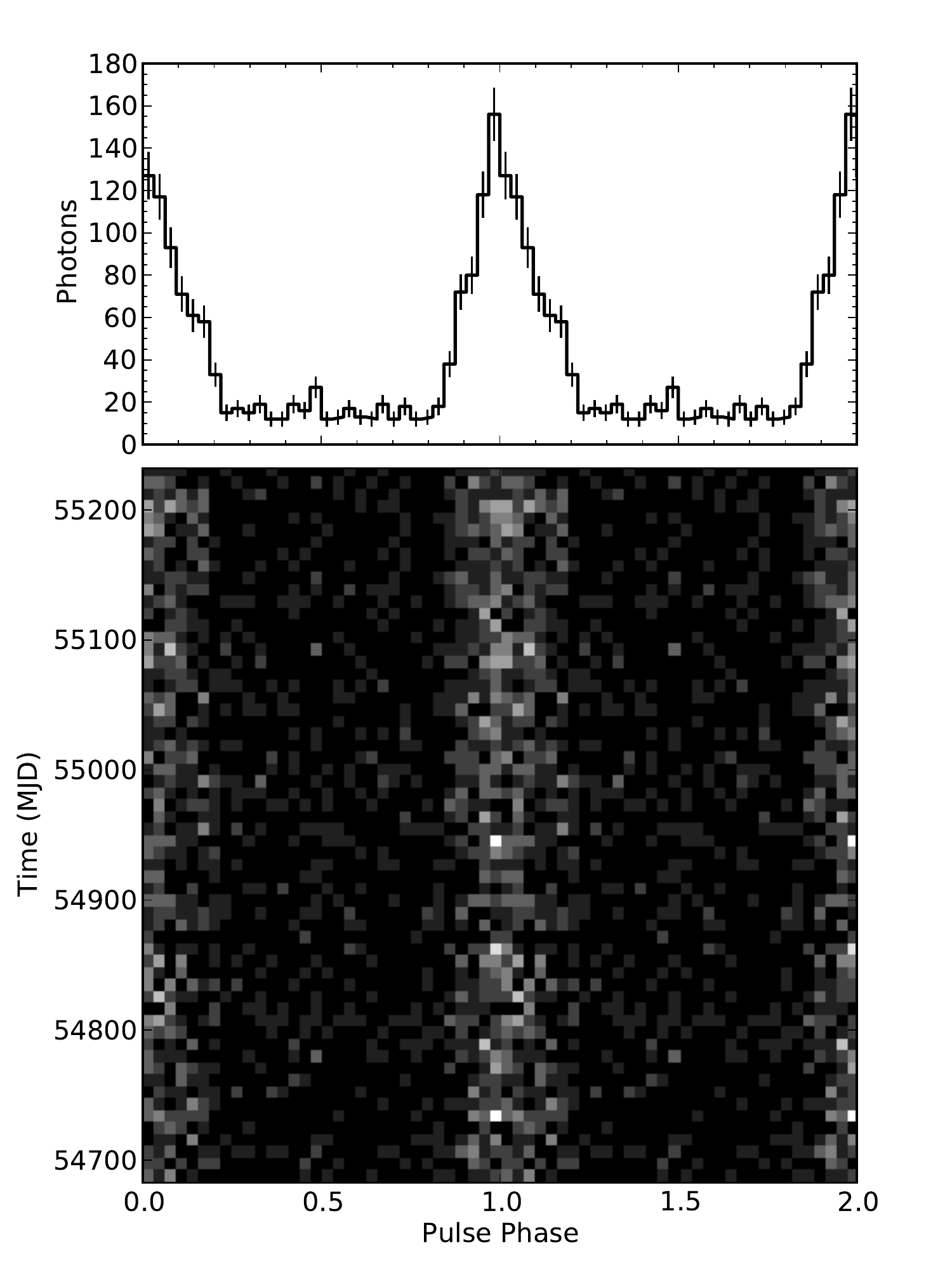}
\caption{2-D phaseogram and pulse profile of PSR J0357+3205.  Two rotations are shown on the X-axis. The photons were selected according to the ROI and $E_\mathrm{min}$ in Table~\ref{tab:0357}. The fiducial point corresponding to TZRMJD is phase 0.0.\label{phaseogram:0357}}
\end{figure}

The timing model parameters for this pulsar are displayed in Table~\ref{tab:0357} and the timing position determination, post-fit residuals, 2-D phaseogram, and folded pulse profile for this pulsar are shown in Figures \ref{pos:0357}, \ref{resid:0357}, and \ref{phaseogram:0357}, respectively.

PSR J0357+3205 is the slowest spin period (444 ms), and lowest $\dot{E}$ ($5.8\times 10^{33}$ erg s$^{-1}$) pulsar in our sample. In the discovery paper \citep{LATBlindSearch}, it was flagged as having a potentially large systematic error in the $\dot{\nu}$ and the parameters derived from it, because of the uncertain position.  The long period, low count rate, and relatively broad pulse profile still limit the timing precision to an RMS of 5.3~ms, but nevertheless the frequency derivative is now determined to an accuracy of $\sim 0.2$ percent.

For this low $\dot{E}$, the distance is constrained to be $<870$ pc, assuming the flux correction factor $f_\Omega = 1$ \citep{wrwj09} and using the LAT $\gamma$-ray flux ($G_{100}$) from \citet{PulsarCatalog} to keep the $\gamma$-ray efficiency $<1$.  As seen in Figure \ref{pos:0357}, no X-ray counterpart is apparent in a \textit{Swift} image of the region, which is not surprising in such a shallow exposure.  However, as the pulsar is at such a small distance, this is a promising target for deeper \textit{XMM-Newton} or \textit{Chandra} follow up. Using the \textsc{Tempo2} simulation capability, we predict a 1-$\sigma$ uncertainty on the timing position of 2\arcsec\ after 5 years of observation.

\subsection{PSR J0633+0632}

\begin{figure}
\includegraphics[width=3.5in]{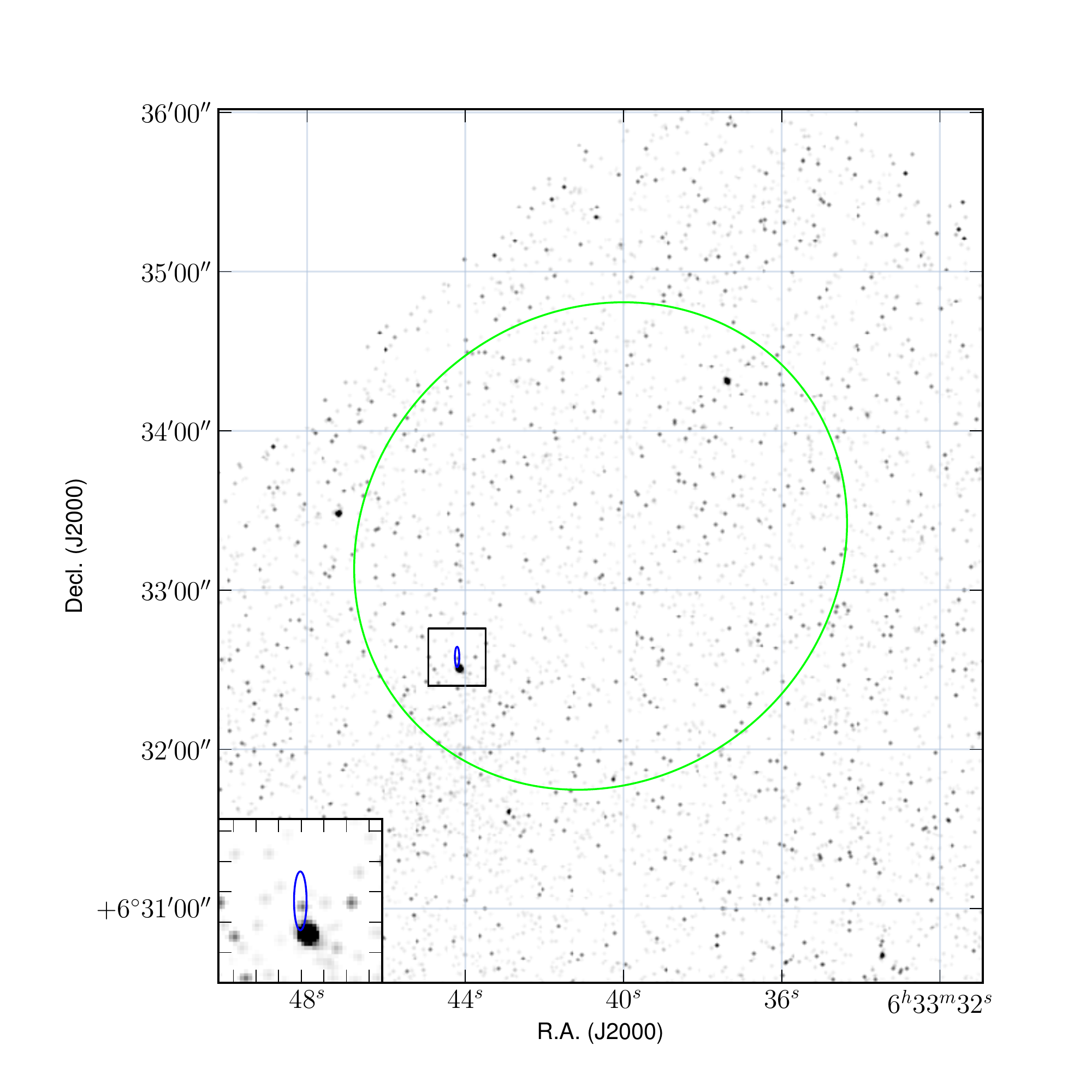} 
\caption{Timing position for PSR J0633+0632 (blue ellipse). The large ellipse is the LAT position of 1FGL J0633.7+0632, based on 18 months of data. The background 0.5--8 keV X-ray image is a 20 ks \textit{Chandra} ACIS-S image (ObsID 11123), smoothed with a gaussian with $\sigma = 0.5$\arcsec. The inset shows a 10\arcsec\  region around the timing location. \label{pos:0633} }
\end{figure}

\begin{figure}
\includegraphics[width=3.0in]{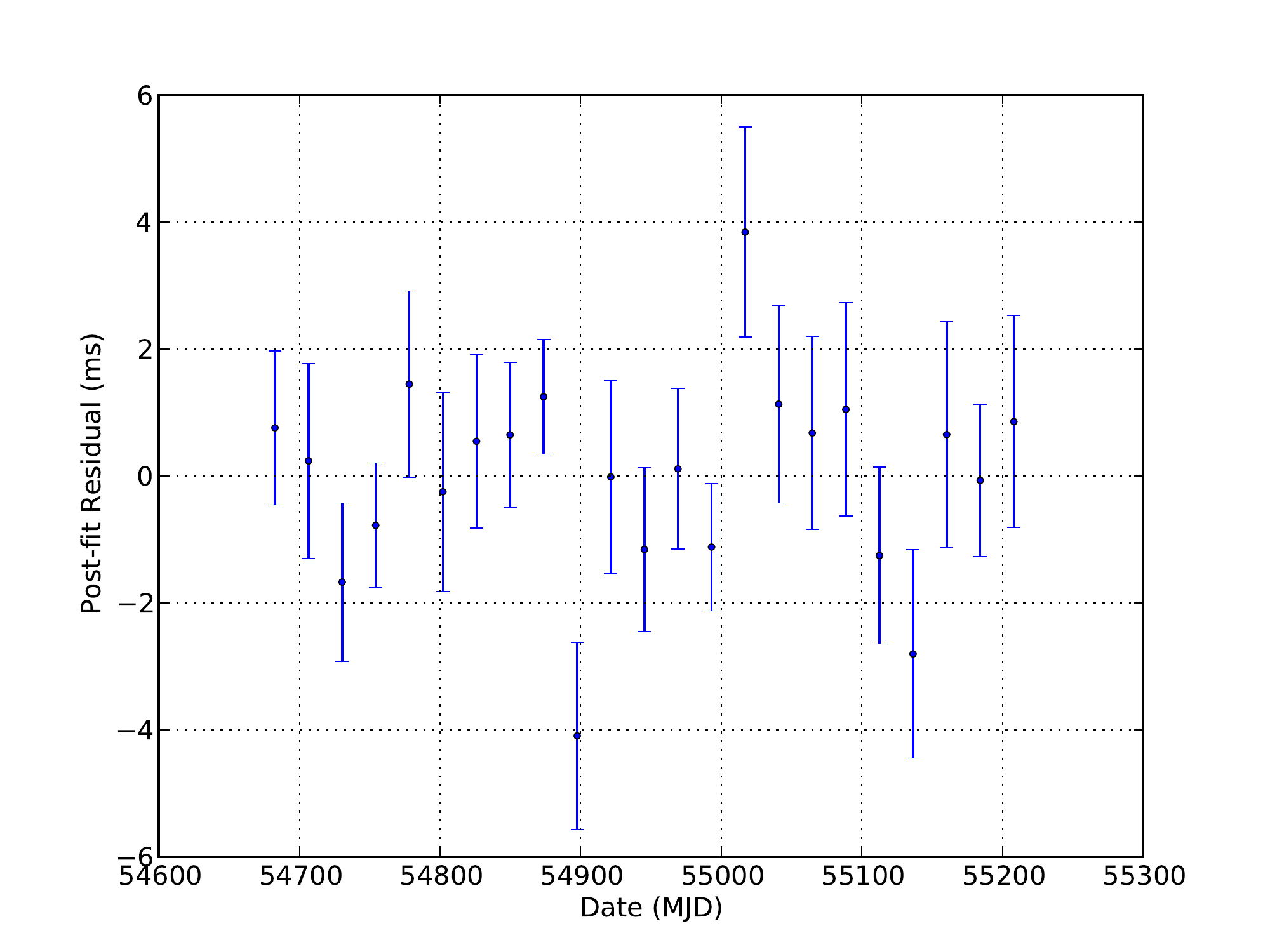}
\caption{Post-fit timing residuals for PSR J0633+0632.\label{resid:0633}}
\end{figure}

\begin{figure}
\includegraphics[width=3.0in]{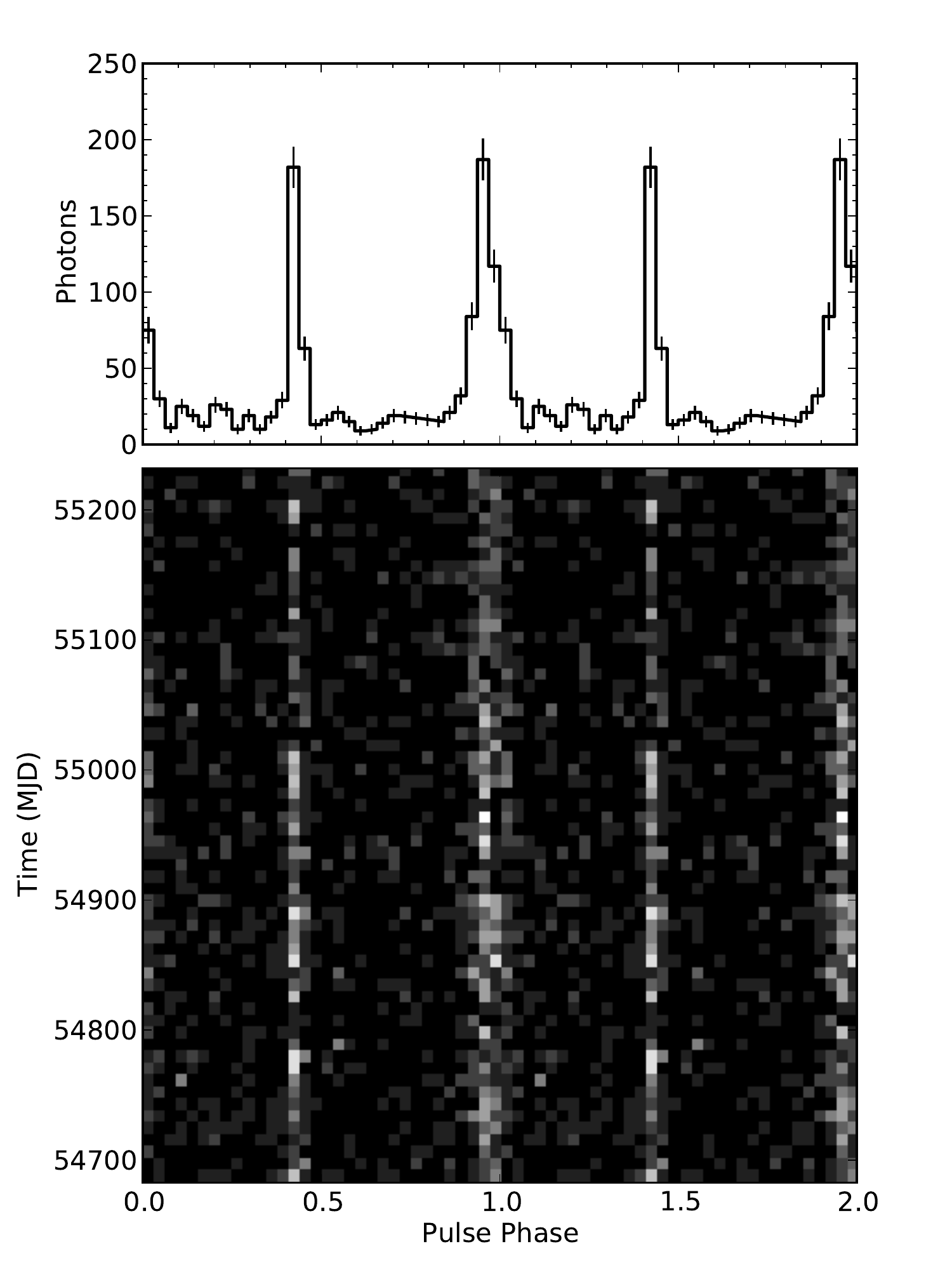}
\caption{2-D phaseogram and pulse profile of PSR J0633+0632.  Two rotations are shown on the X-axis. The photons were selected according to the ROI and $E_\mathrm{min}$ in Table~\ref{tab:0633}. The fiducial point corresponding to TZRMJD is phase 0.0.\label{phaseogram:0633}}
\end{figure}

The timing model parameters for this pulsar are displayed in Table~\ref{tab:0633} and the timing position determination, post-fit residuals, 2-D phaseogram, and folded pulse profile for this pulsar are shown in Figures \ref{pos:0633}, \ref{resid:0633}, and \ref{phaseogram:0633}, respectively.

This pulsar is also rather slow (297 ms) and faint (only 815 photons detected per year), but the timing is still quite good (RMS = $1.4$ ms), as a result of the very narrow pulses. The timing localization is close to the X-ray point source Swift J063343.8+063223 which was proposed as the counterpart by \citet{LATBlindSearch}.

To further study the X-ray counterpart, we obtained a 20 ks Chandra ACIS-S image of the region on 2009 December 11 (ObsID 11123).  
The X-ray point source counterpart to the pulsar is clearly visible in Figure \ref{pos:0633}.  We measure a position of 06:33:44.142, +06:32:30.40, which is 4.6\arcsec\ from the best fit timing position. To fit the spectrum of this source, 
we analyzed the data using CIAO version 4.3 with the latest calibrations (CALDB 4.1.1) applying the standard particle background subtraction and exposure correction.  We extracted 326 photons from a 3.5 pixel extraction 
region around the source location (for a count rate of $1.63\times10^{-2}$ cts s$^{-1}$. We see no evidence for a compact (arcsecond-scale) PWN in the immediate vicinity of the point source. To fit the spectrum, 
we found that an absorbed blackbody + powerlaw model is required.  We obtain the following parameters from our fits, with 90\% confidence error estimates: $n_H = 0.15^{+0.16}_{-0.10} \times 10^{22}$ cm$^{-2}$, $kT = 0.11^{+0.03}_{-0.02}$ keV, $\Gamma = 1.5\pm 0.6$. This model yields a 0.5--8 keV flux estimate of $9.2^{+1.8}_{-1.2}\times10^{-14}$ erg cm$^{-2}$ s$^{-1}$.  If we instead fit an absorbed neutron star atmosphere (\texttt{nsa}; \citet{zps96}) plus powerlaw model, we find a somewhat higher $n_H$ of $0.24^{+0.12}_{-0.21} \times 10^{22}$ cm$^{-2}$, a lower temperature of $kT = 0.048^{+0.019}_{-0.016}$ keV, and a similar photon index $\Gamma = 1.39^{+0.6}_{-0.3}$.

\begin{figure}
\includegraphics[width=3.5in]{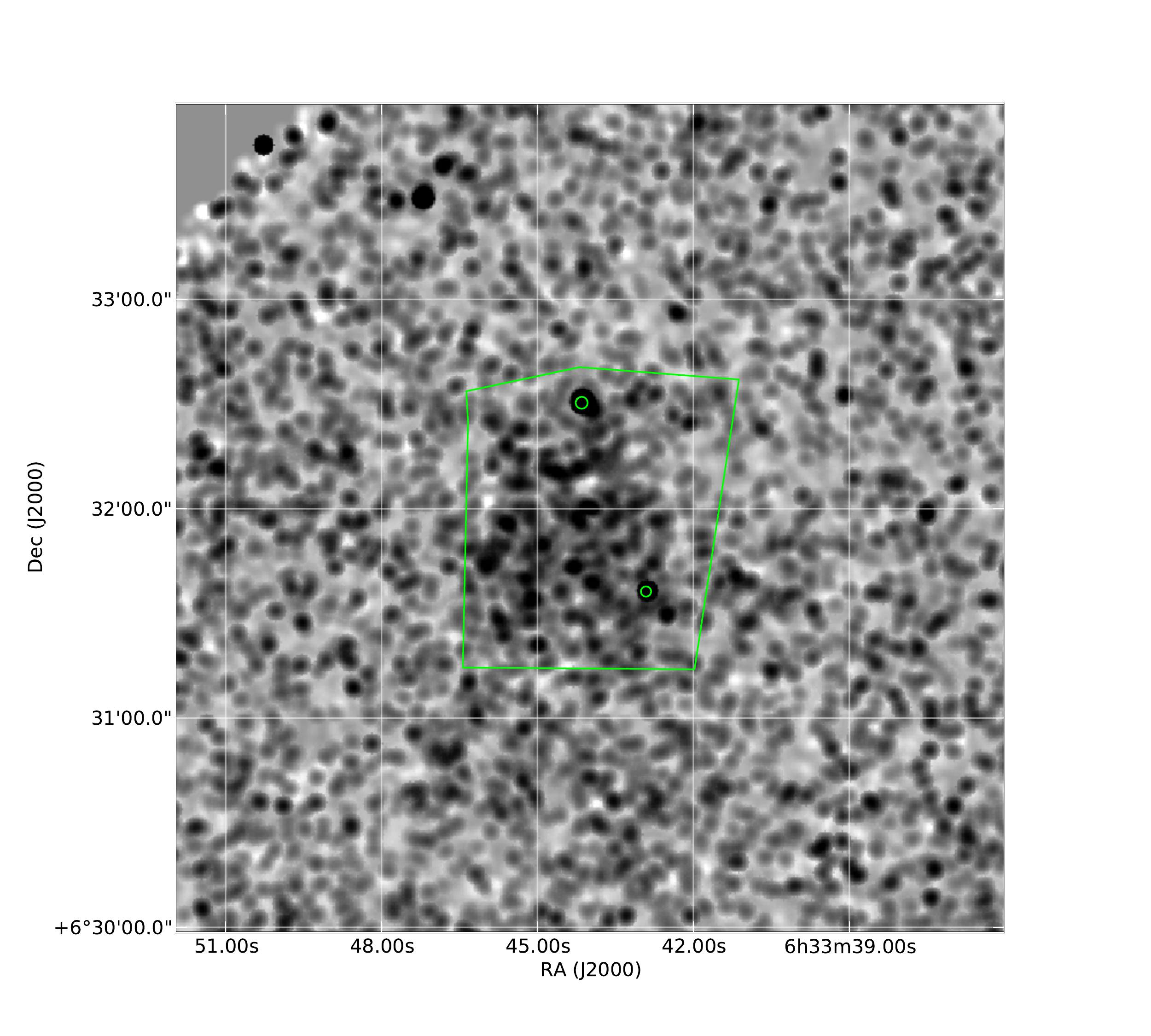}
\caption{Chandra 0.5--7 keV ACIS-S image of PSR J0633+0632, smoothed with a gaussian of 3 pixels (at a scale of 0.5\arcsec\ per pixel) to highlight the extended PWN emission. The extraction region used for the PWN spectral analysis is shown, where the two point sources are excluded from the region. The pulsar is the northernmost of the two point sources in the region.\label{fig:pwn0633}}
\end{figure}

To look for larger-scale extended emission, we smoothed the Chandra image with a gaussian kernel of 1.5\arcsec\ width (see Figure~\ref{fig:pwn0633}) and find a faint X-ray nebula extending about an arcminute south of the pulsar. In the region of the PWN (as shown in the Figure~\ref{fig:pwn0633}), we find an excess of 738 counts on a background of about 1600 counts and have fit the integrated spectrum with an absorbed powerlaw model.  With all parameters free, we find $n_H = 0.1^{+0.3}_{-0.1} \times 10^{22}$ cm$^{-2}$, $\Gamma = 0.9_{-0.4}^{+0.5}$ for a flux in the 0.5--8 keV band of $2.2\pm 0.5\times 10^{-13}$ erg cm$^{-2}$ s$^{-1}$, where the error regions are at the 90\% confidence level.  If instead, we freeze $n_H$ at $0.154\times10^{22}$ cm$^{-2}$, as found in the blackbody + powerlaw spectral fits of the point source, we find a 90\% confidence range for the photon spectral index $\Gamma$ of 0.74--1.29.

\subsection{PSR J1124$-$5916}
\label{sec:1124}

\begin{figure}
\includegraphics[width=3.5in]{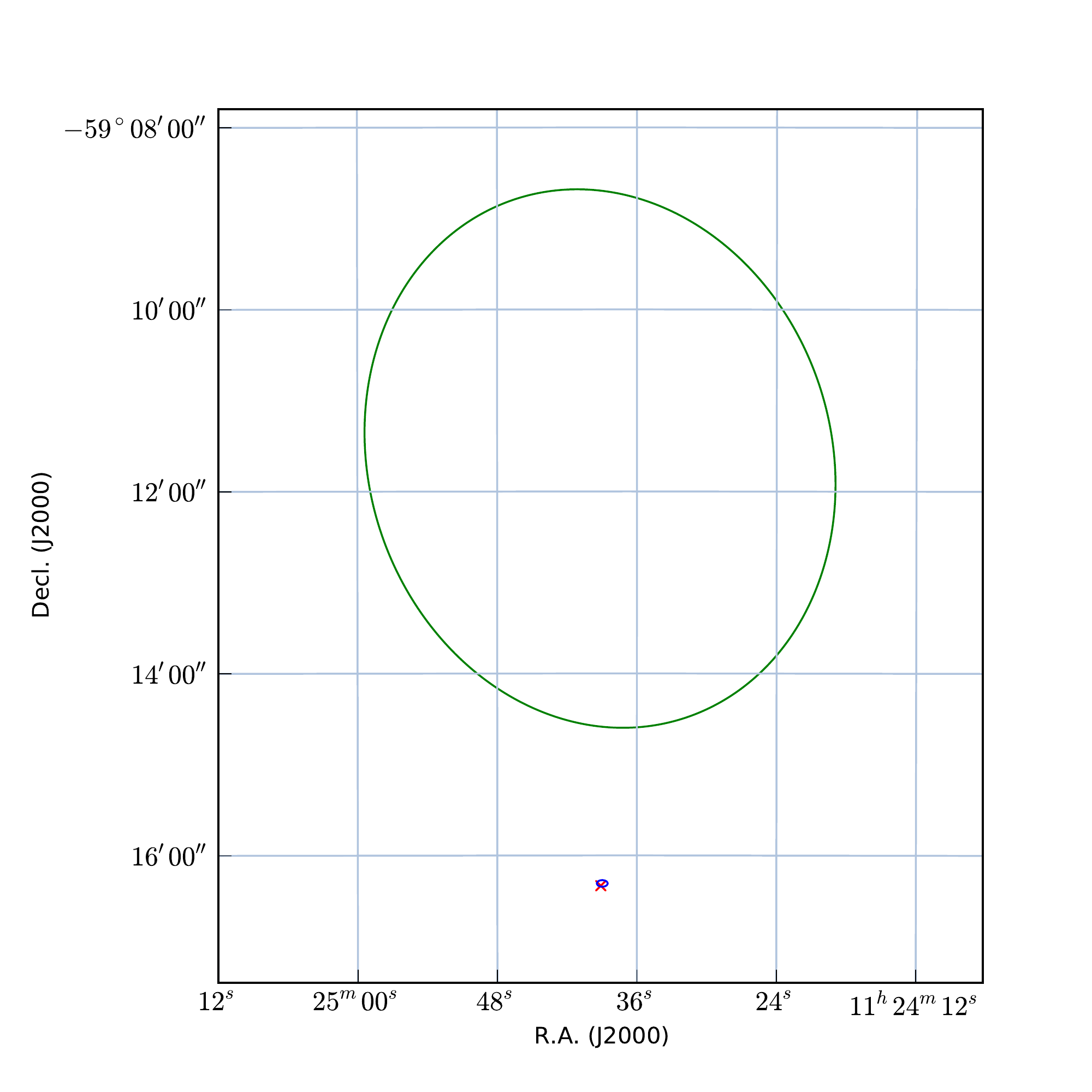} 
\caption{Timing position for PSR J1124$-$5916 (small blue ellipse). The large green ellipse is the LAT position of 1FGL J1124.6$-$5916, based on 18 months of data. The red cross marks the position of the Chandra point source \citep{cmg+02} associated with the pulsar (see \S \ref{sec:1124}).\label{pos:1124}}
\end{figure}

\begin{figure}
\includegraphics[width=3.0in]{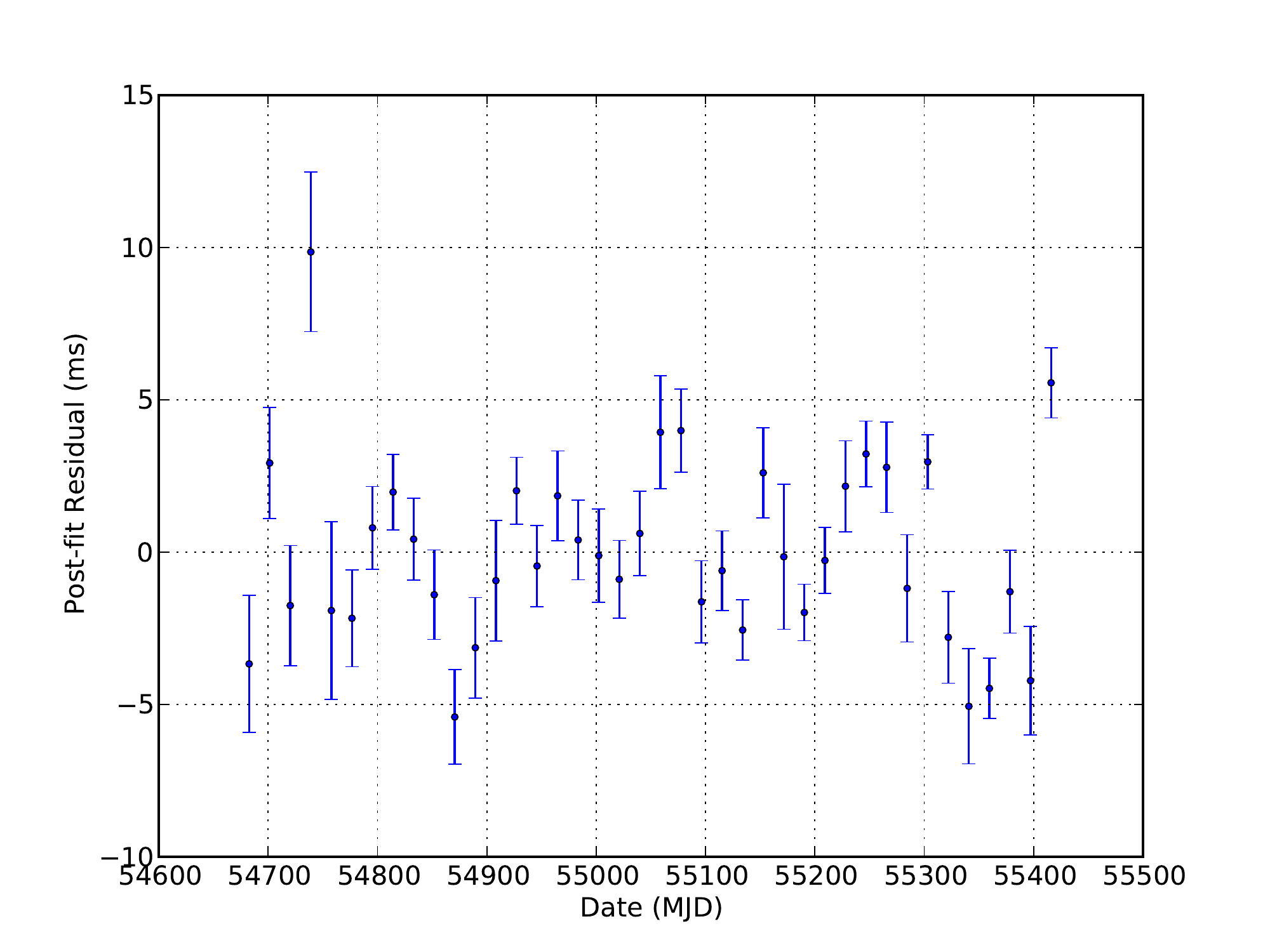}
\caption{Post-fit timing residuals for PSR J1124-5916.\label{resid:1124}}
\end{figure}

\begin{figure}
\includegraphics[width=3.0in]{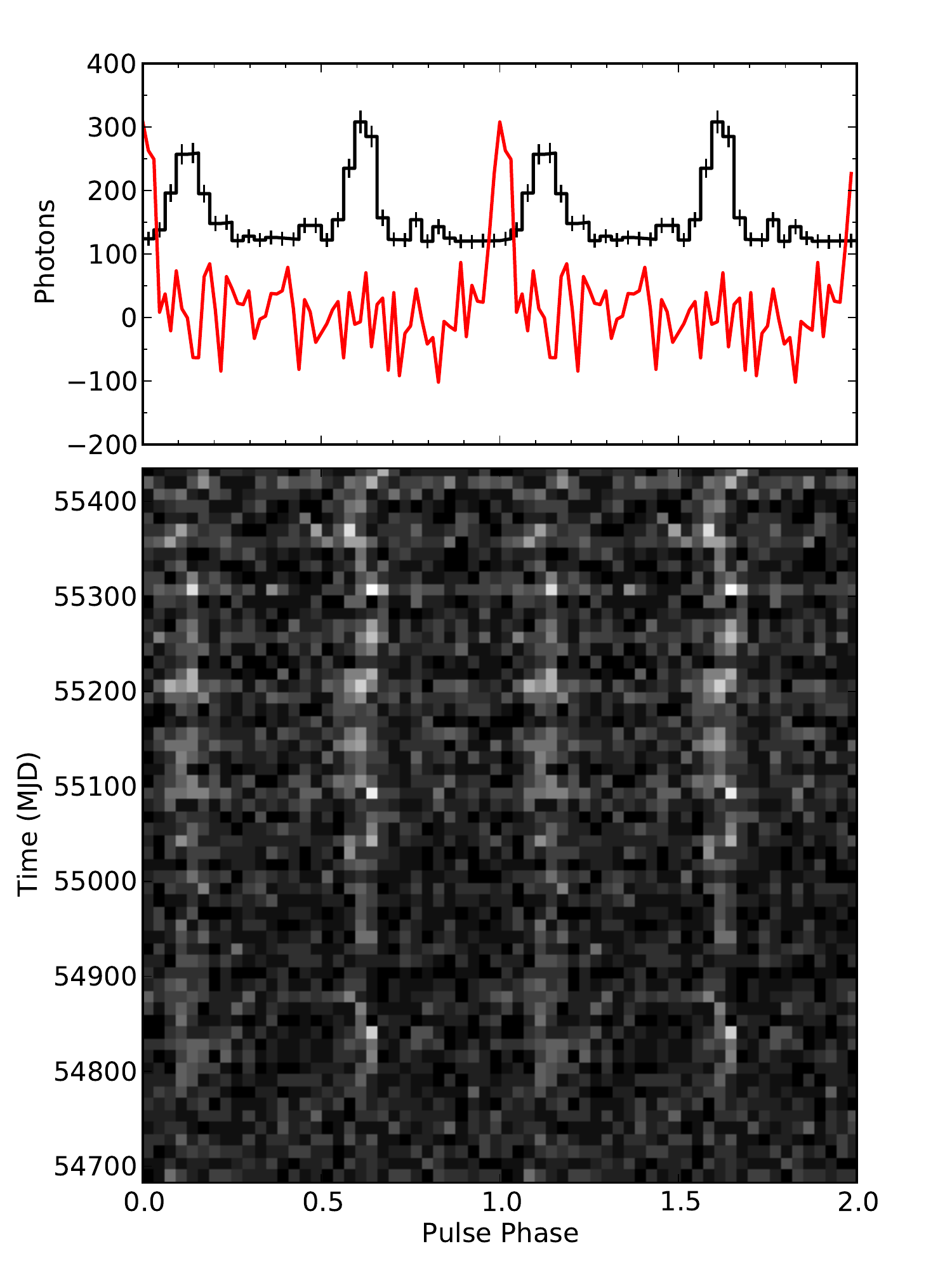}
\caption{2-D phaseogram and pulse profile of PSR J1124$-$5916.  Two rotations are shown on the X-axis. The photons were selected according to the ROI and $E_\mathrm{min}$ in Table~\ref{tab:1124}. The fiducial point corresponding to TZRMJD is phase 0.0. The red line is a 1.4 GHz radio profile from the Parkes radio telescope, with the correct absolute phase alignment.\label{phaseogram:1124}}
\end{figure}

The timing model parameters for this pulsar are displayed in Table~\ref{tab:1124} and the timing position determination, post-fit residuals, 2-D phaseogram, and folded pulse profile for this pulsar are shown in Figures \ref{pos:1124}, \ref{resid:1124}, and \ref{phaseogram:1124}, respectively.

This pulsar with very small characteristic age ($\tau_c = P/2\dot{P} = 2900$ yr) is associated with the supernova remnant G292.0+1.8 and is the only one in this sample that was previously known as a radio pulsar \citep{cmg+02}. It has a very high $\dot{E}$ of 1.2 $\times 10^{37}$ erg s$^{-1}$ and exhibits a great deal of timing noise. It is also is very faint, with a 1.4 GHz flux density of only 0.08 mJy \citep{cmg+02} and far enough south that it can only be timed with the Parkes Telescope, where it requires several hours of integration to even get a detection. Therefore, it has not been regularly timed with radio observations since its discovery. Because there was no contemporaneous radio ephemeris available, the LAT pulsations from this source were discovered using a limited blind search around the known spin parameters. With our LAT timing, we are able to obtain TOA uncertainties of 1.5--3.2 ms every two weeks and obtain a phase-connected timing model.  The timing position is 1.8\arcsec\ from the \textit{Chandra} source CXOU J112439.1$-$591620 \citep{cmg+02}. The pulsar exhibited a glitch of magnitude $\Delta\nu/\nu = 1.6\times 10^{-8}$ around MJD 55191. A small $\Delta\dot{\nu}/\dot{\nu}$ of $-$0.00472(3) was also observed at the glitch. 

The very large measured $\ddot{\nu}$ results in a Monte Carlo position error estimate of about 1\arcmin, but given the good agreement between the timing position and the Chandra position, this must be a large overestimate, perhaps because the assumptions inherent in the Monte Carlo estimate are violated. The measured $\ddot{\nu}$ results in a braking index $n = \frac{\nu\ddot{\nu}}{\dot{\nu}^2} = -3.78$, which is of comparable magnitude, but \textit{opposite} in sign to the $n=3$ expected for vacuum dipole braking \citep{lk05}. This measurement, along with the substantial red noise still present in the timing residuals (see Figure \ref{resid:1124}) all suggest that the spindown of this pulsar is rather noisy.

To measure the radio to $\gamma$-ray phase alignment of this pulsar, we made a 5-hour observation with the Parkes Radio Telescope at a frequency of 1.4 GHz. Since this pulsar is not timed routinely in the radio, we required a new, contemporaneous, observation because the extreme timing noise in this system prevents the timing model from being extrapolated forward or backwards in time. This radio light curve was presented previously in the First Fermi LAT Catalog of Gamma-ray Pulsars \citep{PulsarCatalog}, but the absolute phase alignment presented there was incorrect. The version presented here correctly accounts for the delay from interstellar dispersion using DM = 330 pc cm$^{-3}$.  The new value for the lag from the radio peak to the first $\gamma$-ray peak ($\delta$) is 0.128(3). This correction was also made to the catalog in an erratum \citep{erratum}.

\subsection{PSR J1418$-$6058}
\label{sec:1418}

\begin{figure}
\includegraphics[width=3.5in]{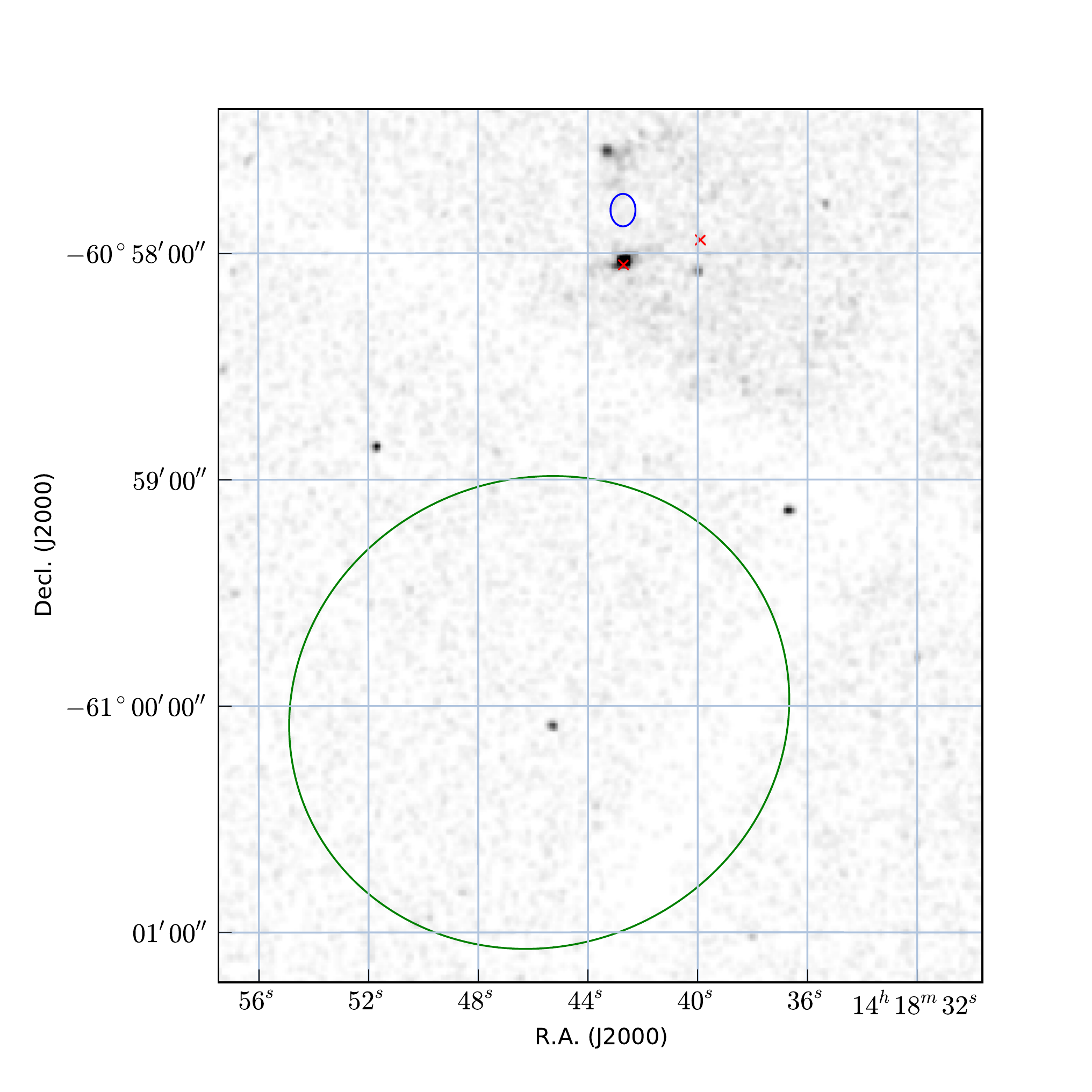} 
\caption{Timing position for PSR J1418$-$6058 (blue ellipse). The large green ellipse is the LAT position of 1FGL J1418.7$-$6057, based on 18 months of data. Red  crosses mark the positions of the sources R1 and R2 (see \S \ref{sec:1418}). The X-ray image is \label{pos:1418} a 70 ks \textit{Chandra} ACIS observation from 2007 July 14 (ObsID 7640), first published by \citet{r08} before the pulsar itself was detected.}
\end{figure}

\begin{figure}
\includegraphics[width=3.0in]{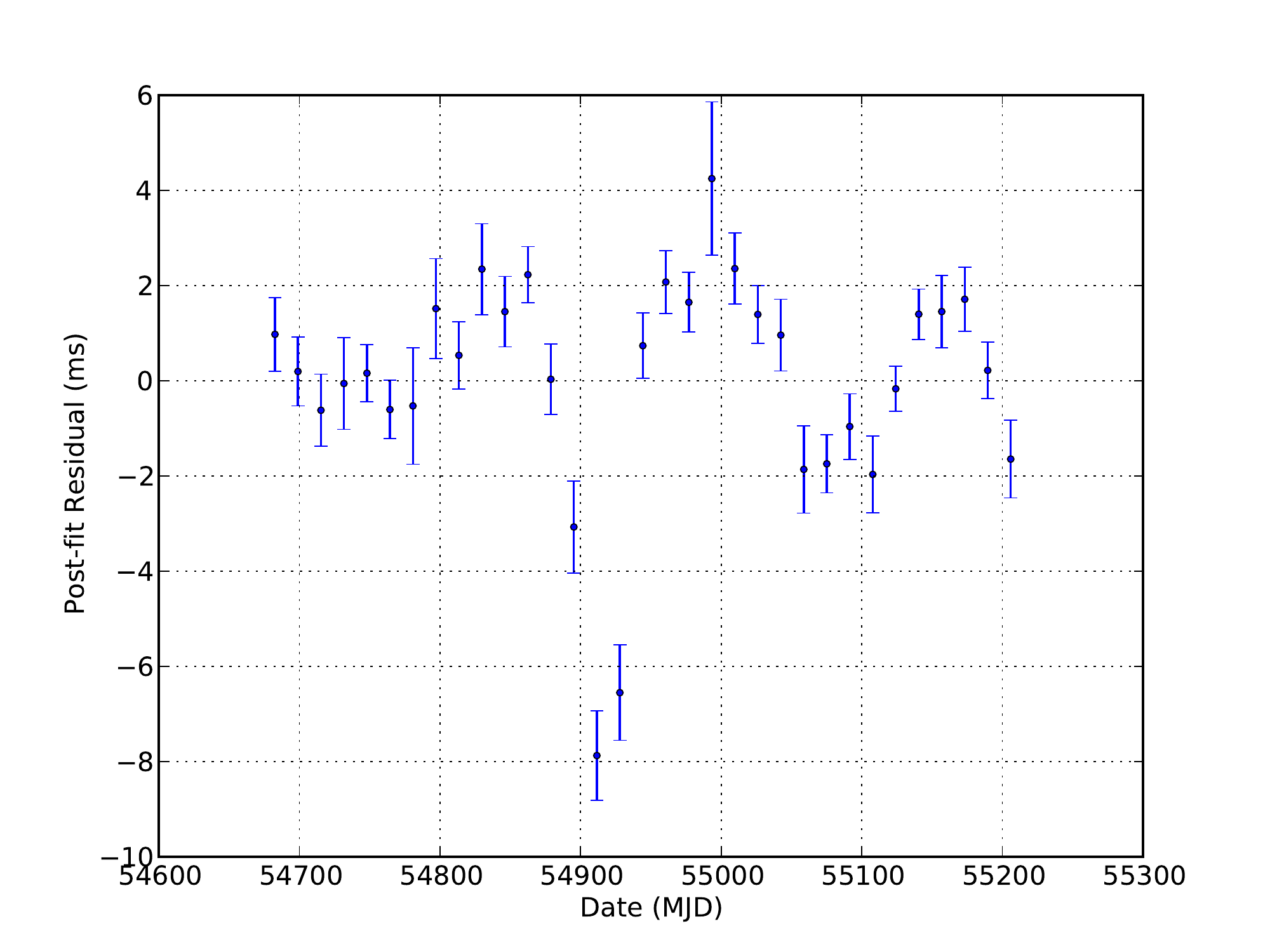}
\caption{Post-fit timing residuals for PSR J1418-6058.\label{resid:1418}}
\end{figure}

\begin{figure}
\includegraphics[width=3.0in]{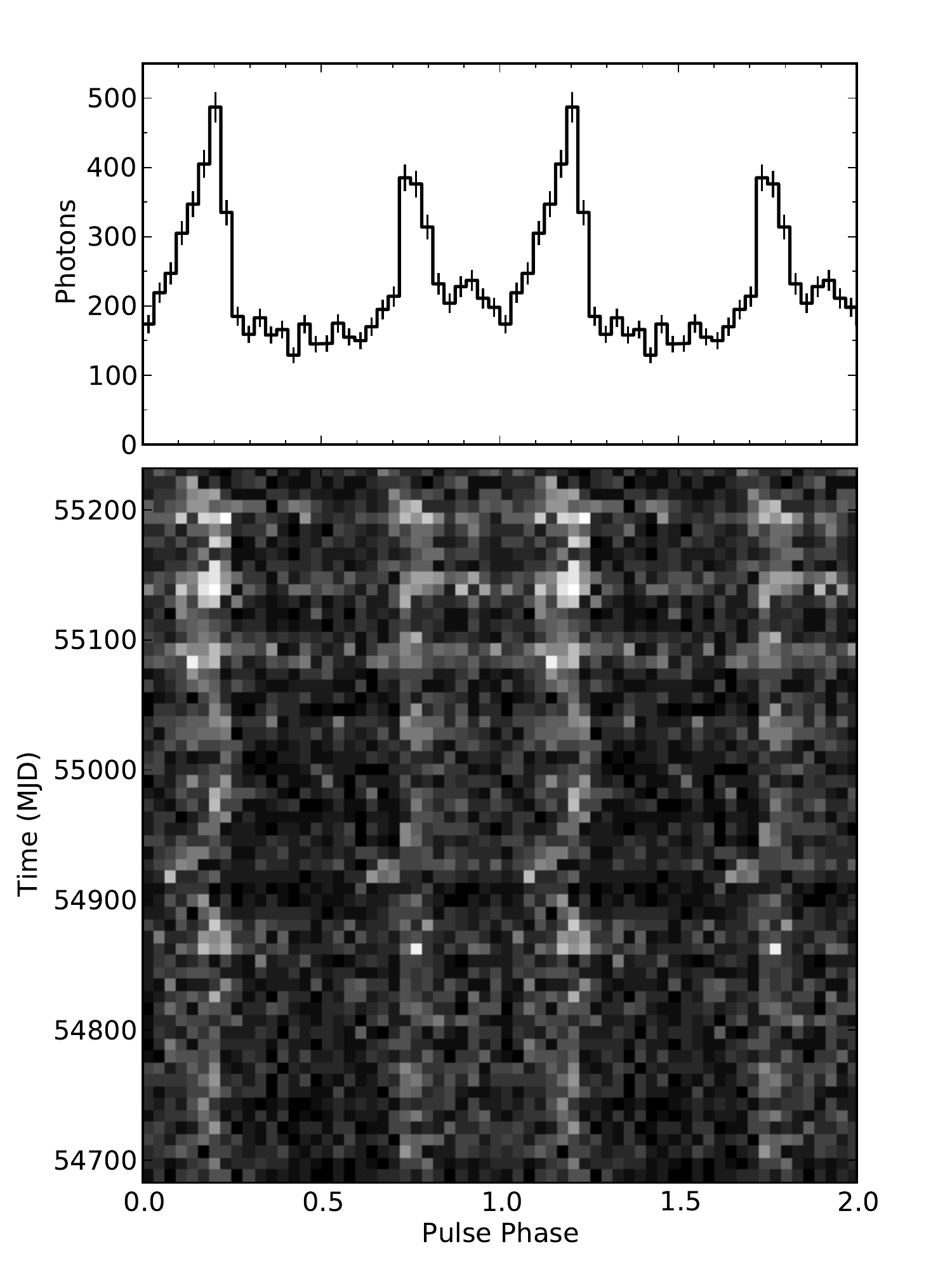}
\caption{2-D phaseogram and pulse profile of PSR J1418-6058.  Two rotations are shown on the X-axis. The photons were selected according to the ROI and $E_\mathrm{min}$ in Table~\ref{tab:1418}. The fiducial point corresponding to TZRMJD is phase 0.0.\label{phaseogram:1418}}
\end{figure}

The timing model parameters for this pulsar are displayed in Table~\ref{tab:1418} and the timing position determination, post-fit residuals, 2-D phaseogram, and folded pulse profile for this pulsar are shown in Figures \ref{pos:1418}, \ref{resid:1418}, and \ref{phaseogram:1418}, respectively.

In the discovery paper this pulsar was proposed to be associated with the PWN G313.3+0.1 (``The Rabbit'').  \citet{nrr05} find two point sources (R1 and R2) in \textit{Chandra} and XMM observations of the region. Recently, \citet{rob09} reported a weak detection of X-ray pulsations at the period of PSR J1418$-$6058 in XMM data, so it is believed that R1 is the correct counterpart. However, the timing position is 14\arcsec\ from the position of R1 (RA. = 14:18:42.7, Decl. = $-$60:58:03), which is significantly larger than the 2\arcsec\ statistical error on the position.  It is important to note that this pulsar is very noisy and the timing model, which includes terms up to the frequency third derivative, clearly does not fully describe the data (see Figures \ref{resid:1418} and \ref{phaseogram:1418}). This causes the statistical error to be underestimated and there is a significant systematic error on the position as well.  For example, adding a fourth frequency derivative to the model causes the position to shift by 10\arcsec. As seen in Figure~\ref{pos:1418}, there are three X-ray point sources near the nominal timing position, but none are coincident with the timing position.  The X-ray point source R2 is not apparent in this image, so it is likely variable, and probably not the counterpart to the pulsar.  A Monte Carlo estimation of the systematic error (see \S \ref{sec:results}) on the position induced by the timing noise seen in this pulsar is 7\arcsec\ for the polynomial model for timing noise and 40\arcsec\ using the red noise model with an RMS of 62 ms (as computed from the measured $\ddot{\nu}$ and $\dddot{\nu}$ for the pulsar). Therefore, we can't exclude R1 as being the counterpart based on the positional disagreement.  In addition, the faint X-ray source just north of the pulsar, which we call CXOU J141843.3-605734, is equally consistent with the timing and so further data, or a confirmation of the X-ray pulsations from R1, will be required to confirm the association with either source.

\subsection{PSR J1459$-$6053}

\begin{figure}
\includegraphics[width=3.5in]{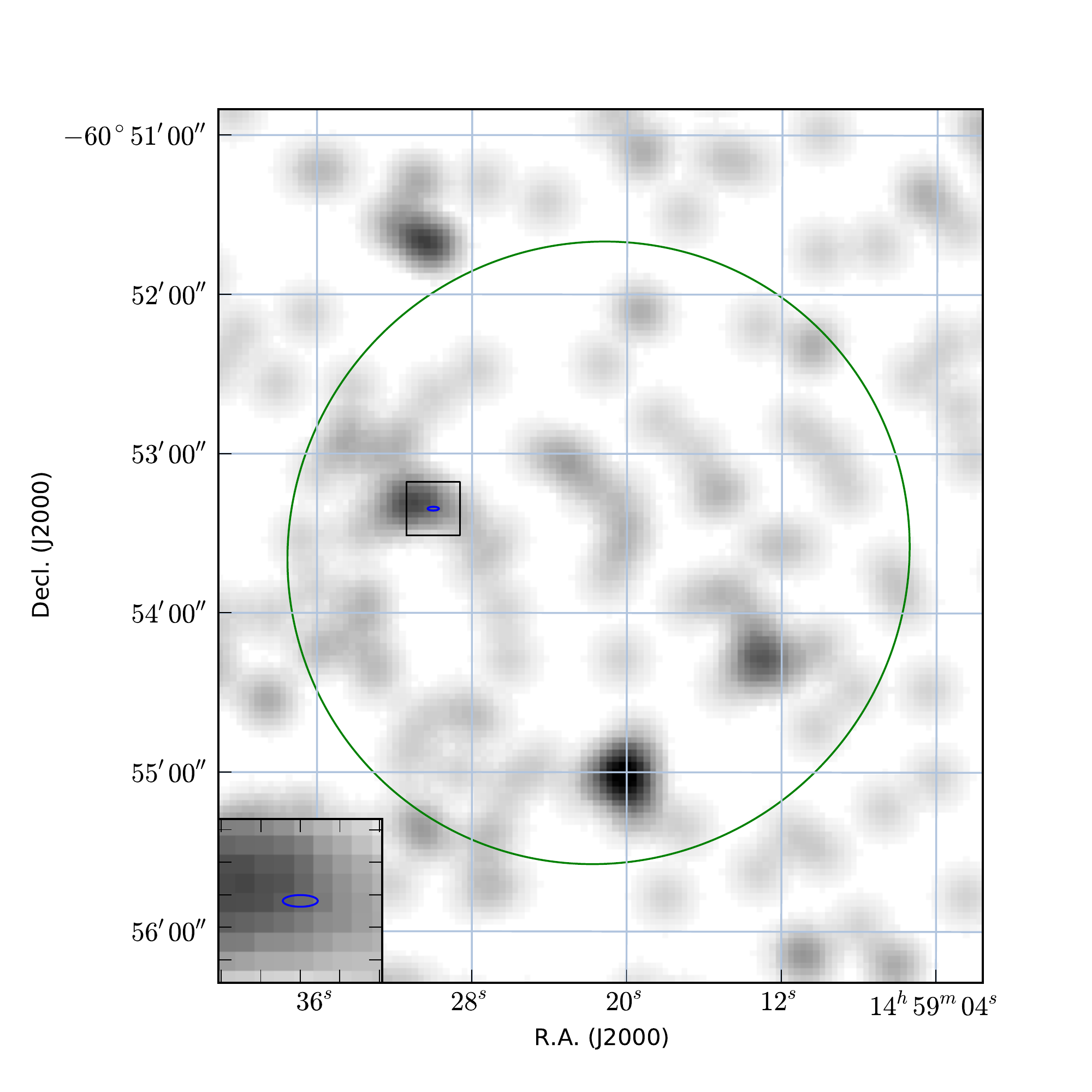} 
\caption{Timing position for PSR J1459$-$6053 (blue ellipse). The large green ellipse is the LAT position of 1FGL J1459.4$-$6053, based on 18 months of data. The background 0.2--10 keV X-ray image is a 6.8 ks \textit{Swift} image (ObsID 00031359002), smoothed with a gaussian with $\sigma = 7$\arcsec. \label{pos:1459} A $10\times 10$\arcsec\ region around the timing position is indicated with the black square and shown in the inset at the lower left.}
\end{figure}

\begin{figure}
\includegraphics[width=3.0in]{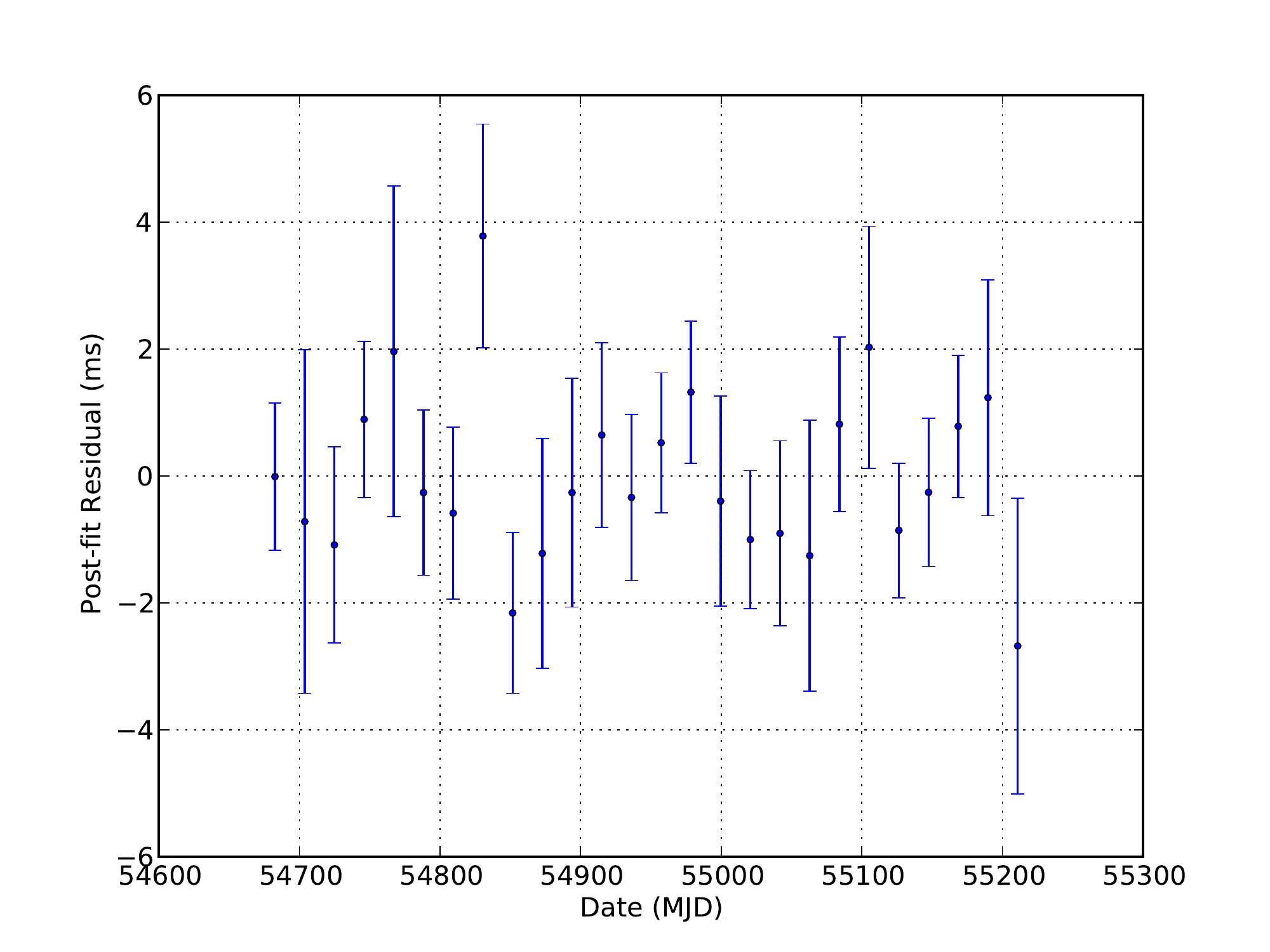}
\caption{Post-fit timing residuals for PSR J1459-6053.\label{resid:1459}}
\end{figure}

\begin{figure}
\includegraphics[width=3.0in]{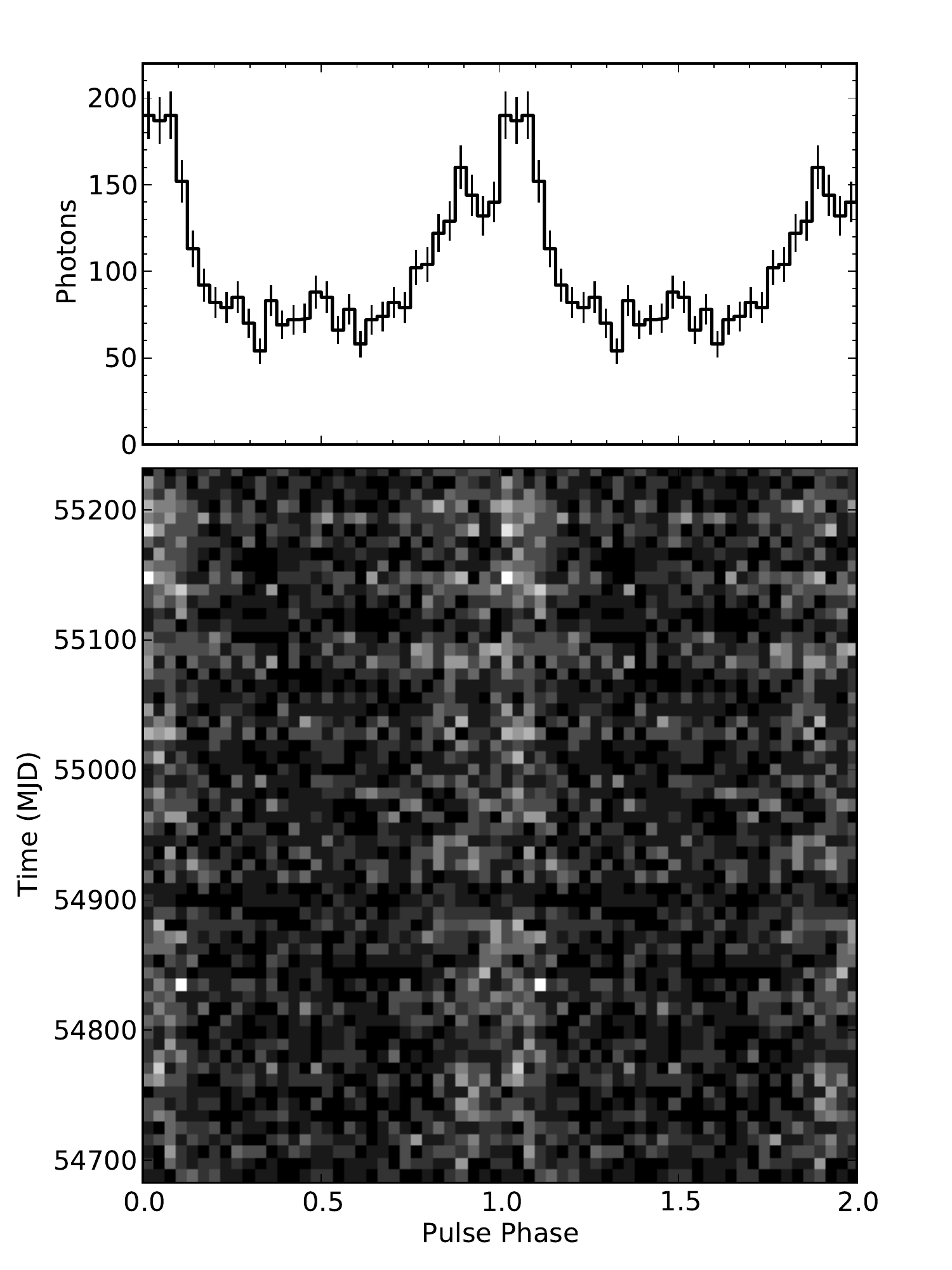}
\caption{2-D phaseogram and pulse profile of PSR J1459$-$6053.  Two rotations are shown on the X-axis. The photons were selected according to the ROI and $E_\mathrm{min}$ in Table~\ref{tab:1459}. The fiducial point corresponding to TZRMJD is phase 0.0.\label{phaseogram:1459}}
\end{figure}

The timing model parameters for this pulsar are displayed in Table~\ref{tab:1459} and the timing position determination, post-fit residuals, 2-D phaseogram, and folded pulse profile for this pulsar are shown in Figures \ref{pos:1459}, \ref{resid:1459}, and \ref{phaseogram:1459}, respectively.

In the discovery paper, no counterpart was known for this pulsar. With the high precision timing now available, we see that the \textit{Swift} image shows an apparent faint point source near (9.8\arcsec\ offset from) the timing position (see Figure~\ref{pos:1459}). We call this source Swift J145931.3$-$605319, but its properties are not well constrained because of the faintness in the 6 ks Swift image.  A deeper X-ray image is required to confirm this source.

\subsection{PSR J1732$-$3131}

\begin{figure}
\includegraphics[width=3.5in]{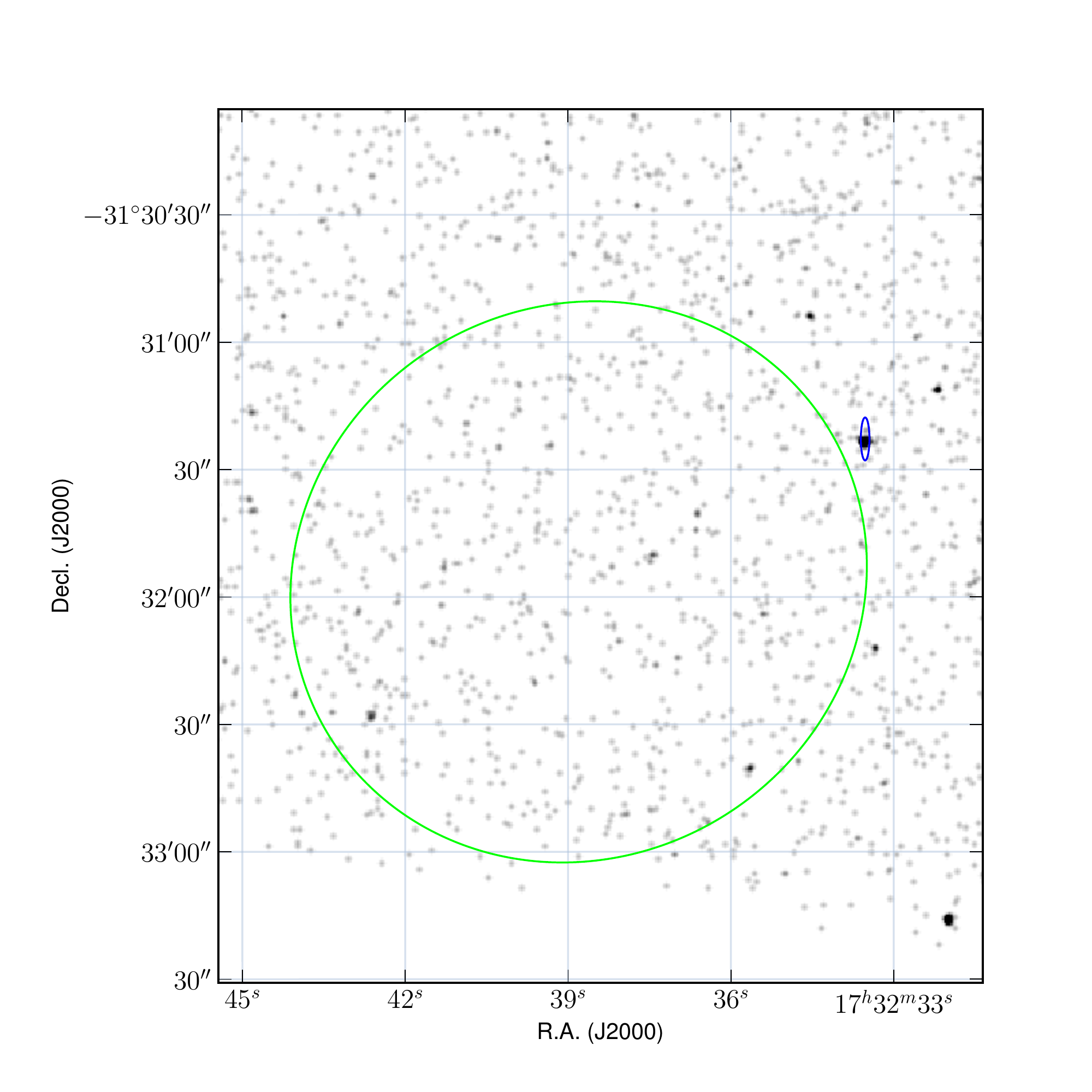} 
\caption{Timing position for PSR J1732$-$3131 (blue ellipse). The large green ellipse is the LAT position of 1FGL J1732.5$-$3131, based on 18 months of data. The background 0.5--8 keV X-ray image is a 20 ks \textit{Chandra} ACIS-S image (ObsID 11125), smoothed with a gaussian with $\sigma = 0.5$\arcsec.  \label{pos:1732} }
\end{figure}

\begin{figure}
\includegraphics[width=3.0in]{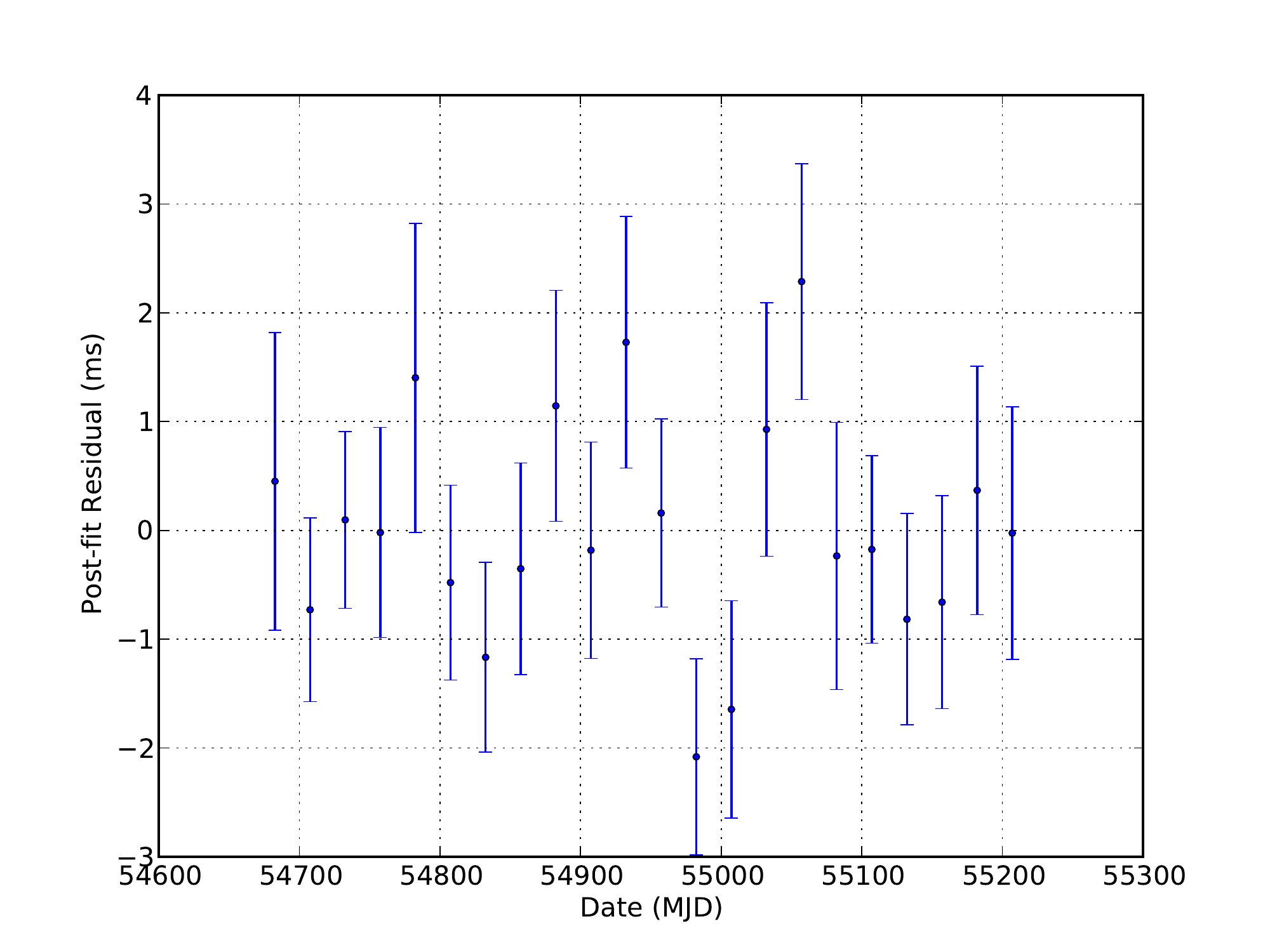}
\caption{Post-fit timing residuals for PSR J1732$-$3131.\label{resid:1732}}
\end{figure}

\begin{figure}
\includegraphics[width=3.0in]{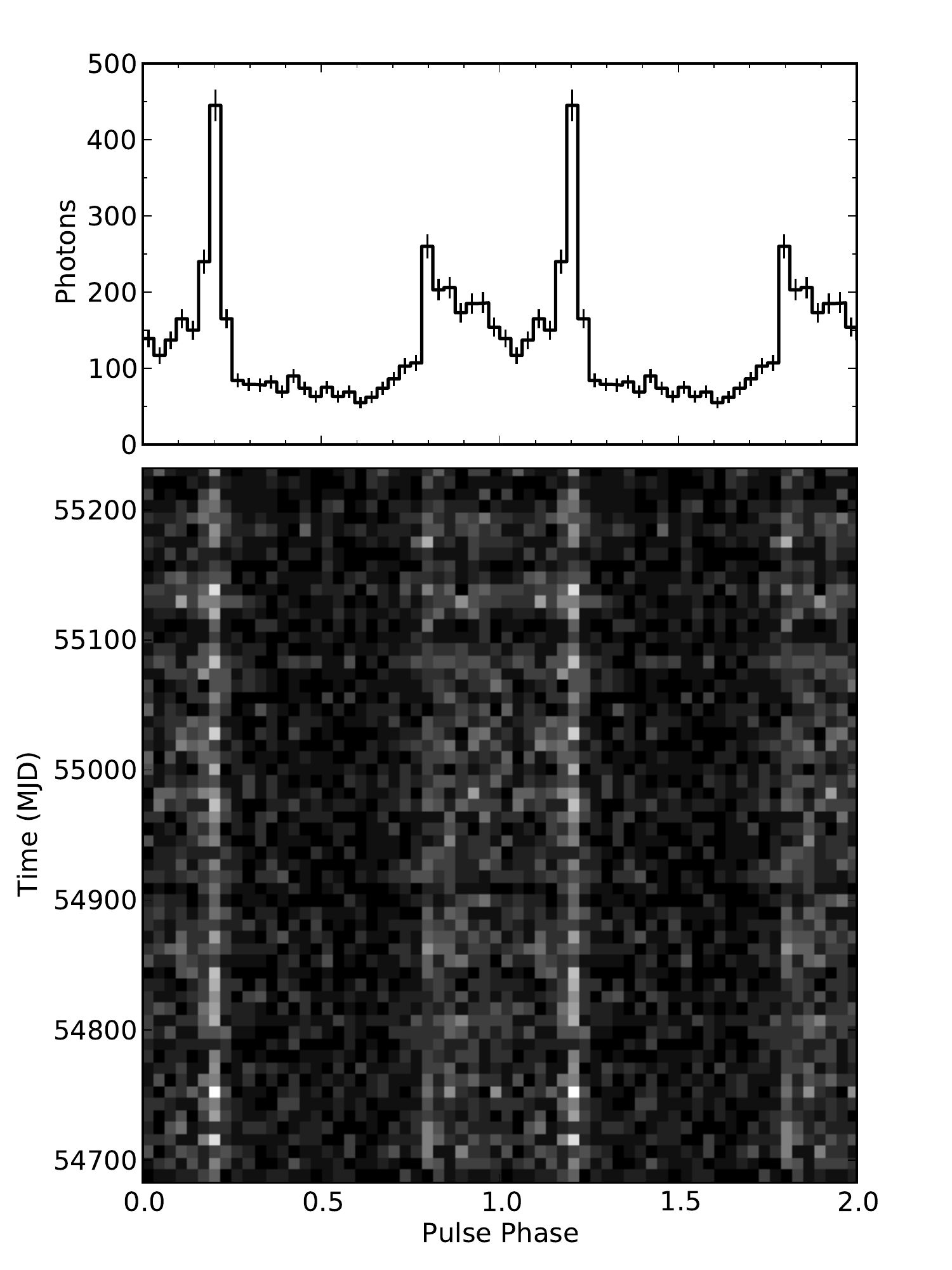}
\caption{2-D phaseogram and pulse profile of PSR J1732$-$3131.  Two rotations are shown on the X-axis. The photons were selected according to the ROI and $E_\mathrm{min}$ in Table~\ref{tab:1732}. The fiducial point corresponding to TZRMJD is phase 0.0.\label{phaseogram:1732}}
\end{figure}

The timing model parameters for this pulsar are displayed in Table~\ref{tab:1732} and the timing position determination, post-fit residuals, 2-D phaseogram, and folded pulse profile for this pulsar are shown in Figures \ref{pos:1732}, \ref{resid:1732}, and \ref{phaseogram:1732}, respectively.

This source shows minimal timing noise, with only an upper limit of $2\times10^{-23}$ s$^{-3}$ on $|\ddot{\nu}|$. The timing error ellipse is significantly elongated because of the low ecliptic latitude of the source ($\beta = -8.2^\circ$), causing the declination to be more poorly constrained than the R.A. Earlier \textit{Swift} imaging showed no significantly-detected X-ray source at the pulsar location, so we pursued a deeper observations with \textit{Chandra}. Our 20\ ks Chandra ACIS-S image (ObsID 11125) reveals an X-ray point source consistent with the timing position (see Figure \ref{pos:1732}). We measure the position as 17:32:33.551, $-$31:31:23.92, which is 0.9\arcsec\ from the timing position, well within the 95\% confidence region.

We performed a spectral analysis of the source based on 79 photons from the point source (with $\lesssim 1$ count from the background). Since the source is still detected even with a low energy cut of 3.5 keV, it is clear that a non-thermal component is required. However, with the small number of counts, the power law photon index cannot be constrained, so we freeze it at $\Gamma = 1.5$ in the fits.  The spectral parameters from our fits to an absorbed blackbody + powerlaw are (with 90\% confidence error regions) $n_H = 0.22_{-0.22}^{+0.50} \times 10^{22}$ cm$^{-2}$, $kT = 0.19_{-0.07}^{+0.20}$ keV. The implied 0.5--8 keV flux estimate is $(2.8 \pm 0.7) \times 10^{-14}$ erg cm$^{-2}$ s$^{-1}$ with the error at the 68\% confidence interval.  This corresponds to an unabsorbed flux of $4\times 10^{-14}$ erg cm$^{-2}$ s$^{-1}$.

\subsection{PSR J1741$-$2054}

\begin{figure}
\includegraphics[width=3.5in]{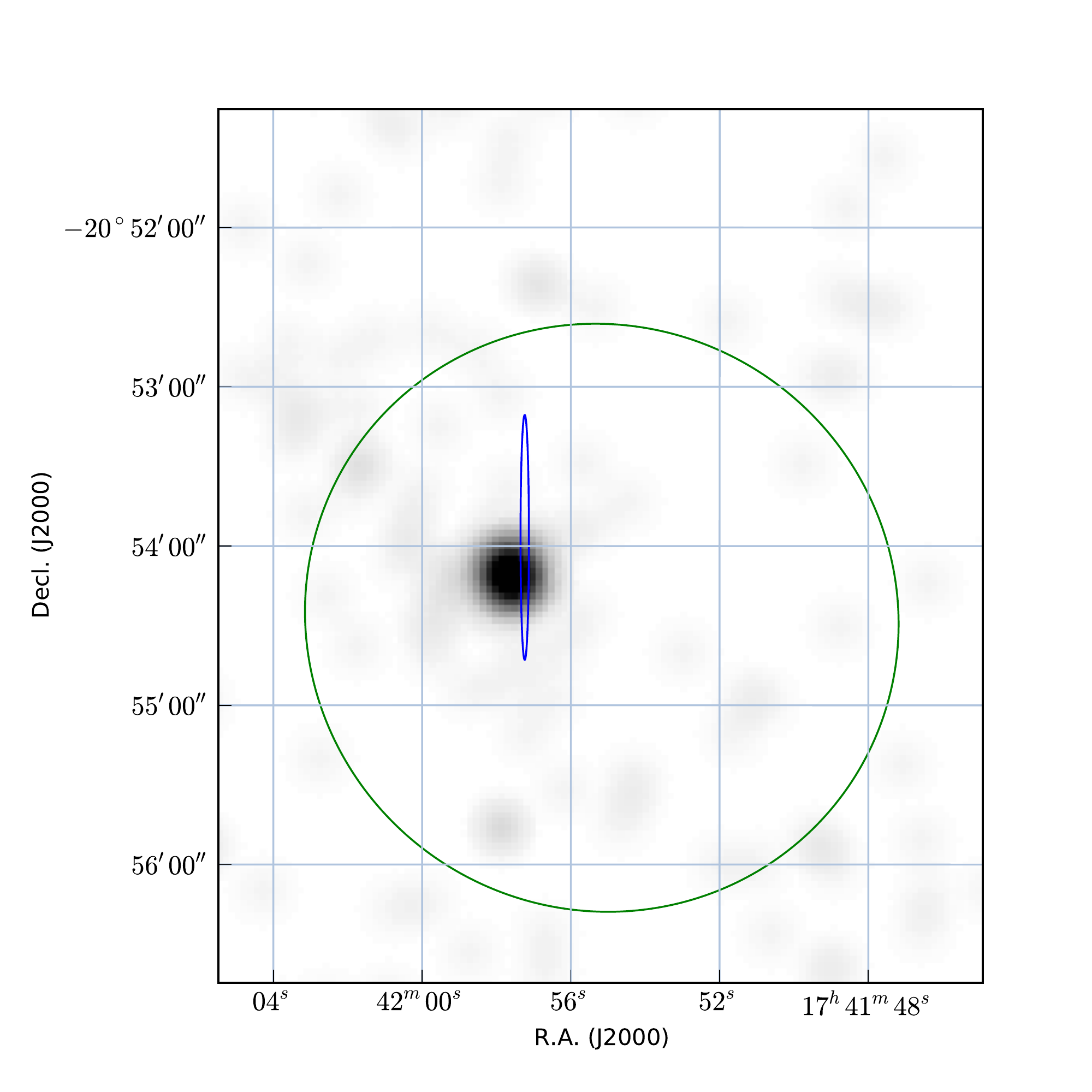} 
\caption{Timing position for PSR J1741$-$2054 (blue ellipse). The large green ellipse is the LAT position of 1FGL J1741.8$-$2101, based on 18 months of data. The background 0.2--10 keV X-ray image is a 4.3 ks \textit{Swift} XRT image (ObsID 00031277001), smoothed with a gaussian with $\sigma = 7$\arcsec. \label{pos:1741} }
\end{figure}

\begin{figure}
\includegraphics[width=3.0in]{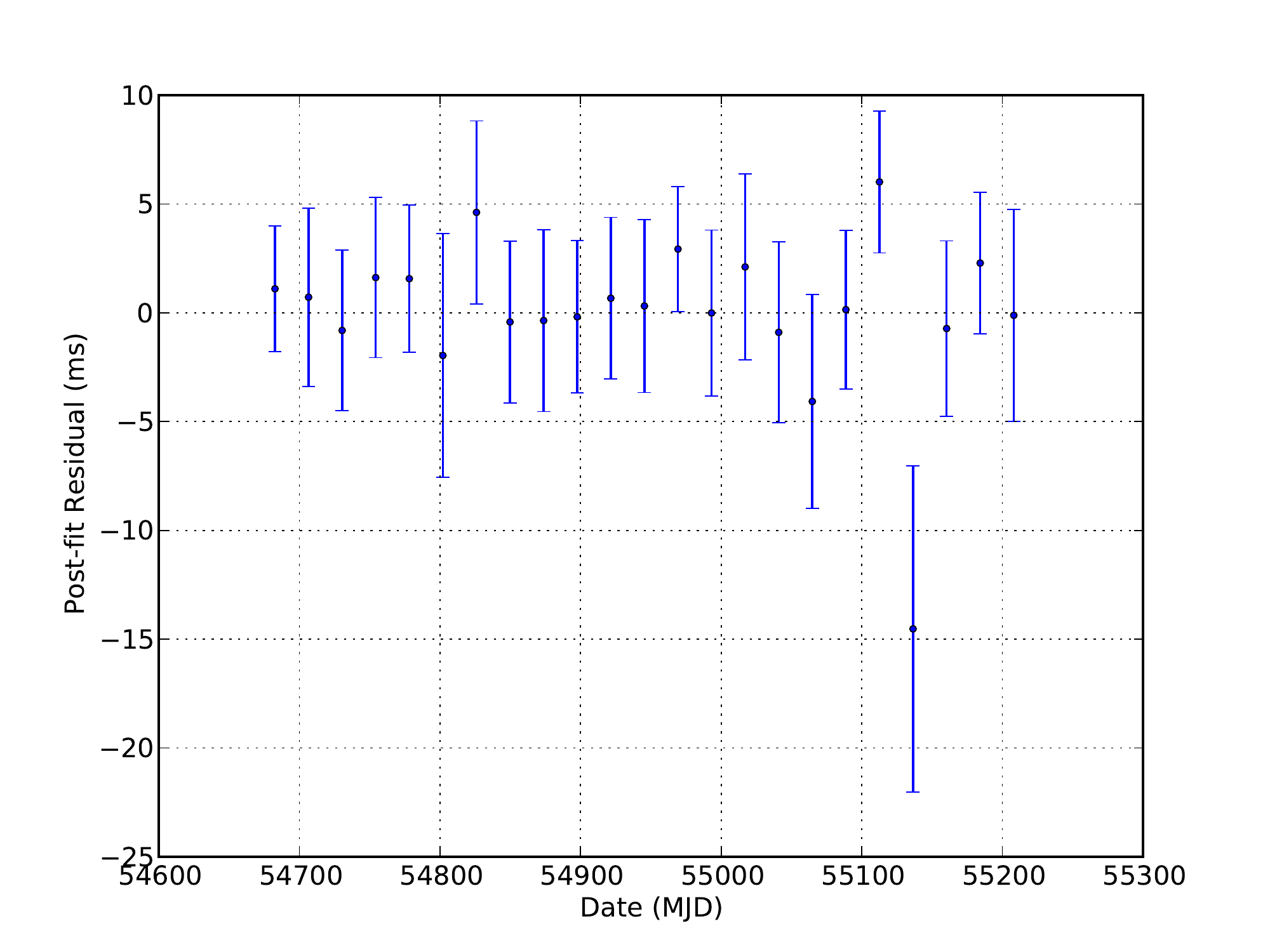}
\caption{Post-fit timing residuals for PSR J1741$-$2054.\label{resid:1741}}
\end{figure}

\begin{figure}
\includegraphics[width=3.0in]{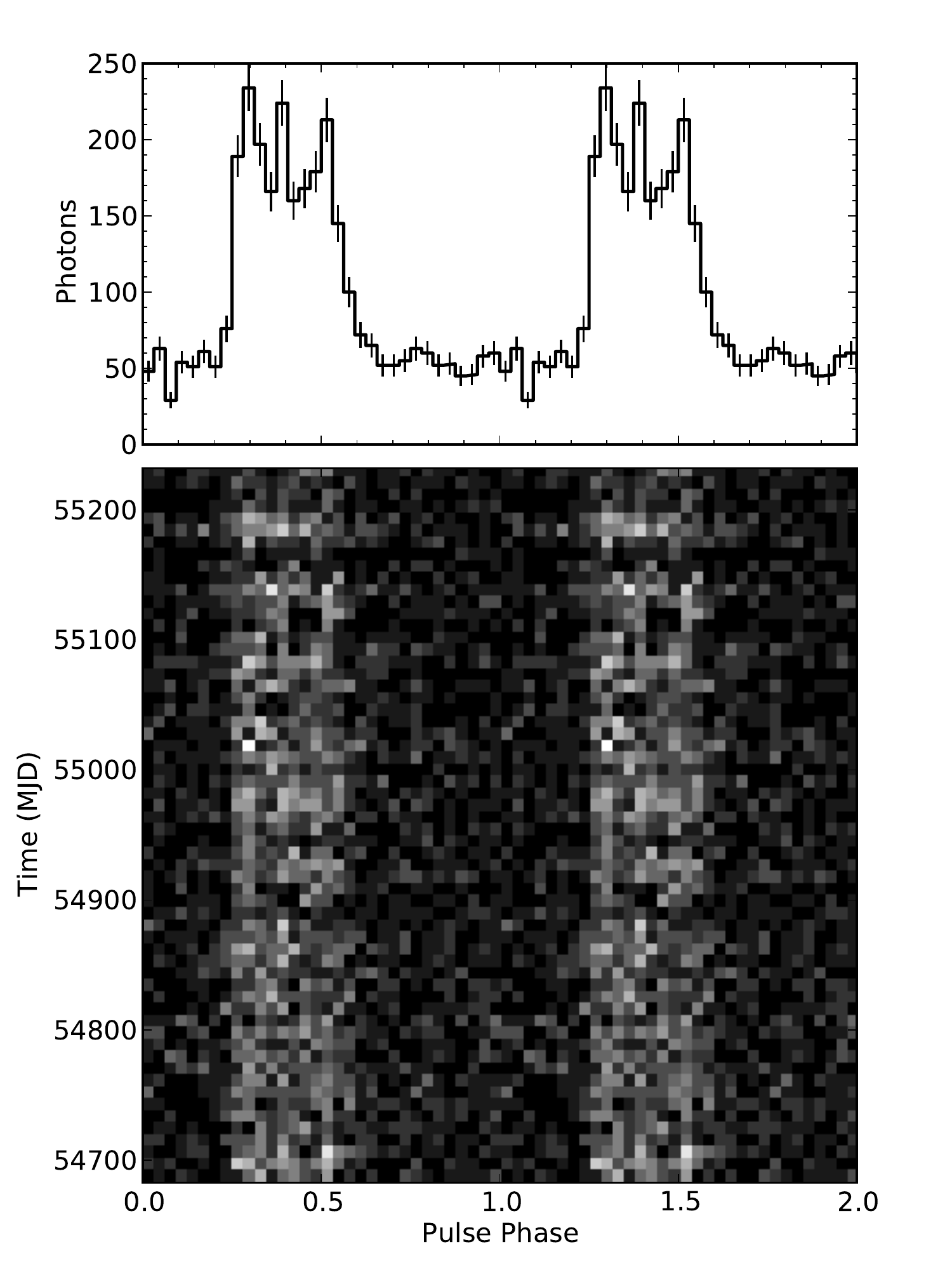}
\caption{2-D phaseogram and pulse profile of PSR J1741-2054.  Two rotations are shown on the X-axis. The photons were selected according to the ROI and $E_\mathrm{min}$ in Table~\ref{tab:1741}. The fiducial point corresponding to TZRMJD is phase 0.0.\label{phaseogram:1741}}
\end{figure}

The timing model parameters for this pulsar are displayed in Table~\ref{tab:1741} and the timing position determination, post-fit residuals, 2-D phaseogram, and folded pulse profile for this pulsar are shown in Figures \ref{pos:1741}, \ref{resid:1741}, and \ref{phaseogram:1741}, respectively.

The bright X-ray counterpart (Swift J174157.6$-$205411) seen in Figure \ref{pos:1741} was proposed as the likely counterpart to this pulsar in the discovery paper.  Subsequently, a LAT timing position presented by \citet{crr+09}, who also reported the discovery of radio pulsations from this pulsar, added confidence to this proposal, and the model we present here strengthens the case.  The position error is still highly elongated in the declination direction because of the very low ecliptic latitude of the source. The X-ray source properties are studied in detail in \citet{crr+09}. The larger span of data included in this model results in significantly more counts in the light curve, confirming the apparent 3-peak nature as proposed by \citet{crr+09}, in contrast to the peak multiplicity of 2 assigned by \citet{PulsarCatalog}.

\subsection{PSR J1809$-$2332}

\begin{figure}
\includegraphics[width=3.5in]{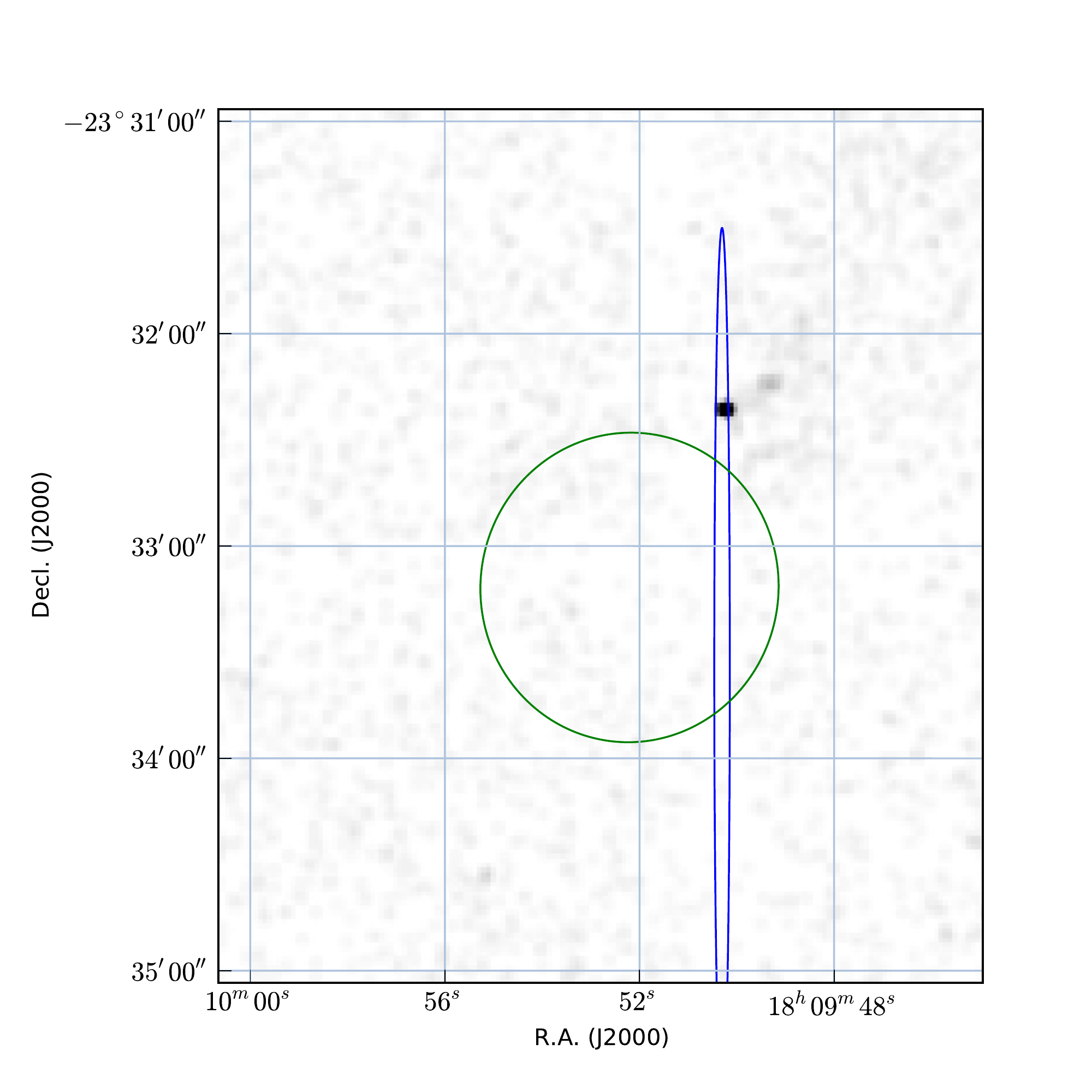} 
\caption{Timing position for PSR J1809$-$2332 (blue ellipse). The large green ellipse is the LAT position of 1FGL J1809.8$-$2332, based on 18 months of data.  The background image is a 9.8 ks \textit{Chandra} ACIS-I image (ObsId 739) showing the bright point source CXOU J180950.2$-$233223.\label{pos:1809} }
\end{figure}

\begin{figure}
\includegraphics[width=3.0in]{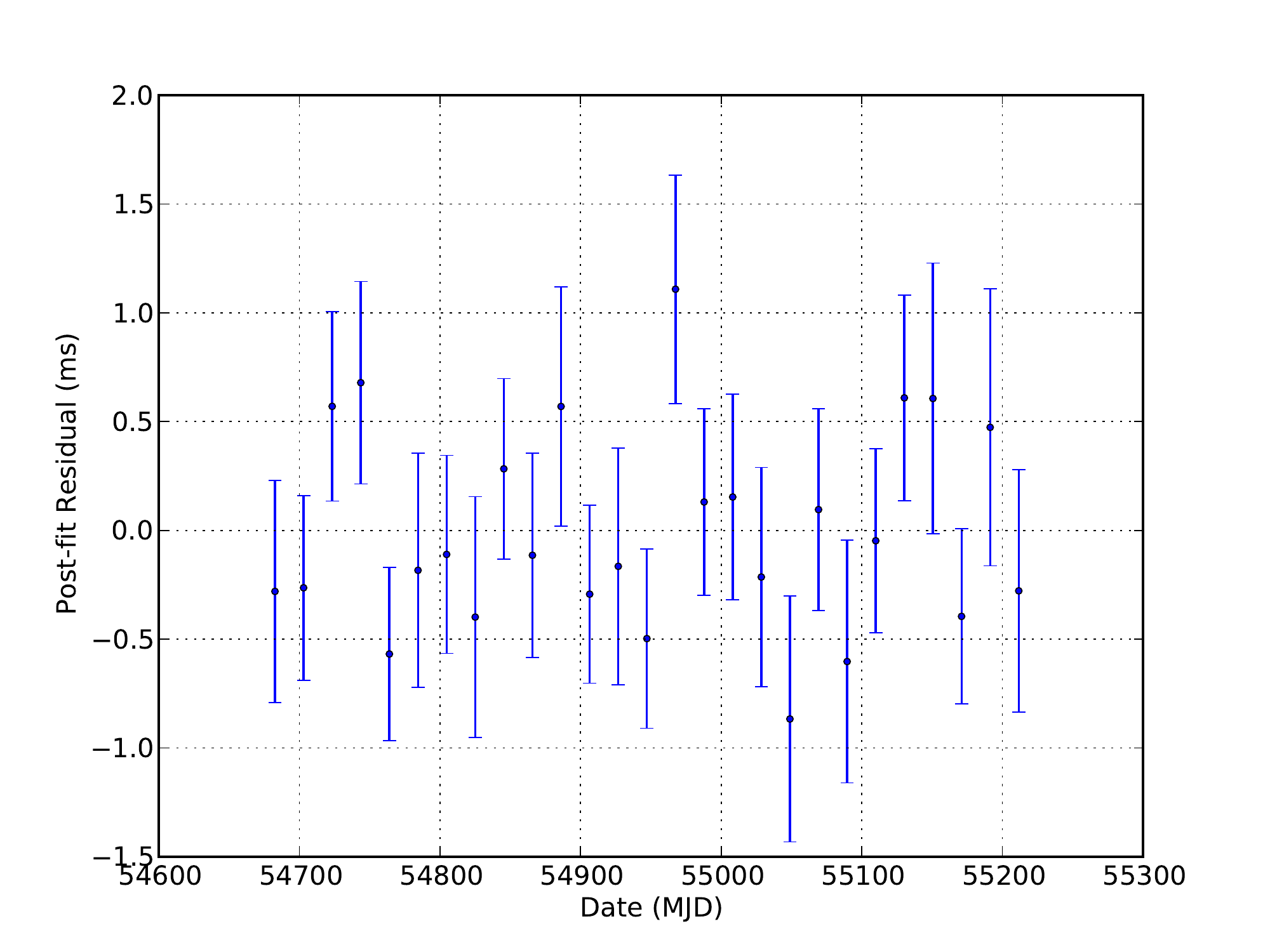}
\caption{Post-fit timing residuals for PSR J1809$-$2332.\label{resid:1809}}
\end{figure}

\begin{figure}
\includegraphics[width=3.0in]{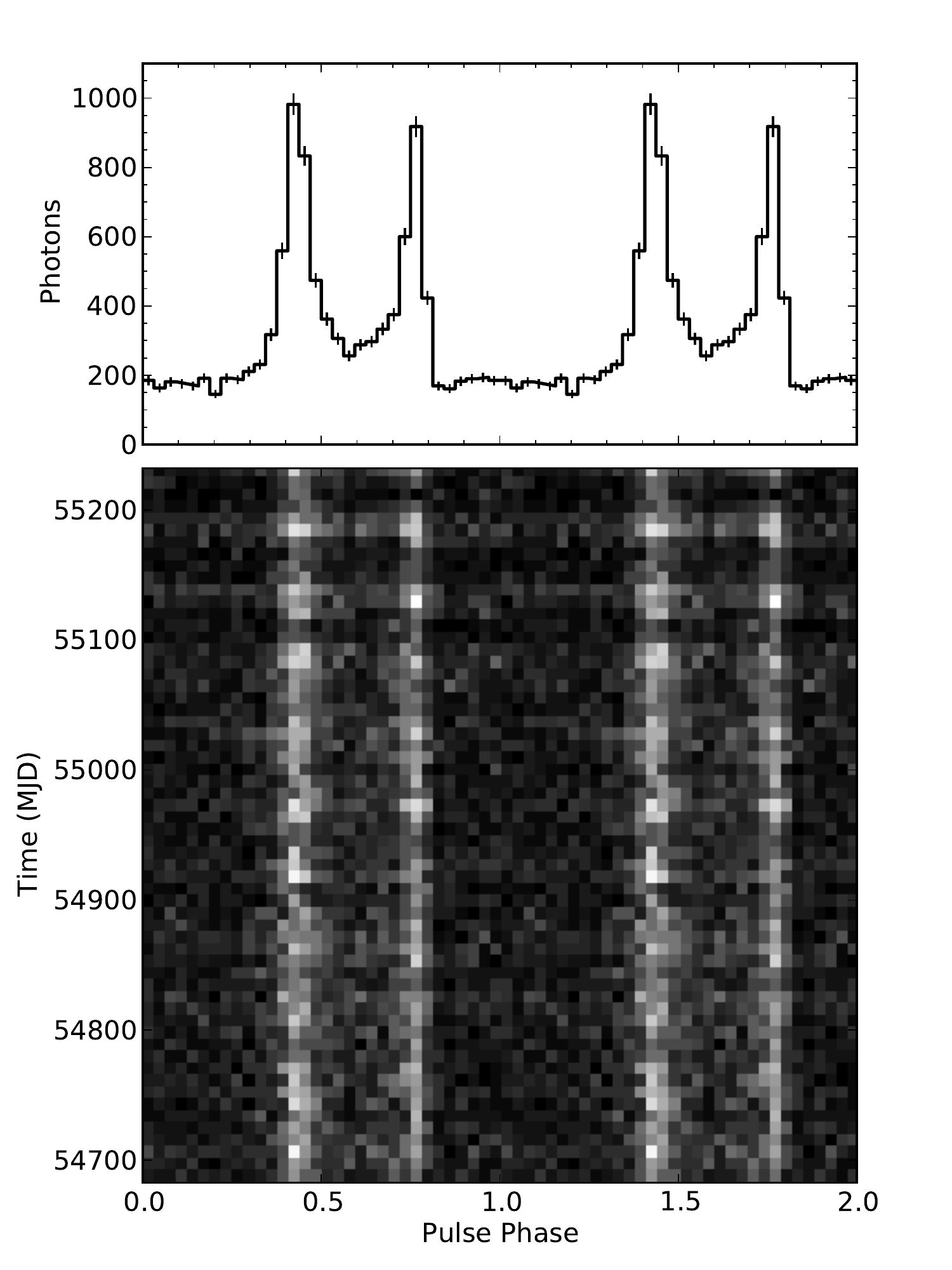}
\caption{2-D phaseogram and pulse profile of PSR J1809$-$2332.  Two rotations are shown on the X-axis. The photons were selected according to the ROI and $E_\mathrm{min}$ in Table~\ref{tab:1809}. The fiducial point corresponding to TZRMJD is phase 0.0.\label{phaseogram:1809}}
\end{figure}

The timing model parameters for this pulsar are displayed in Table~\ref{tab:1809} and the timing position determination, post-fit residuals, 2-D phaseogram, and folded pulse profile for this pulsar are shown in Figures \ref{pos:1809}, \ref{resid:1809}, and \ref{phaseogram:1809}, respectively.

This pulsar was discovered in the direction of the Galactic unidentified $\gamma$-ray source GeV J1809$-$2327. \textit{Chandra} observations revealed a probable pulsar with PWN that was proposed as the source of the $\gamma$-rays \citep{brrk02}.  The point source, CXOU J180950.2$-$233223, was assumed as the counterpart by \citet{LATBlindSearch}.  As seen in Figure~\ref{pos:1809}, the position error ellipse is very strongly elongated, again because of the very low ecliptic latitude of the source.  Nevertheless, the X-ray point source is within the timing error ellipse, strengthening the identification with this point source. If the TOAs are fitted with the position held fixed at the location of the \textit{Chandra} point source, there are significant correlated residuals, which require a frequency third derivative term in the model to give a reasonable $\chi^2$ for the fit.

\subsection{PSR J1813$-$1246}

\begin{figure}
\includegraphics[width=3.5in]{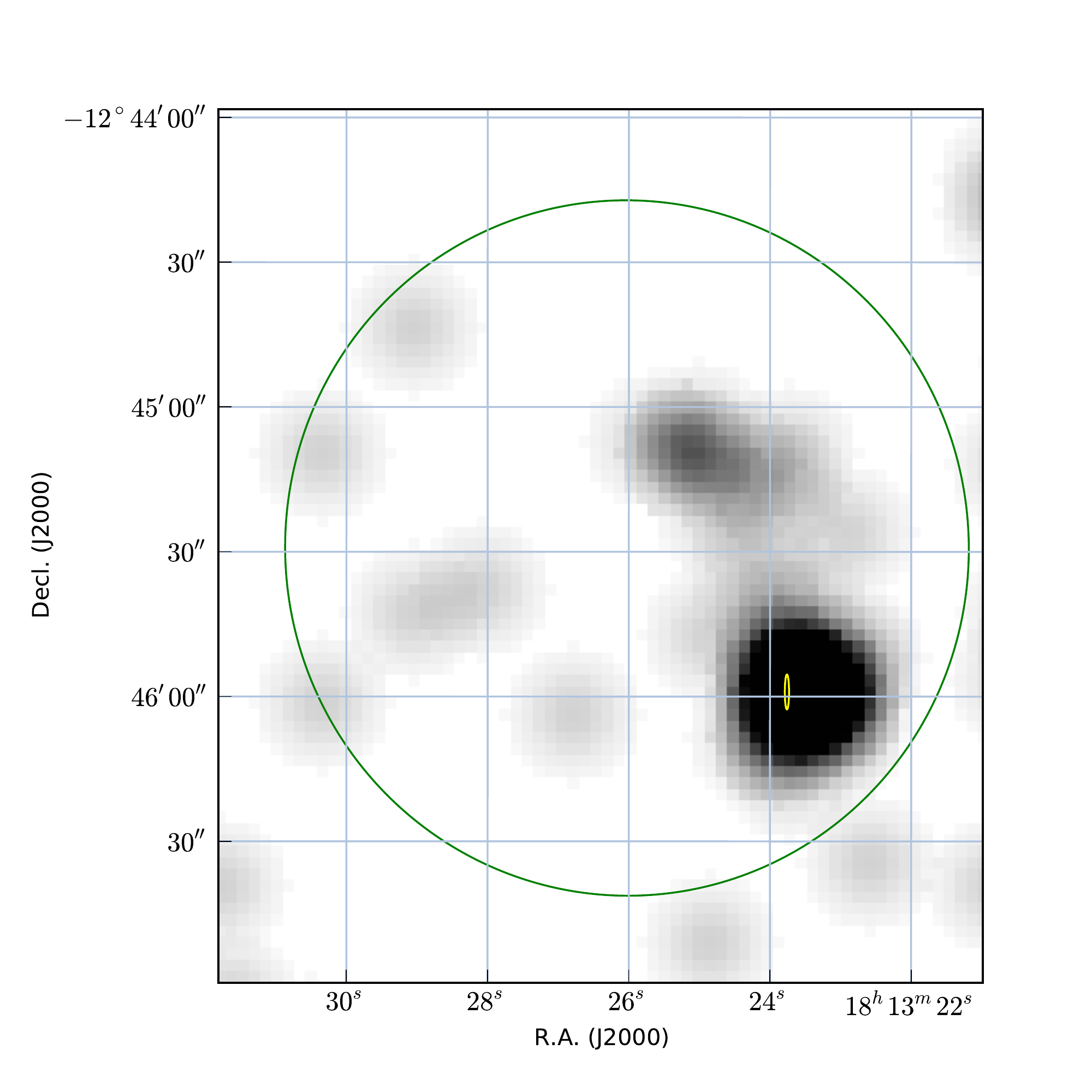} 
\caption{Timing position for PSR J1813$-$1246 (yellow ellipse). The large green ellipse is the LAT position of 1FGL J1813.3$-$1246, based on 18 months of data. The background X-ray image is a 3.2 ks \textit{Swift} XRT observation (ObsID 00031381001), smoothed with a gaussian of width 7\arcsec, showing the bright point source \textit{Swift} J181323.4$-$124600.\label{pos:1813} }
\end{figure}

\begin{figure}
\includegraphics[width=3.0in]{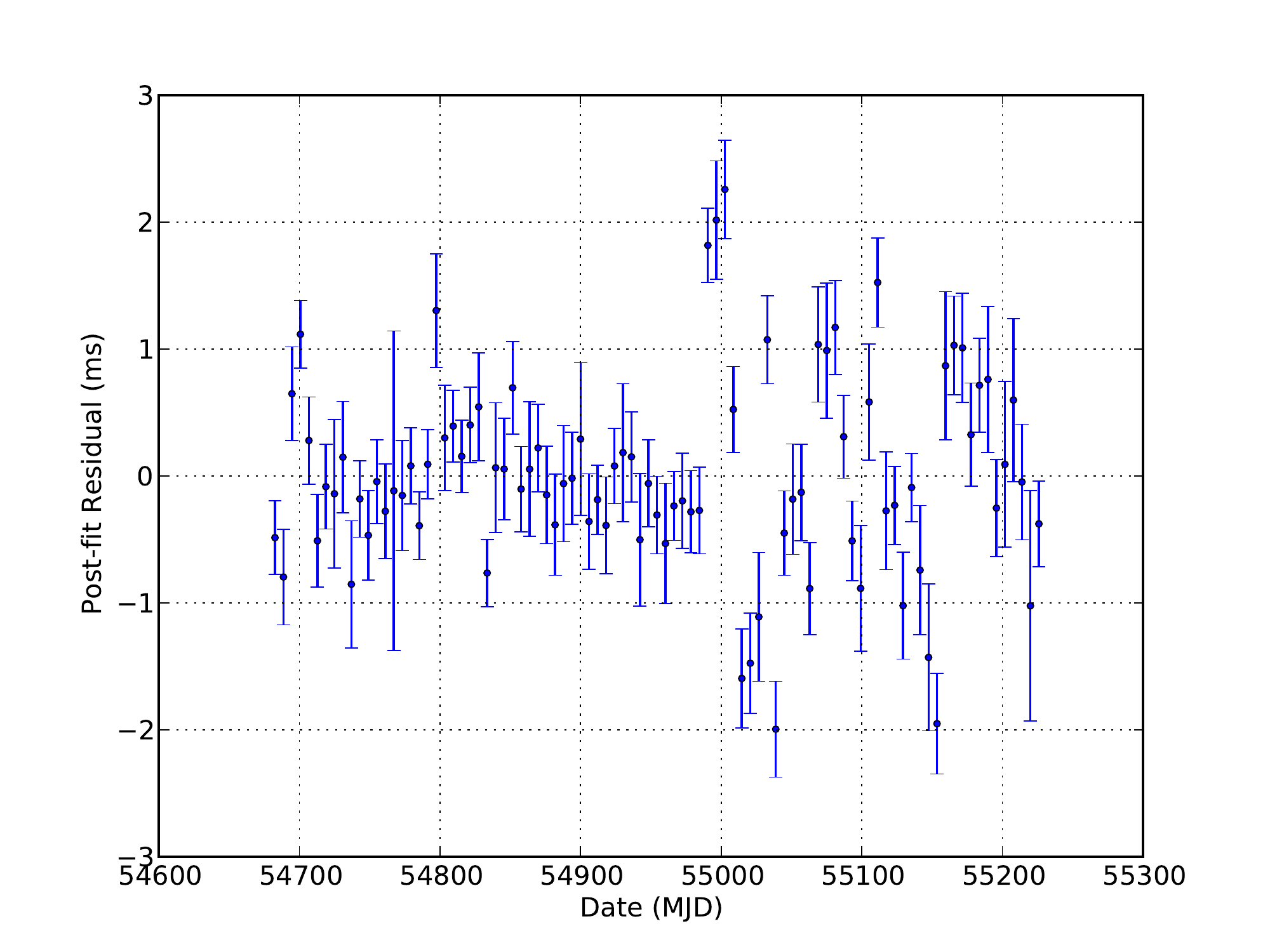}
\caption{Post-fit timing residuals for PSR J1813$-$1246.\label{resid:1813}}
\end{figure}

\begin{figure}
\includegraphics[width=3.0in]{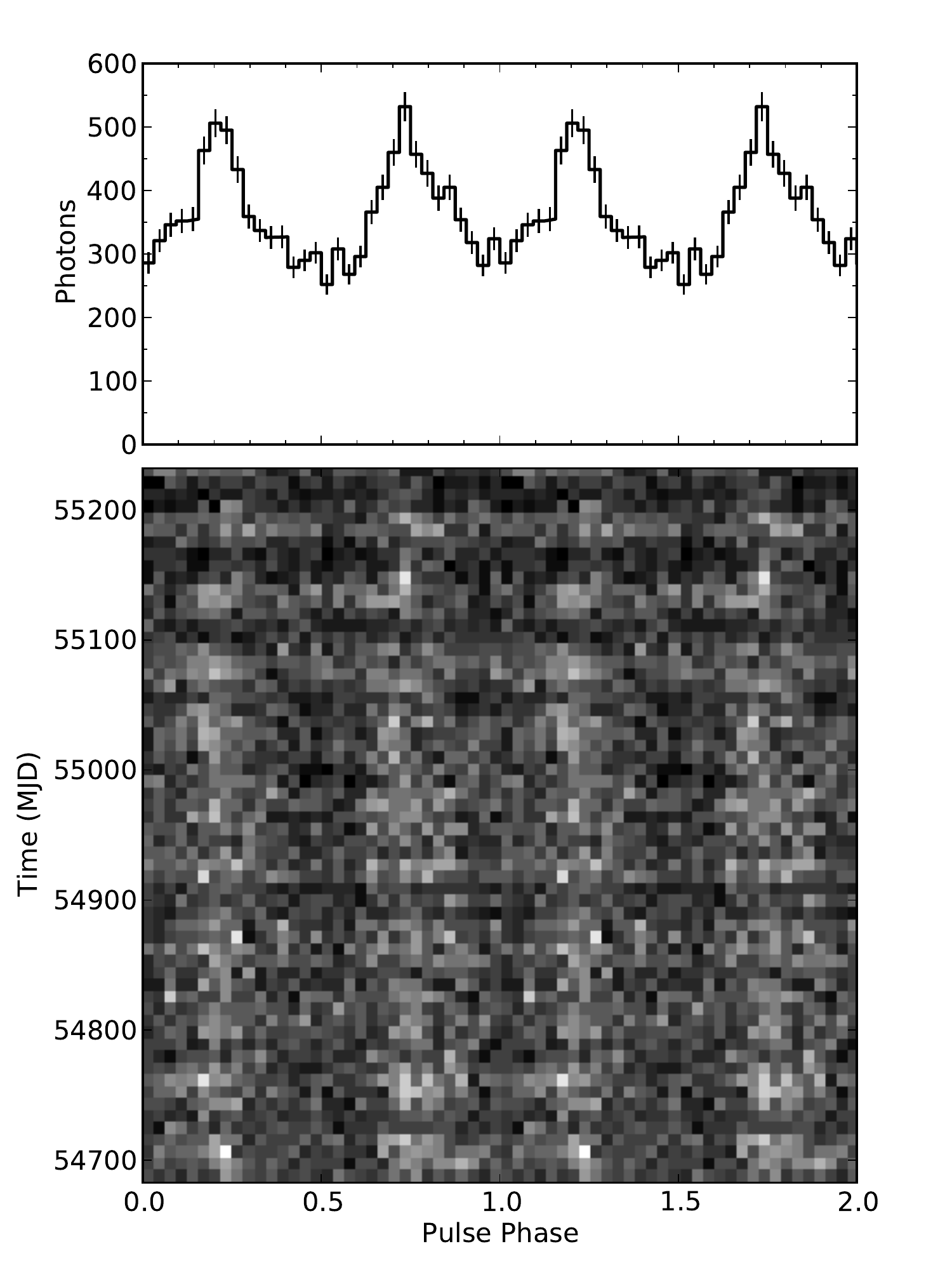}
\caption{2-D phaseogram and pulse profile of PSR J1813$-$1246.  Two rotations are shown on the X-axis. The photons were selected according to the ROI and $E_\mathrm{min}$ in Table~\ref{tab:1813}. The fiducial point corresponding to TZRMJD is phase 0.0.\label{phaseogram:1813}}
\end{figure}

The timing model parameters for this pulsar are displayed in Table~\ref{tab:1813} and the timing position determination, post-fit residuals, 2-D phaseogram, and folded pulse profile for this pulsar are shown in Figures \ref{pos:1813}, \ref{resid:1813}, and \ref{phaseogram:1813}, respectively.

This pulsar, which has the highest spindown luminosity of the first 16 blind search pulsars discovered with the LAT, exhibited a glitch on about 2009 September 20 with magnitude $\Delta\nu/\nu = 1.17 \times 10^{-6}$. The timing model for the glitch presented here represents the data fairly well, but is not unique.  Our model includes an instantaneous and permanent jump in the pulsar frequency and frequency derivative at the glitch.  Other solutions are possible with slightly different glitch epochs (within a day or so) and different parameters, or with decaying transient changes in the spin parameters at the glitch.  With 6 days between TOA measurements and significant timing noise seen in this pulsar, we are not able to distinguish between these possibilities. It is clear in Figure~\ref{resid:1813} that there are significant non-white residuals remaining in the data, indicating that the model is not fully accounting for the spindown behavior of the pulsar.  With a longer span of post-glitch data a more definitive model may be possible.

The bright X-ray source, Swift J181323.4$-$124600, was noted as the counterpart to this pulsar in the discovery paper and our timing confirms the association.

\subsection{PSR J1826$-$1256}

\begin{figure}
\includegraphics[width=3.5in]{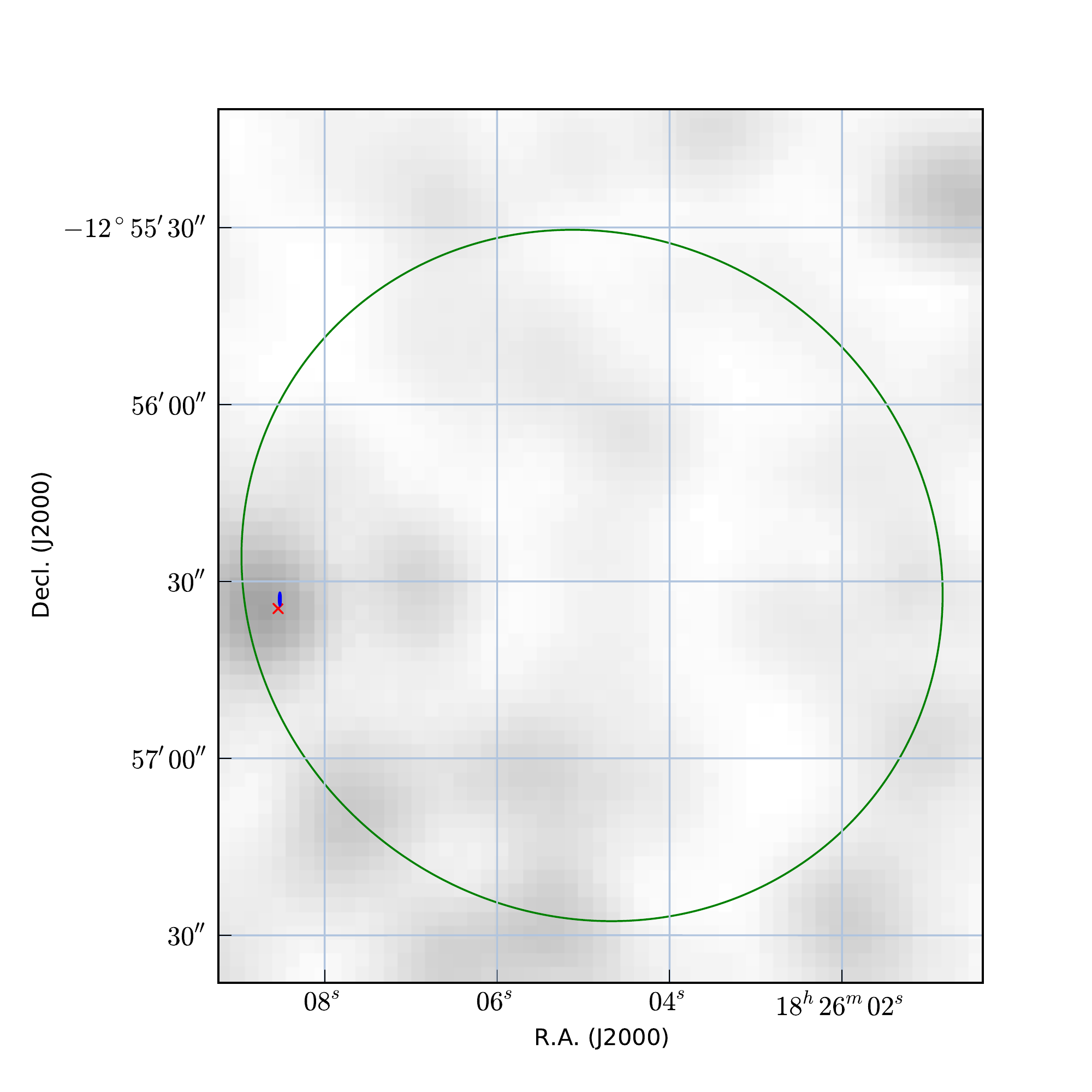} 
\caption{Timing position for PSR J1826$-$1256 (blue ellipse). The large green ellipse is the LAT position of 1FGL J1826.1$-$1256, based on 18 months of data. The background 0.2--10 keV X-ray image is a 4.3 ks \textit{Swift} XRT image (ObsID 00035179002), smoothed with a gaussian with $\sigma = 7$\arcsec. The red cross marks the Chandra position of X-ray point source AX J1826.1$-$1257. \label{pos:1826} }
\end{figure}

\begin{figure}
\includegraphics[width=3.0in]{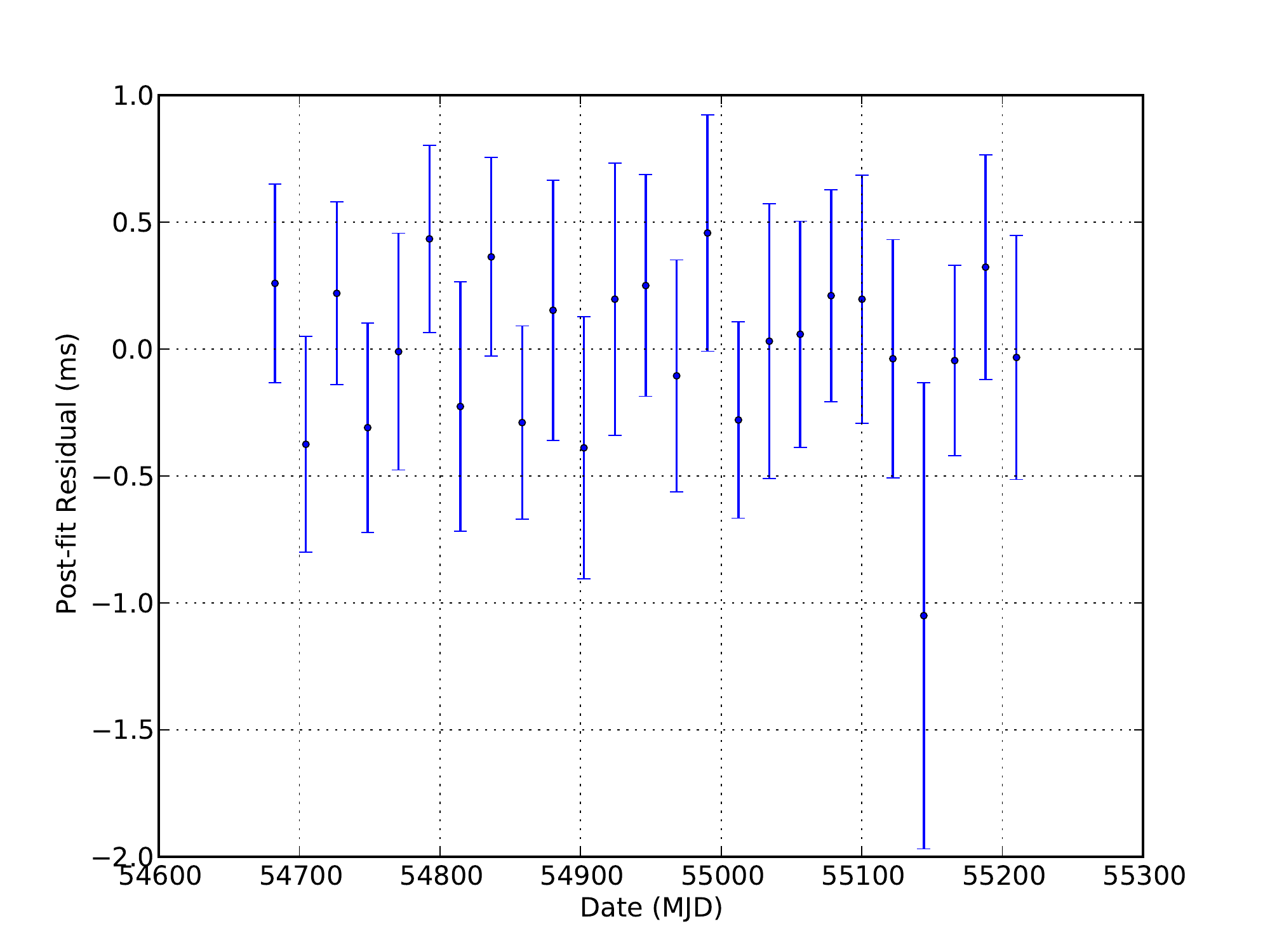}
\caption{Post-fit timing residuals for PSR J1826$-$1256\label{resid:1826}}
\end{figure}

\begin{figure}
\includegraphics[width=3.0in]{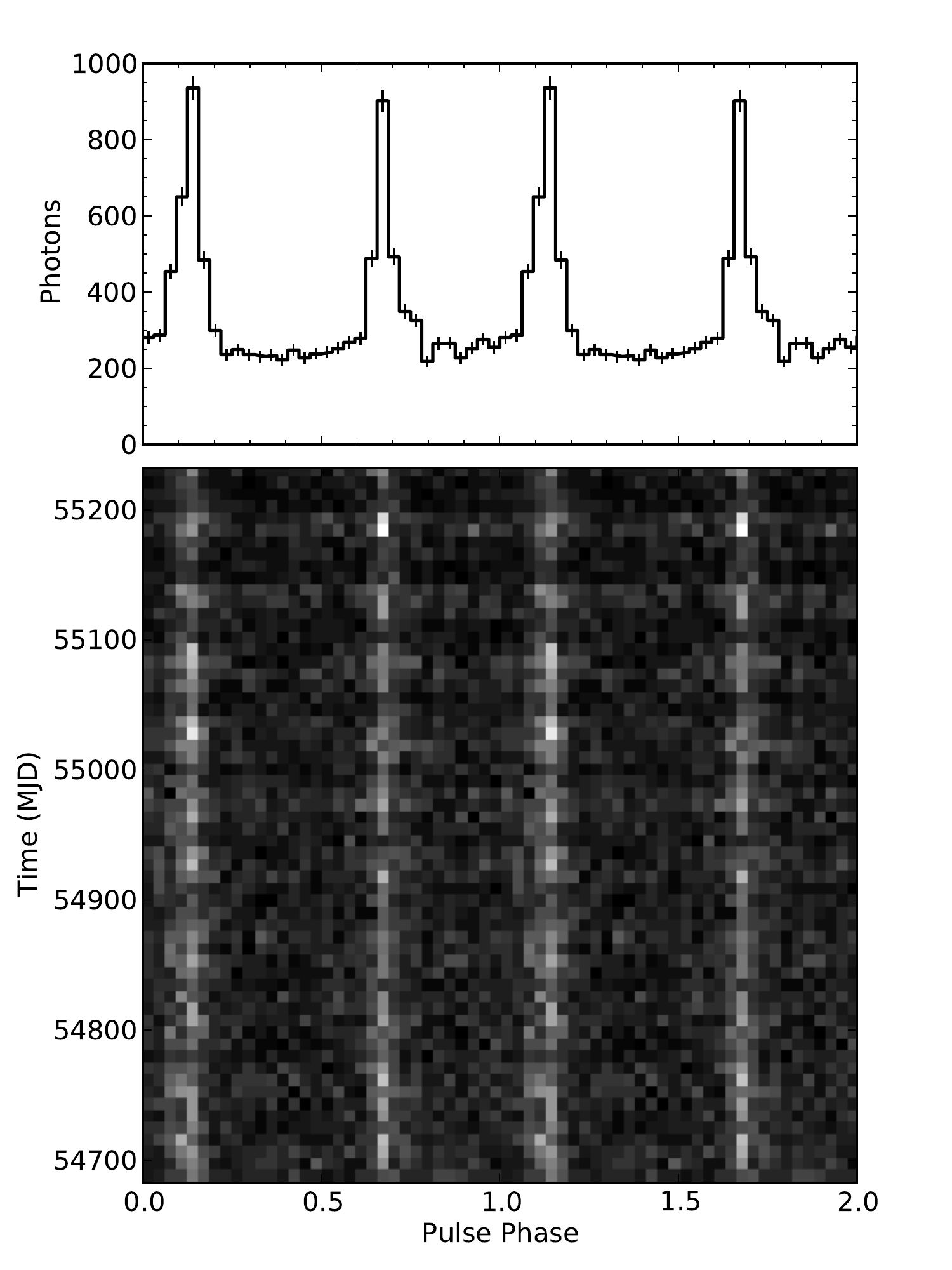}
\caption{2-D phaseogram and pulse profile of PSR J1826$-$1256.  Two rotations are shown on the X-axis. The photons were selected according to the ROI and $E_\mathrm{min}$ in Table~\ref{tab:1826}. The fiducial point corresponding to TZRMJD is phase 0.0.\label{phaseogram:1826}}
\end{figure}

The timing model parameters for this pulsar are displayed in Table~\ref{tab:1826} and the timing position determination, post-fit residuals, 2-D phaseogram, and folded pulse profile for this pulsar are shown in Figures \ref{pos:1826}, \ref{resid:1826}, and \ref{phaseogram:1826}, respectively.

The fast spin and narrow pulse profile allow this pulsar to be localized to about 1\arcsec. The timing position is consistent with the X-ray point source AX J1826.1$-$1257 (R.A. = 18:26:08.2, Decl. = $-12$:56:46), which was discovered in ASCA observations of the EGRET $\gamma$-ray source GeV J1825$-$1310 \citep{rrk01}. An improved position of the X-ray point source (R.A. = 18:26:08.54, Decl. = $-12$:56:34.6) was derived from a Chandra image (M. Roberts, private communication), which is 1.6\arcsec\ from the timing position.  The measured $\ddot{\nu}$ for this pulsar is quite large, and the Monte Carlo error estimate yields 17\arcsec, which is much larger than the offset seen between the timing position and the X-ray counterpart. 

\subsection{PSR J1836+5925}

\begin{figure}
\includegraphics[width=3.5in]{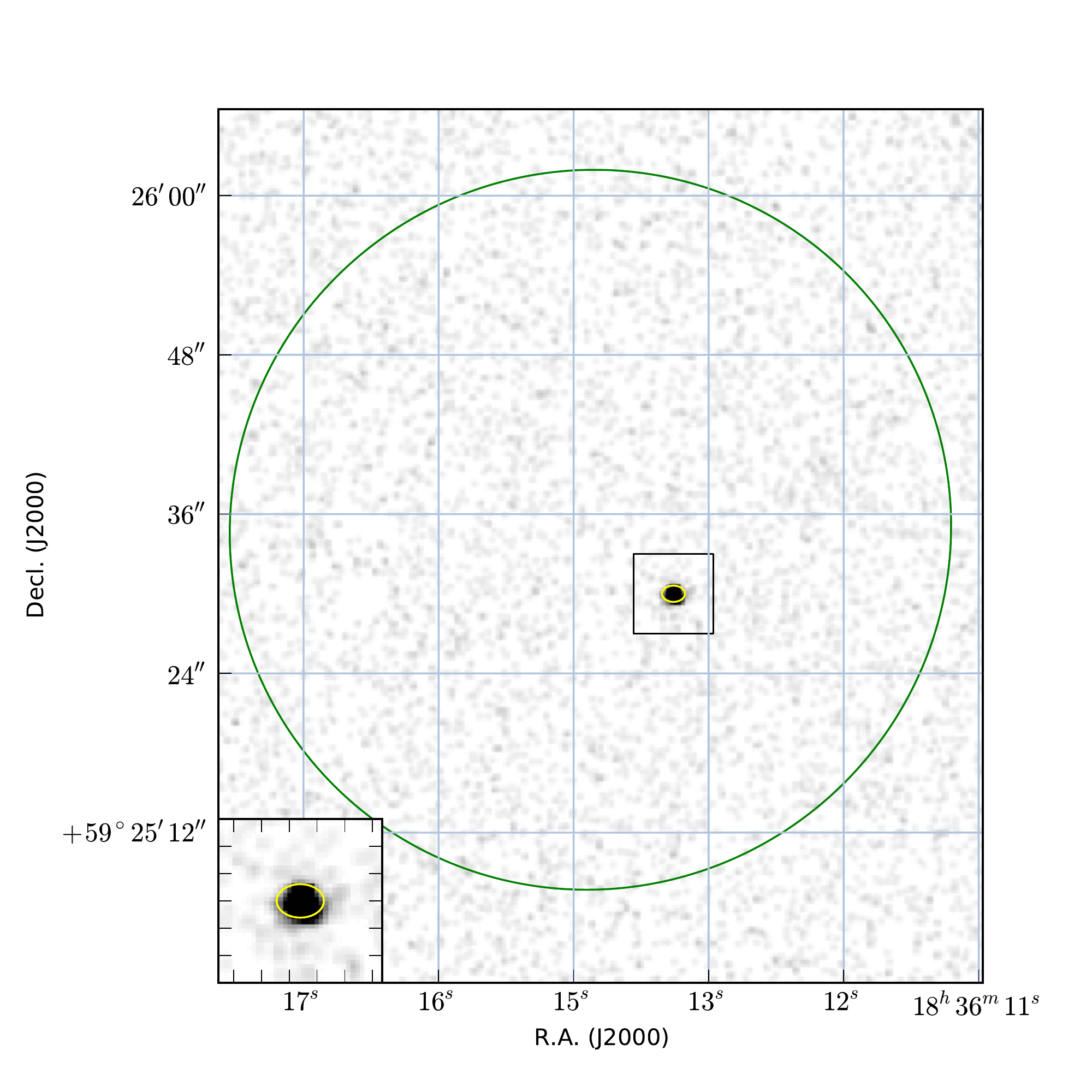} 
\caption{Timing position for PSR J1836+5925 (yellow ellipse). The large green ellipse is the LAT position of 1FGL J1836.2+5925, based on 18 months of data. The X-ray image is a 46 ks \textit{Chandra} HRC image (ObsId 6182) and the point source at the timing position is RX J1836.2+5925. The inset (3.0\arcsec\ in width) shows the region of the source in more detail.\label{pos:1836} }
\end{figure}

\begin{figure}
\includegraphics[width=3.0in]{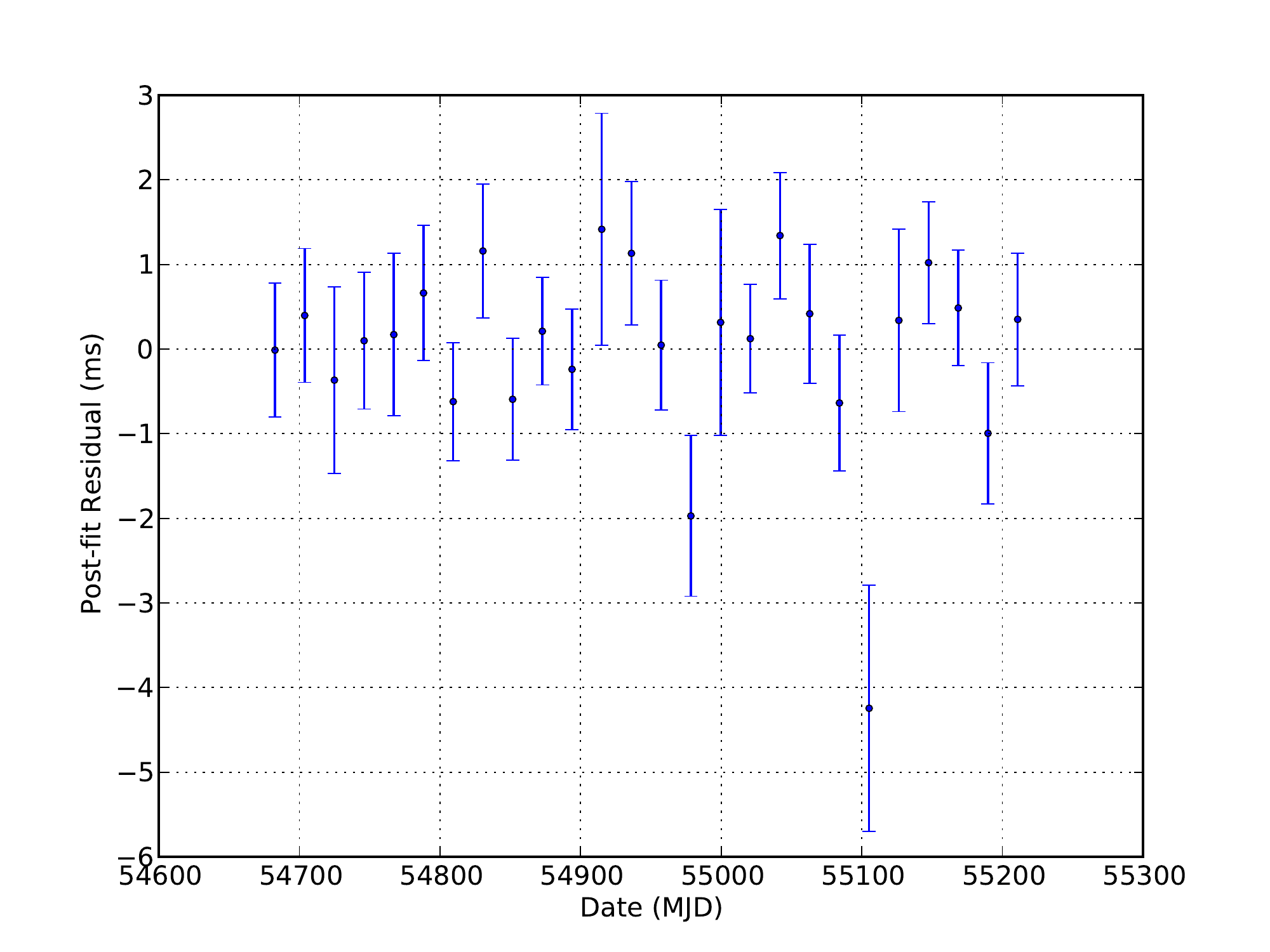}
\caption{Post-fit timing residuals for PSR J1836+5925.\label{resid:1836}}
\end{figure}

\begin{figure}
\includegraphics[width=3.0in]{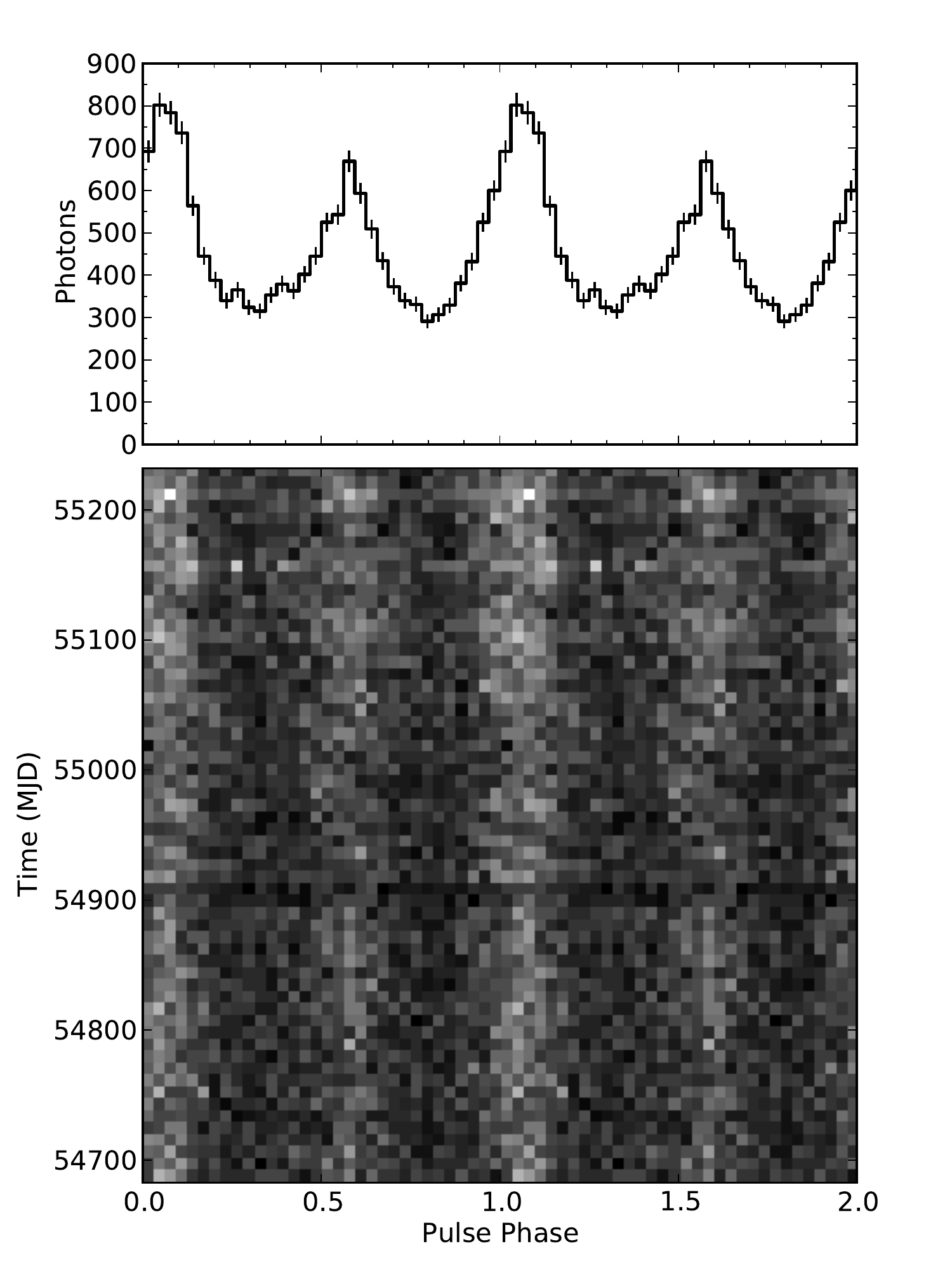}
\caption{2-D phaseogram and pulse profile of PSR J1836+5925.  Two rotations are shown on the X-axis. The photons were selected according to the ROI and $E_\mathrm{min}$ in Table~\ref{tab:1836}. The fiducial point corresponding to TZRMJD is phase 0.0.\label{phaseogram:1836}}
\end{figure}

The timing model parameters for this pulsar are displayed in Table~\ref{tab:1836} and the timing position determination, post-fit residuals, 2-D phaseogram, and folded pulse profile for this pulsar are shown in Figures \ref{pos:1836}, \ref{resid:1836}, and \ref{phaseogram:1836}, respectively.

A detailed analysis and earlier timing model for this source were published by \citet{J1836}.  Our model, which includes an additional 6 months of data is consistent with the earlier results with the addition of a weak detection of a frequency second derivative term. As seen in Figure~\ref{pos:1836}, the timing position is fully consistent with the X-ray source RX J1836.2+5925, with the offset being only 0.2\arcsec.

\subsection{PSR J1907+0602}

\begin{figure}
\includegraphics[width=3.5in]{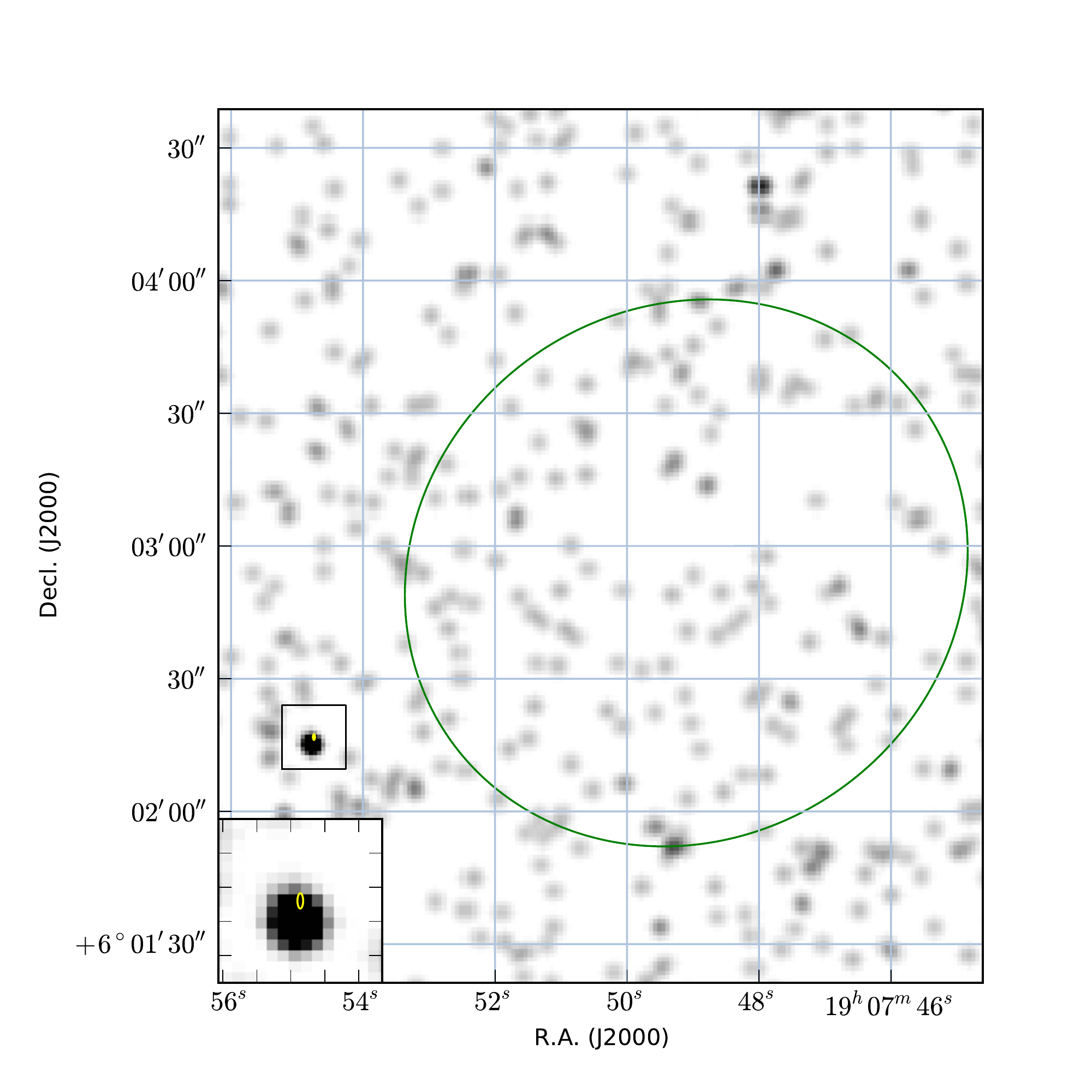} 
\caption{Timing position for PSR J1907+0602 (yellow ellipse). The large green ellipse is the LAT position of 1FGL J1907.9+0602, based on 18 months of data. The X-ray image is a 0.75--2.0 keV \textit{Chandra} image (ObsID 11124). The inset (7.2\arcsec\ in width) shows a detail of the region around the pulsar.  \label{pos:1907} }
\end{figure}

\begin{figure}
\includegraphics[width=3.0in]{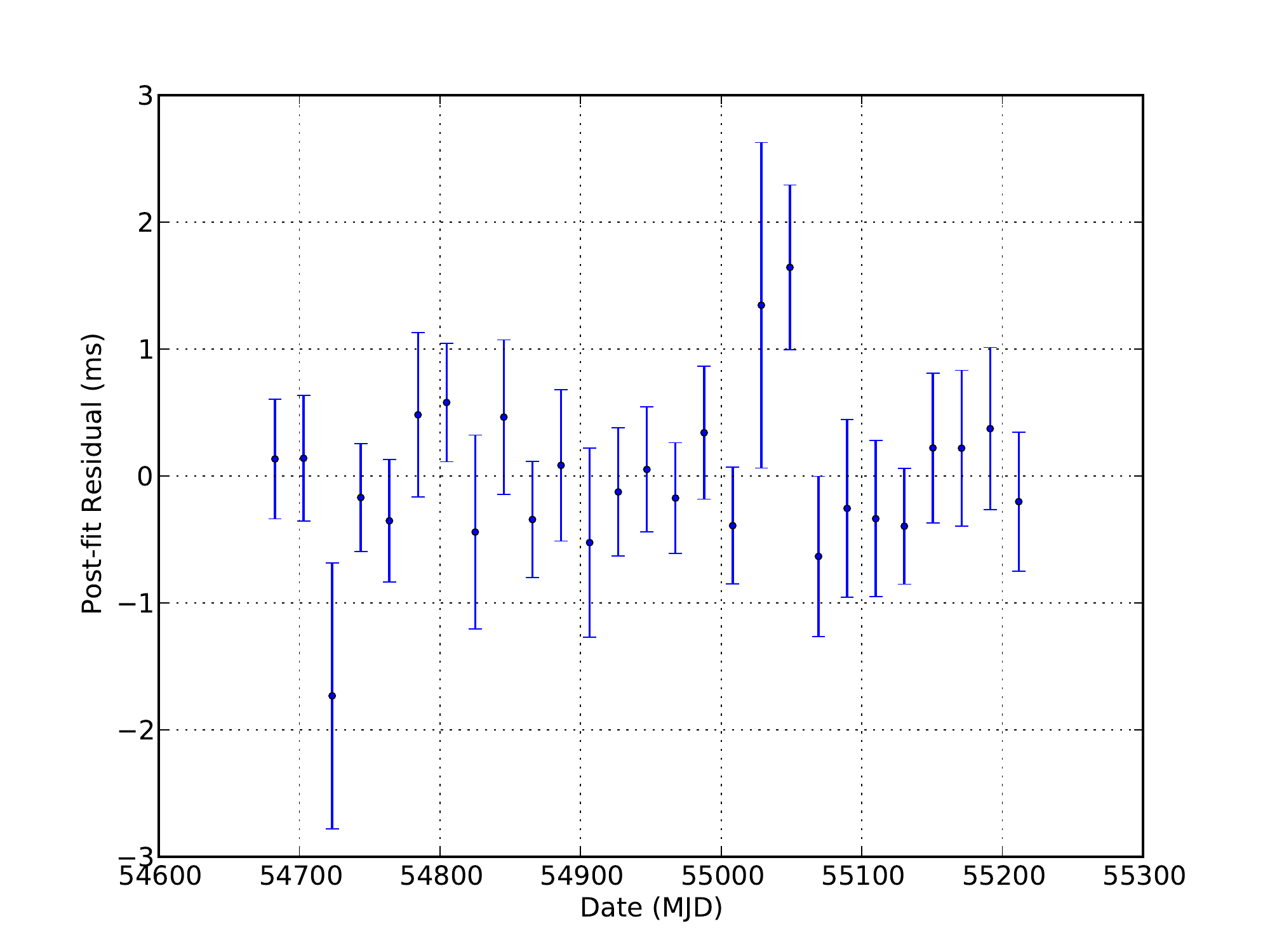}
\caption{Post-fit timing residuals for PSR J1907+0602.\label{resid:1907}}
\end{figure}

\begin{figure}
\includegraphics[width=3.0in]{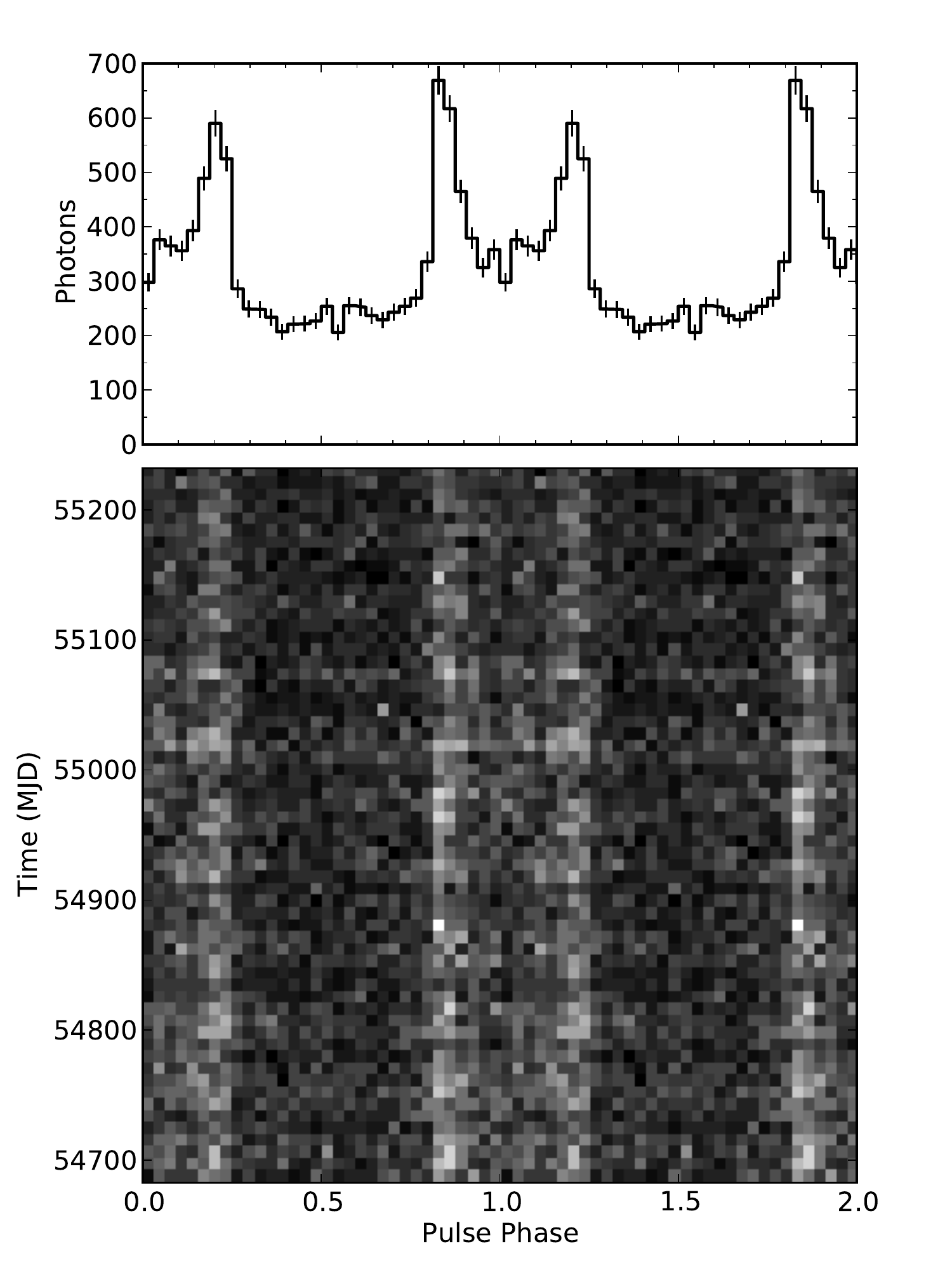}
\caption{2-D phaseogram and pulse profile of PSR J1907+0602.  Two rotations are shown on the X-axis. The photons were selected according to the ROI and $E_\mathrm{min}$ in Table~\ref{tab:1907}. The fiducial point corresponding to TZRMJD is phase 0.0.\label{phaseogram:1907}}
\end{figure}

The timing model parameters for this pulsar are displayed in Table~\ref{tab:1907} and the timing position determination, post-fit residuals, 2-D phaseogram, and folded pulse profile for this pulsar are shown in Figures \ref{pos:1907}, \ref{resid:1907}, and \ref{phaseogram:1907},  respectively.

A detailed analysis of this pulsar, including an earlier timing solution and the discovery of radio pulsations, was presented by \citet{MGROPaper}. Our timing model is fully consistent within the errors to the one they presented, though we now have almost 5 months more data, which significantly reduces the uncertainties in the parameters. The position reported for the \textit{Chandra} point source is 2.3\arcsec\ from our best timing position, which is significantly larger than the 0.6\arcsec\ statistical error in the timing position or the 0.6\arcsec\ error in the \textit{Chandra} position. However, the offset is comparable to the expected systematic error from timing noise, based on the Monte Carlo simulations, so this is not strong evidence against the association.

\subsection{PSR J1958+2846}

\begin{figure}
\includegraphics[width=3.5in]{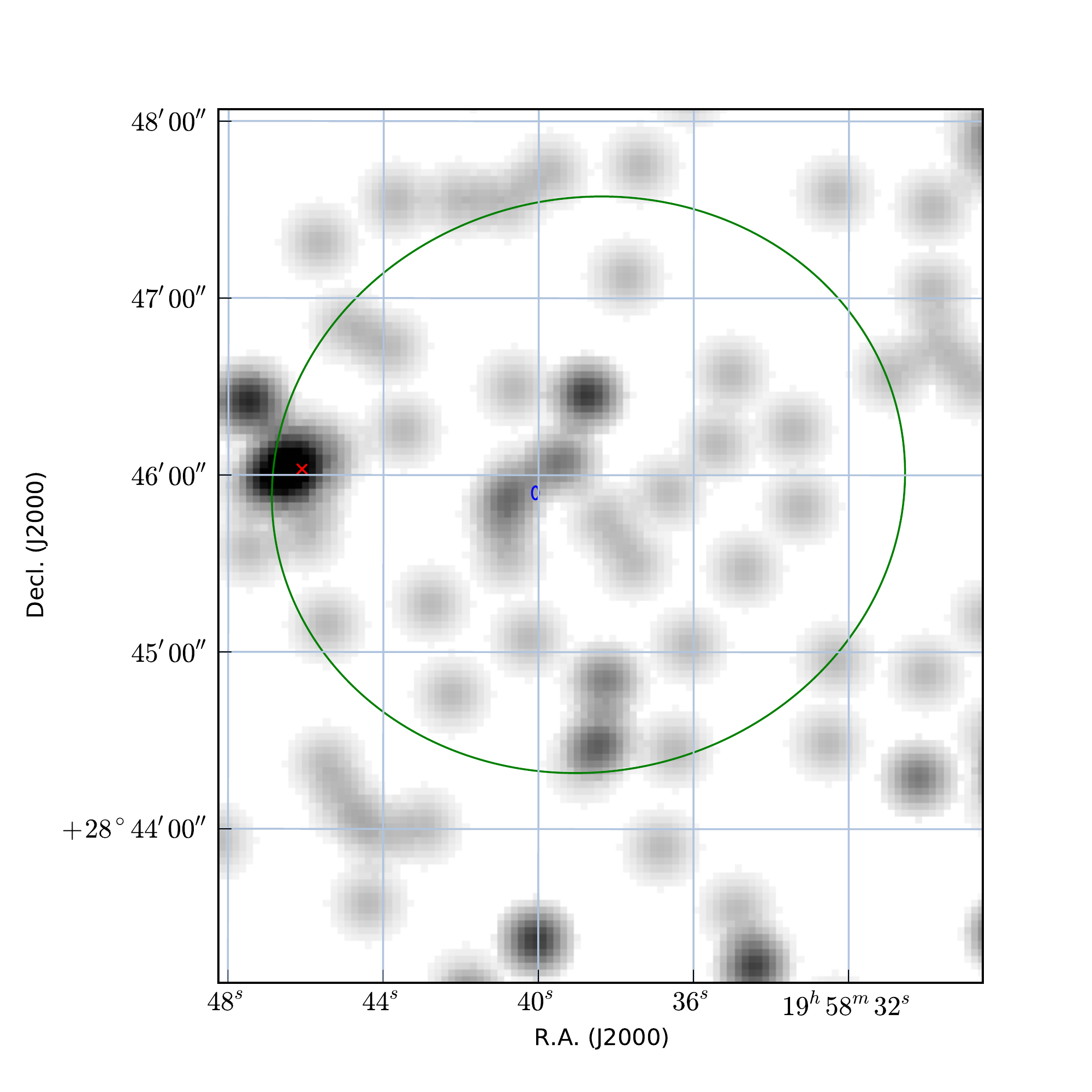} 
\caption{Timing position for PSR J1958+2846 (blue ellipse). The large green ellipse is the LAT position of 1FGL J1958.6+2845, based on 18 months of data. The background 0.2--10 keV X-ray image is a 5.5 ks \textit{Swift} XRT image (ObsID 00031374001), smoothed with a gaussian with $\sigma = 7$\arcsec. The red `x' marks the position of Swift J195846.1+284602. \label{pos:1958} }
\end{figure}

\begin{figure}
\includegraphics[width=3.0in]{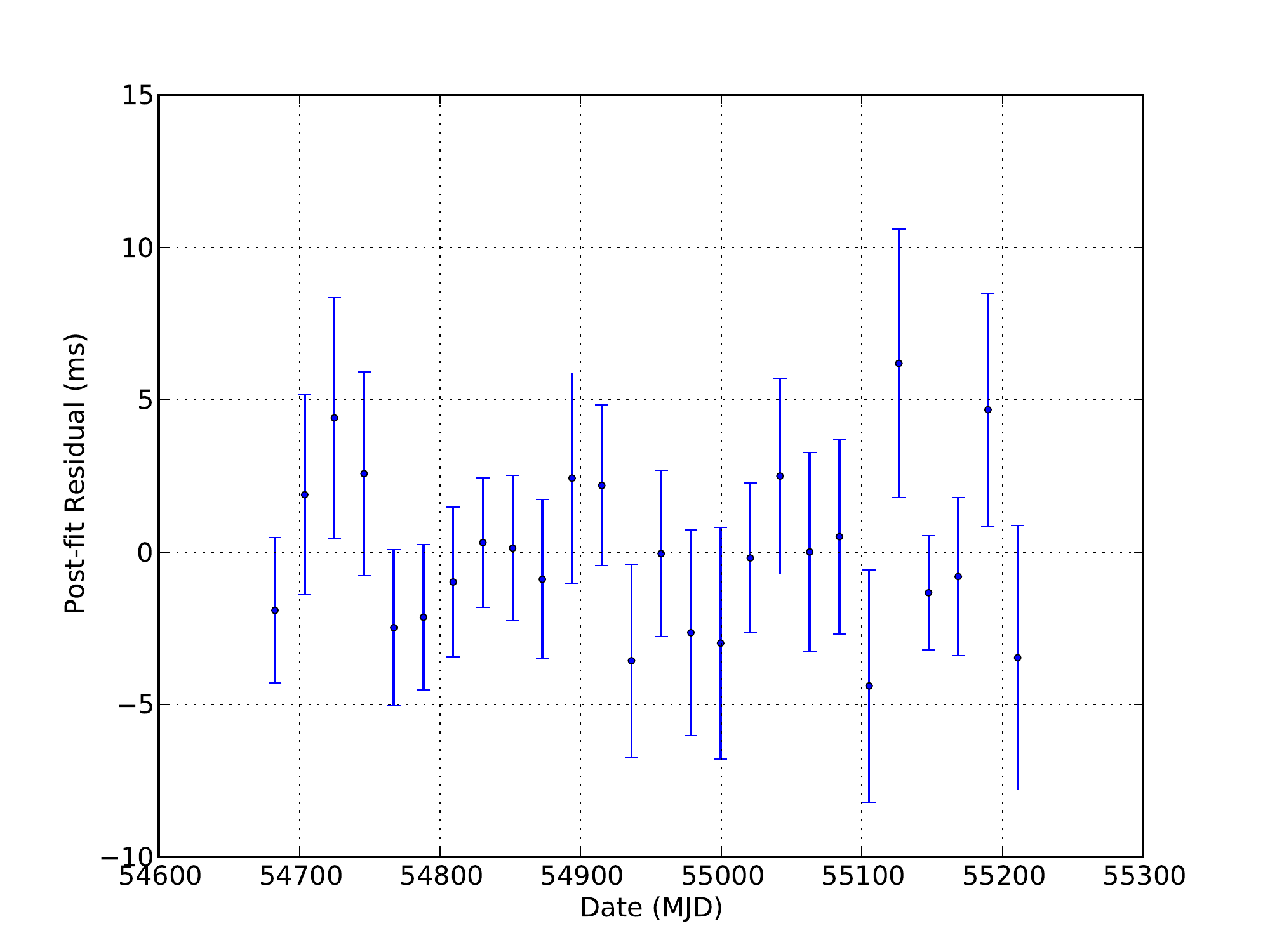}
\caption{Post-fit timing residuals for PSR J1958+2846.\label{resid:1958}}
\end{figure}

\begin{figure}
\includegraphics[width=3.0in]{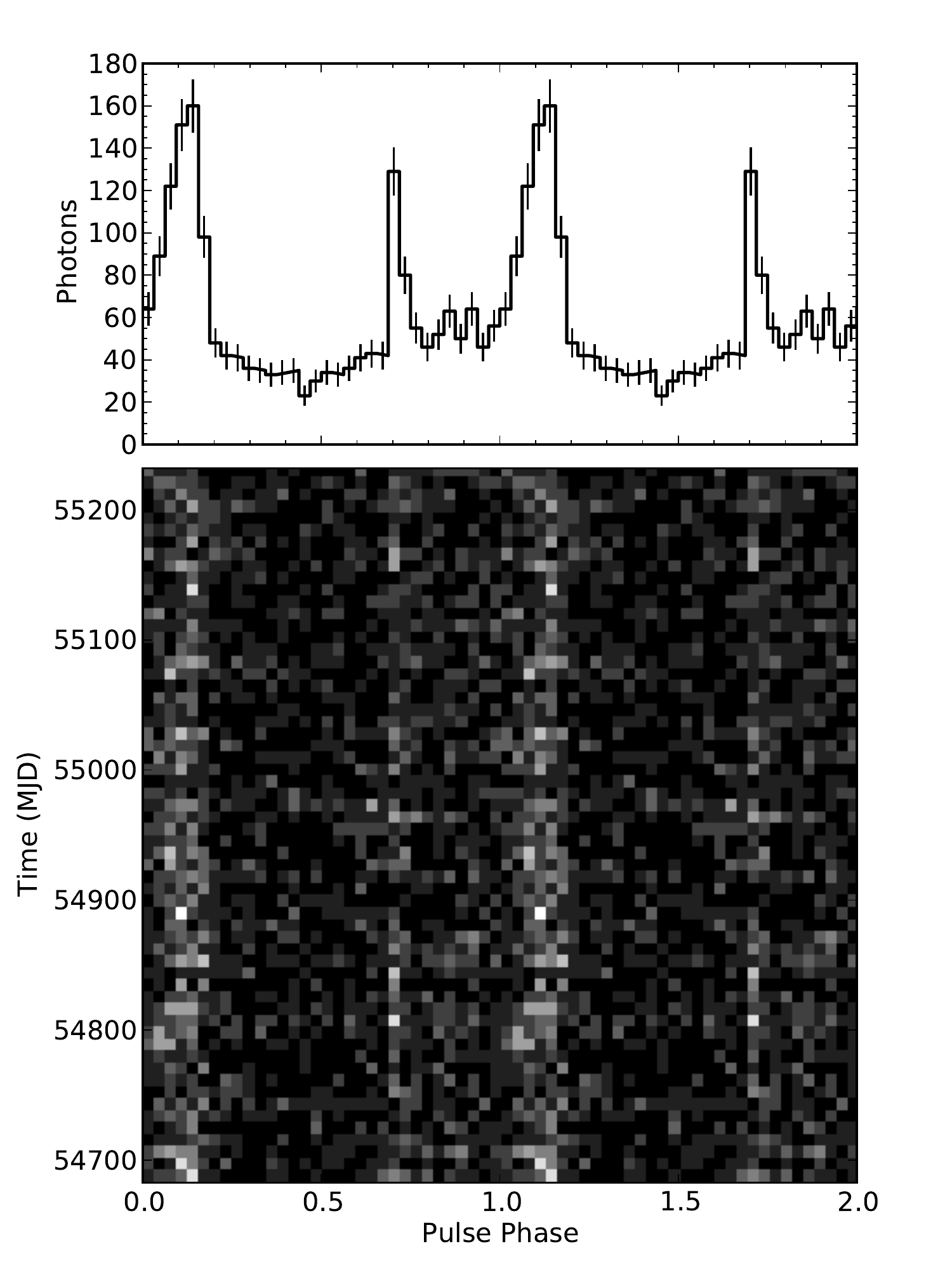}
\caption{2-D phaseogram and pulse profile of PSR J1958+2846.  Two rotations are shown on the X-axis. The photons were selected according to the ROI and $E_\mathrm{min}$ in Table~\ref{tab:1958}. The fiducial point corresponding to TZRMJD is phase 0.0.\label{phaseogram:1958}}
\end{figure}

The timing model parameters for this pulsar are displayed in Table~\ref{tab:1958} and the timing position determination, post-fit residuals, 2-D phaseogram, and folded pulse profile for this pulsar are shown in Figures \ref{pos:1958}, \ref{resid:1958}, and \ref{phaseogram:1958}, respectively.

In the discovery paper \citep{LATBlindSearch}, the X-ray source Swift J195846.1+284602 was proposed as the likely counterpart and used for the name of the pulsar.  As shown in Figure~\ref{pos:1958}, the timing position no longer supports this identification, being offset from the X-ray source by 80\arcsec.  There is no significant X-ray counterpart detected at the timing position. There is also no indication for strong timing noise in this pulsar, which might cause a large systematic error in the timing position. Deeper X-ray observations will be required to detect the true X-ray counterpart for this source.


\subsection{PSR J2021+4026}

\begin{figure}
\includegraphics[width=3.5in]{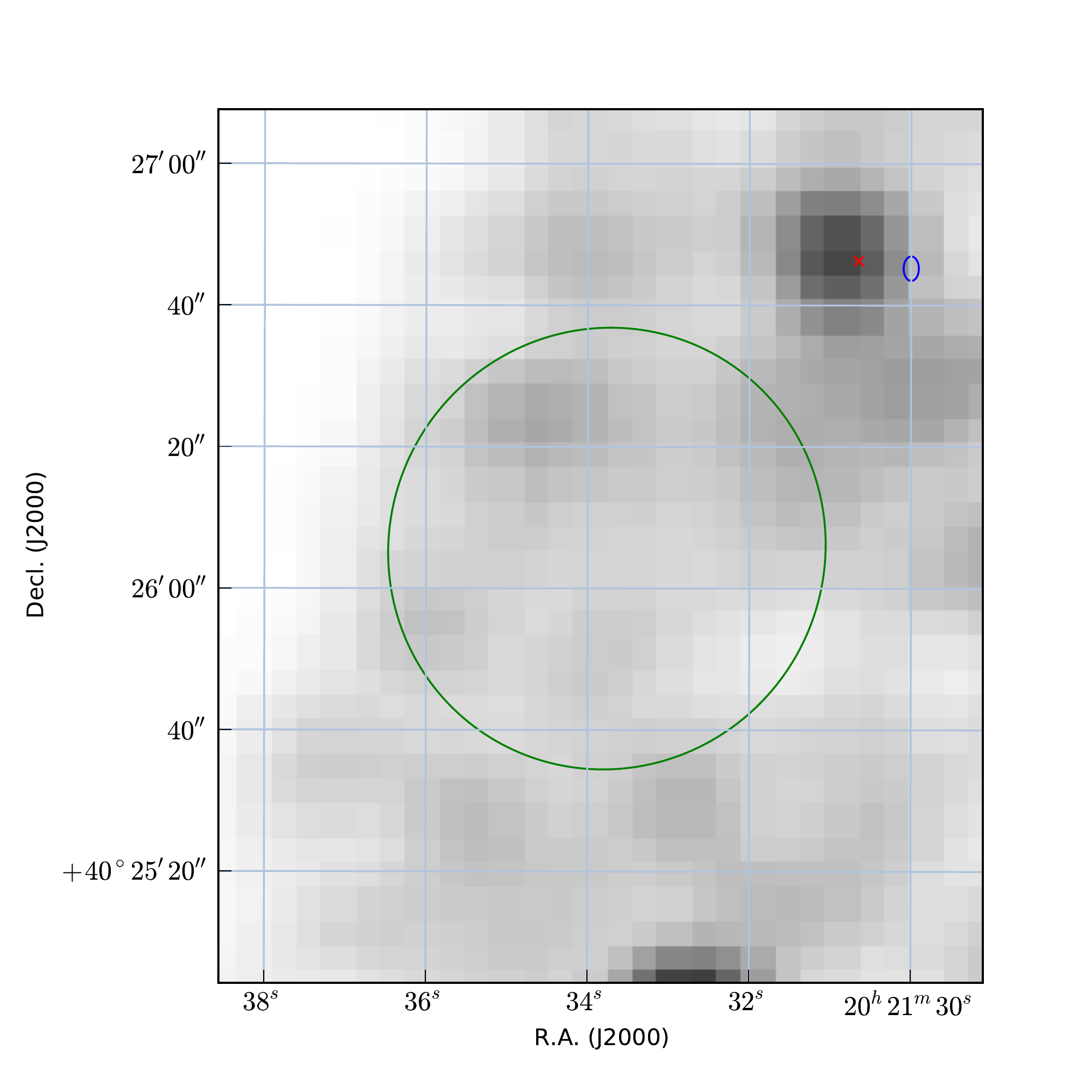} 
\caption{Timing position for PSR J2021+4026 (blue ellipse). The large green ellipse is the LAT position of 1FGL J2021.5+4026, based on 18 months of data. The red `x' marks the position of the source S21 (see text). The background X-ray image is a portion of a \textit{Chandra} ACIS-I image (ObsID 5533), with 3-pixel gaussian smoothing. \label{pos:2021} }
\end{figure}

\begin{figure}
\includegraphics[width=3.0in]{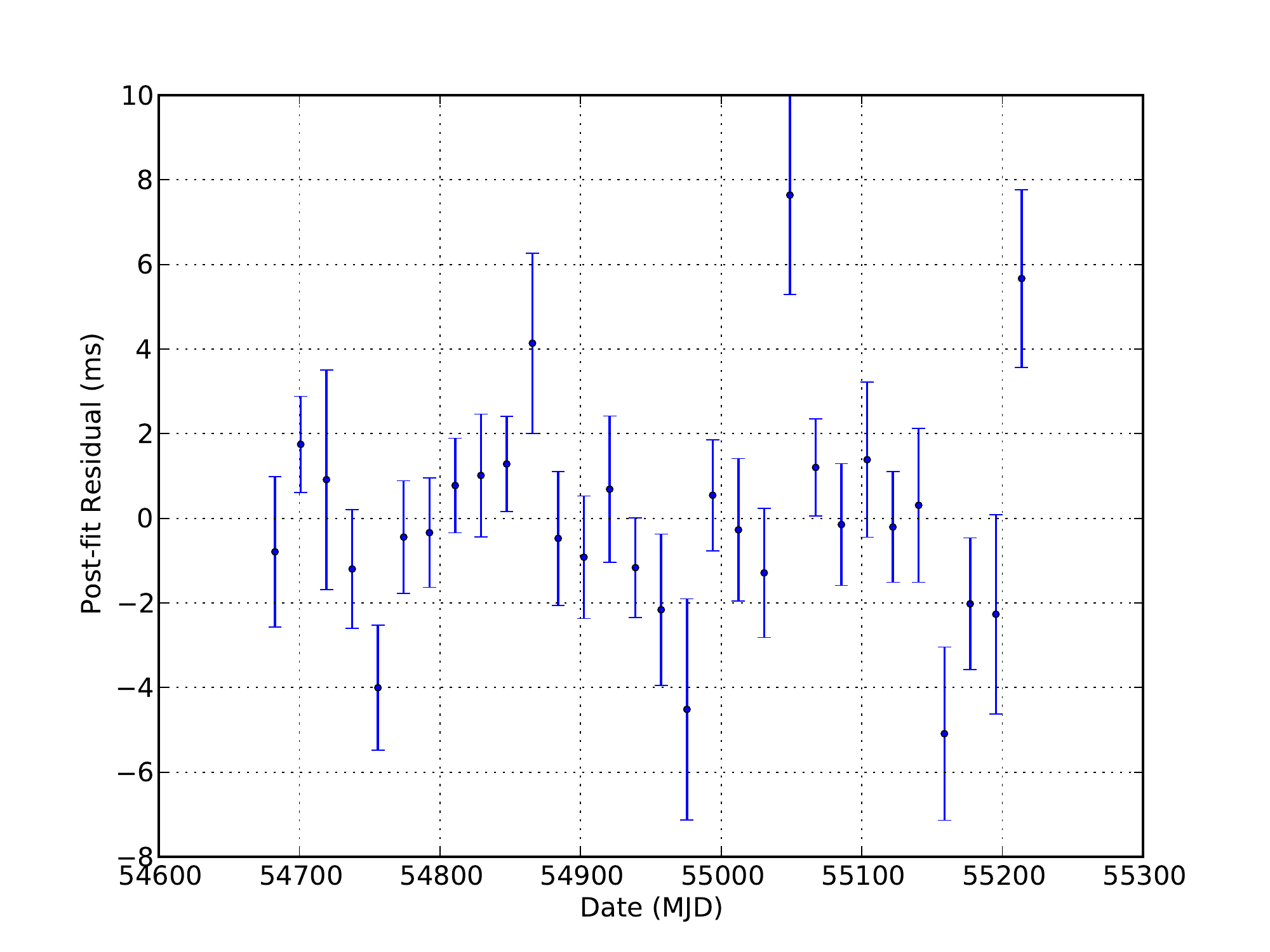}
\caption{Post-fit timing residuals for PSR J2021+4026.\label{resid:2021}}
\end{figure}

\begin{figure}
\includegraphics[width=3.0in]{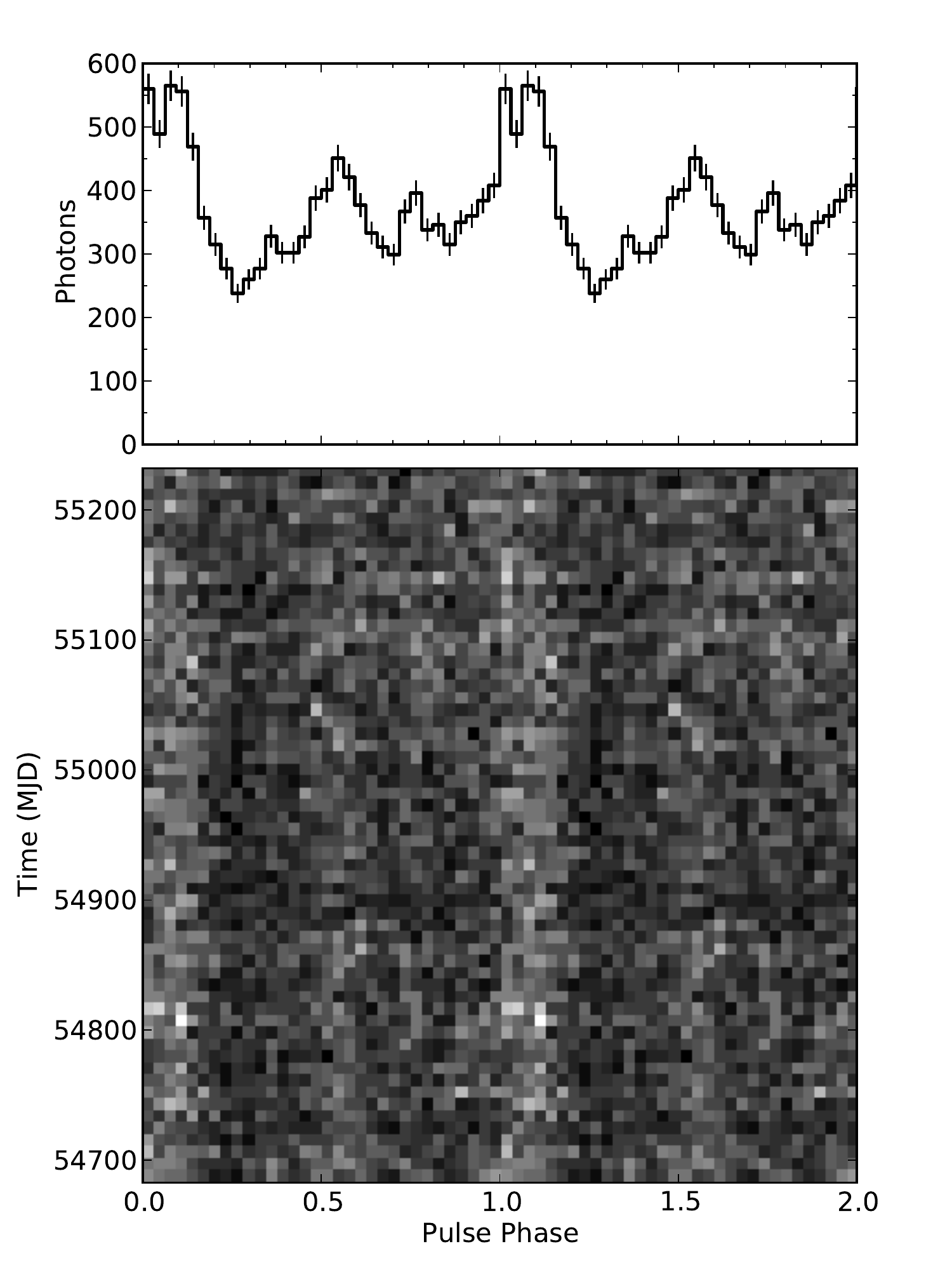}
\caption{2-D phaseogram and pulse profile of PSR J2021+4026.  Two rotations are shown on the X-axis. The photons were selected according to the ROI and $E_\mathrm{min}$ in Table~\ref{tab:2021}. The fiducial point corresponding to TZRMJD is phase 0.0.\label{phaseogram:2021}}
\end{figure}

The timing model parameters for this pulsar are displayed in Table~\ref{tab:2021} and the timing position determination, post-fit residuals, 2-D phaseogram, and folded pulse profile for this pulsar are shown in Figures \ref{pos:2021}, \ref{resid:2021}, and \ref{phaseogram:2021}, respectively.

PSR J2021+4026 is the long-sought pulsar in the $\gamma$ Cygni supernova remnant (SNR G78.2+2.1).  In the discovery paper \citep{LATBlindSearch}, it was pointed out that the X-ray source S21, as identified earlier in Chandra observations \citep{wsc+06}, was the most likely counterpart based on the initial pulsar timing. Our best fit position is 7.7\arcsec\ to the west of S21 (see Figure~\ref{pos:2021}), an offset that is somewhat larger than the predicted systematic error of 2.5\arcsec\ based on our Monte Carlo. However, when $\dddot{\nu}$ is added to the model, the position shifts by 4.6\arcsec, so this is a lower bound on the systematic error from the timing noise.  As S21 is still the closest X-ray source to the timing position, we conclude that it is indeed the likely counterpart. Longer term timing will improve our localization and reduce the systematic error contribution from timing noise. A similar conclusion was reached by \citet{thc+10}.

Both the timing position and the X-ray source S21 are well outside the 95\% confidence localization ellipse of the LAT sources, as seen in Figure~\ref{pos:2021}.  However, this region includes statistical errors only and this source is in the very complicated Cygnus region of the Galaxy. The localization of 0FGL J2021.5+4026 in the LAT Bright Source List \citep{BSL} did include S21, but with the improved statistics using 18 months of data, systematic errors due to improperly modeled diffuse emission or unknown point sources can start to dominate the error budget.  The association of the LAT source with the pulsar is beyond doubt because of the detection of pulsations.

\subsection{PSR J2032+4127}

\begin{figure}
\includegraphics[width=3.5in]{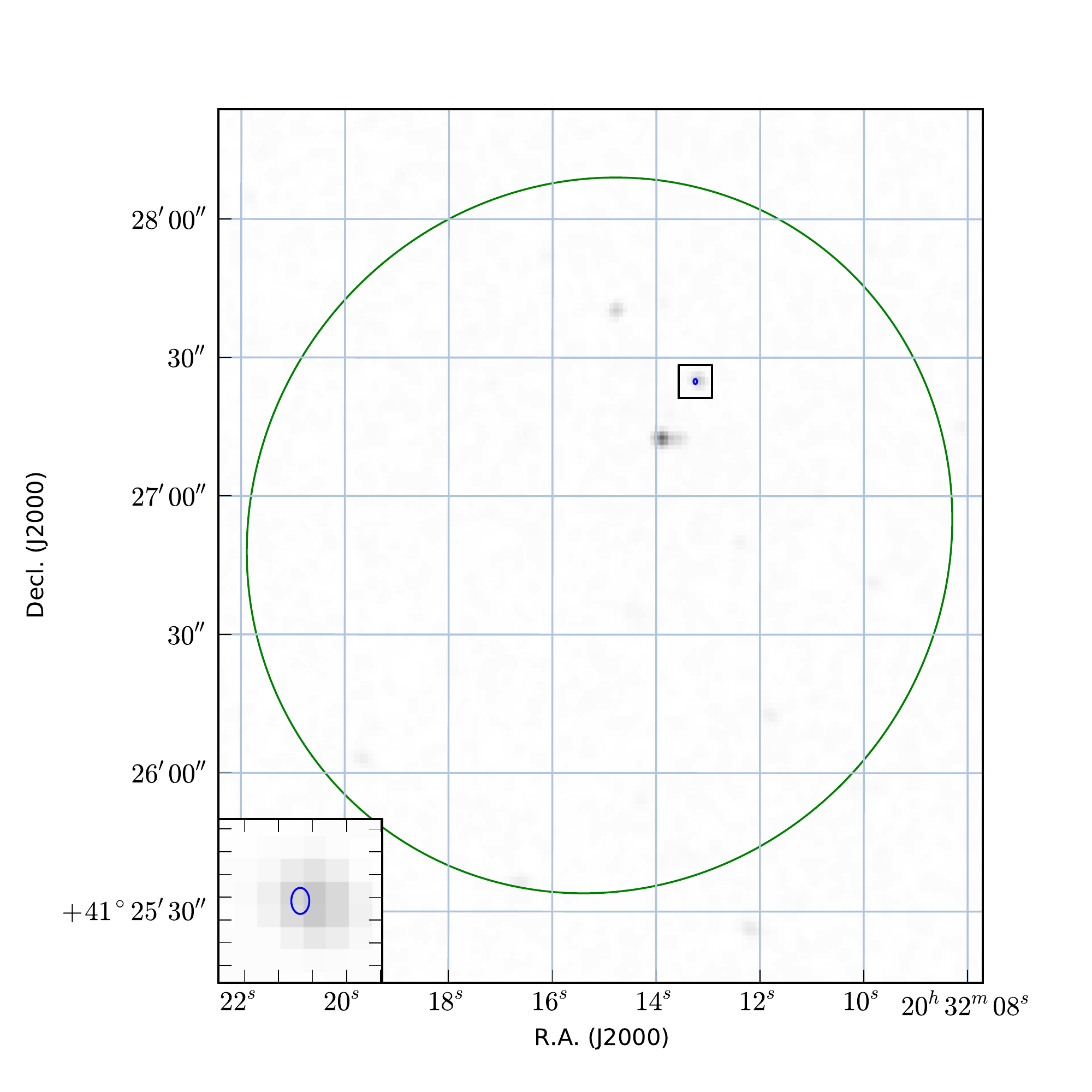} 
\caption{Timing position for PSR J2032+4127 (blue ellipse). The large green ellipse is the LAT position of 1FGL J2032.2+4127, based on 18 months of data. The background image is from a 49 ks Chandra ACIS observation (ObsID 4501). The inset shows a 3.6\arcsec\ region around the timing position in more detail.\label{pos:2032} }
\end{figure}

\begin{figure}
\includegraphics[width=3.0in]{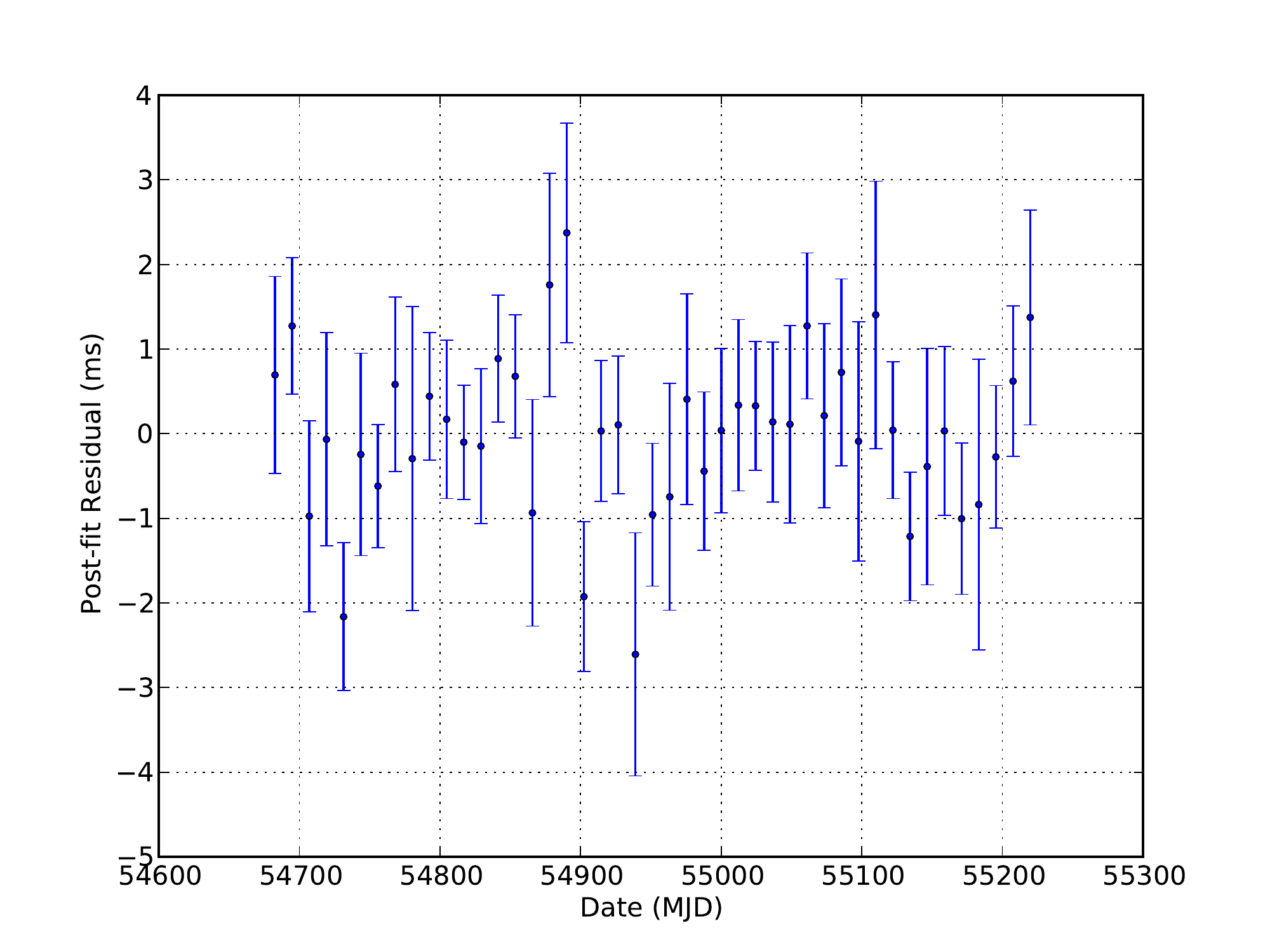}
\caption{Post-fit timing residuals for PSR J2032+4127.\label{resid:2032}}
\end{figure}

\begin{figure}
\includegraphics[width=3.0in]{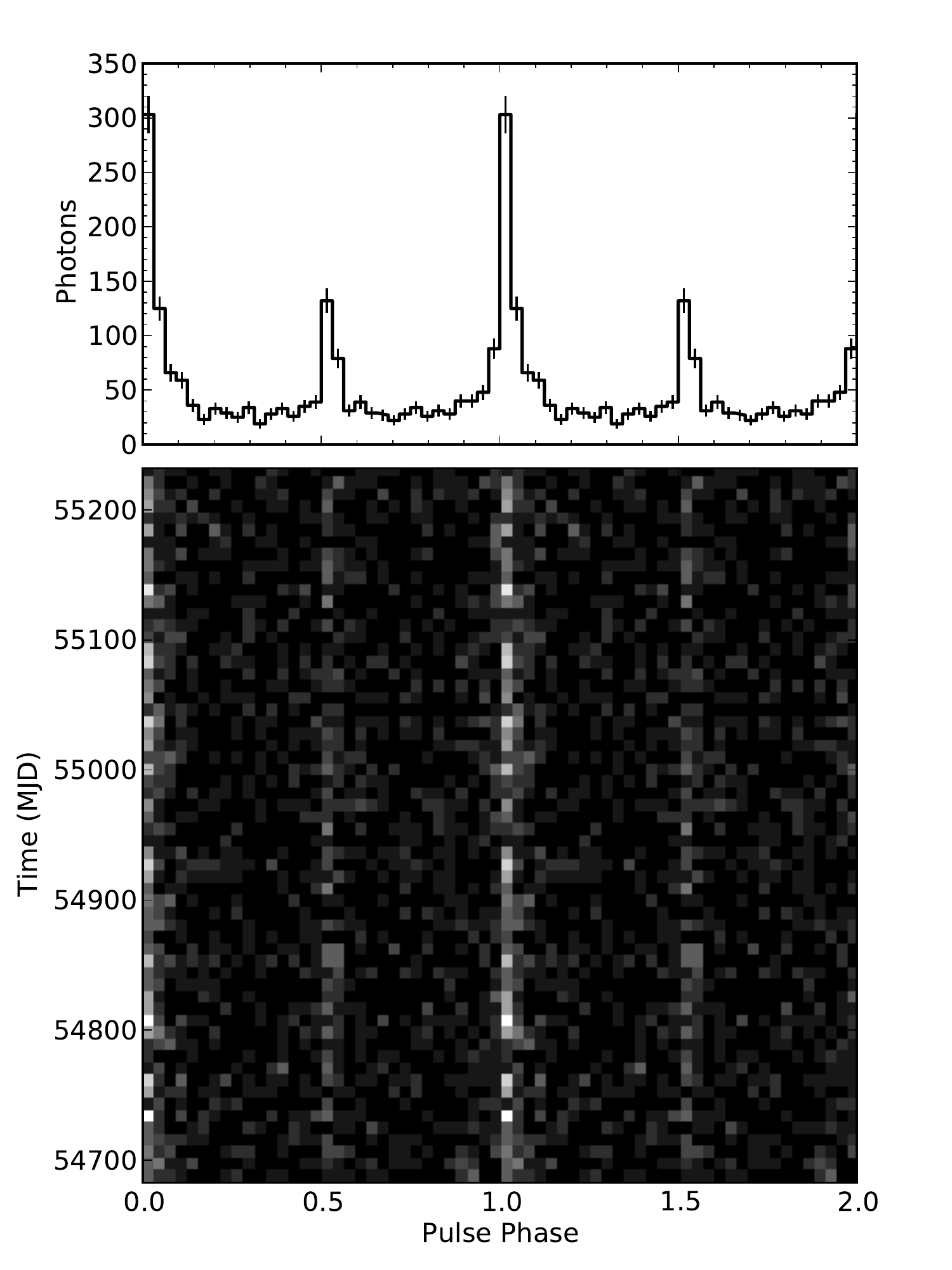}
\caption{2-D phaseogram and pulse profile of PSR J2032+4127.  Two rotations are shown on the X-axis. The photons were selected according to the ROI and $E_\mathrm{min}$ in Table~\ref{tab:2032}. The fiducial point corresponding to TZRMJD is phase 0.0.\label{phaseogram:2032}}
\end{figure}

The timing model parameters for this pulsar are displayed in Table~\ref{tab:2032} and the timing position determination, post-fit residuals, 2-D phaseogram, and folded pulse profile for this pulsar are shown in Figures \ref{pos:2032}, \ref{resid:2032}, and \ref{phaseogram:2032}, respectively.

This pulsar is studied in detail by \citet{crr+09}, who reported the discovery of radio pulsations from this source. The model presented here is consistent with theirs, and the positional association with the \textit{Chandra} point source MT91 213 is confirmed (see Figure \ref{pos:2032}).

\subsection{PSR J2238+5904}

\begin{figure}
\includegraphics[width=3.5in]{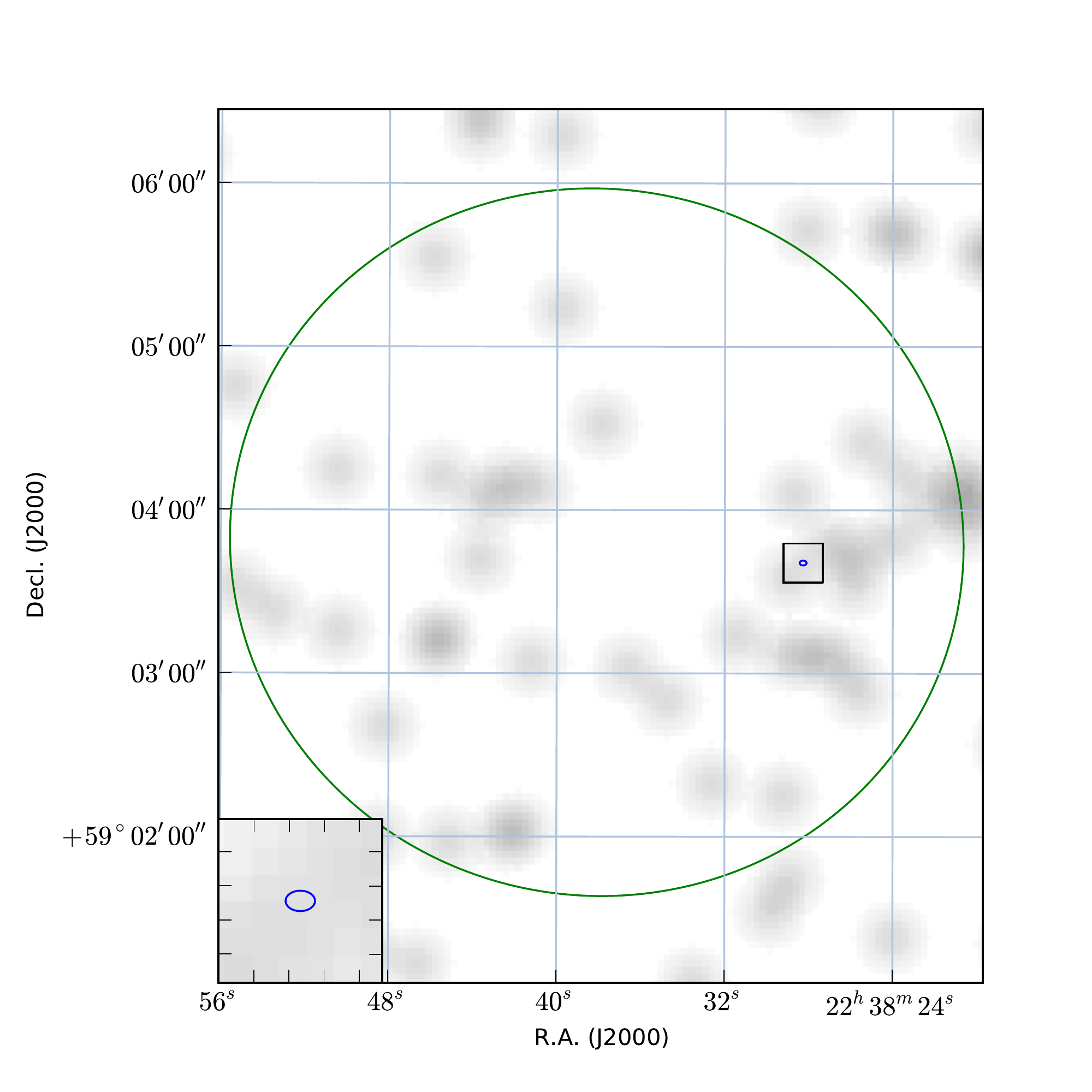} 
\caption{Timing position for PSR J2238+5903 (blue ellipse). The large green ellipse is the LAT position of 1FGL J2238.4+5903, based on 18 months of data.  The background 0.2--10 keV X-ray image is a 4.9 ks \textit{Swift} image (ObsID 00031398001), smoothed with a gaussian with $\sigma = 7$\arcsec. The inset shows a 7.2\arcsec\ region around the pulsar in more detail. \label{pos:2238} }
\end{figure}

\begin{figure}
\includegraphics[width=3.0in]{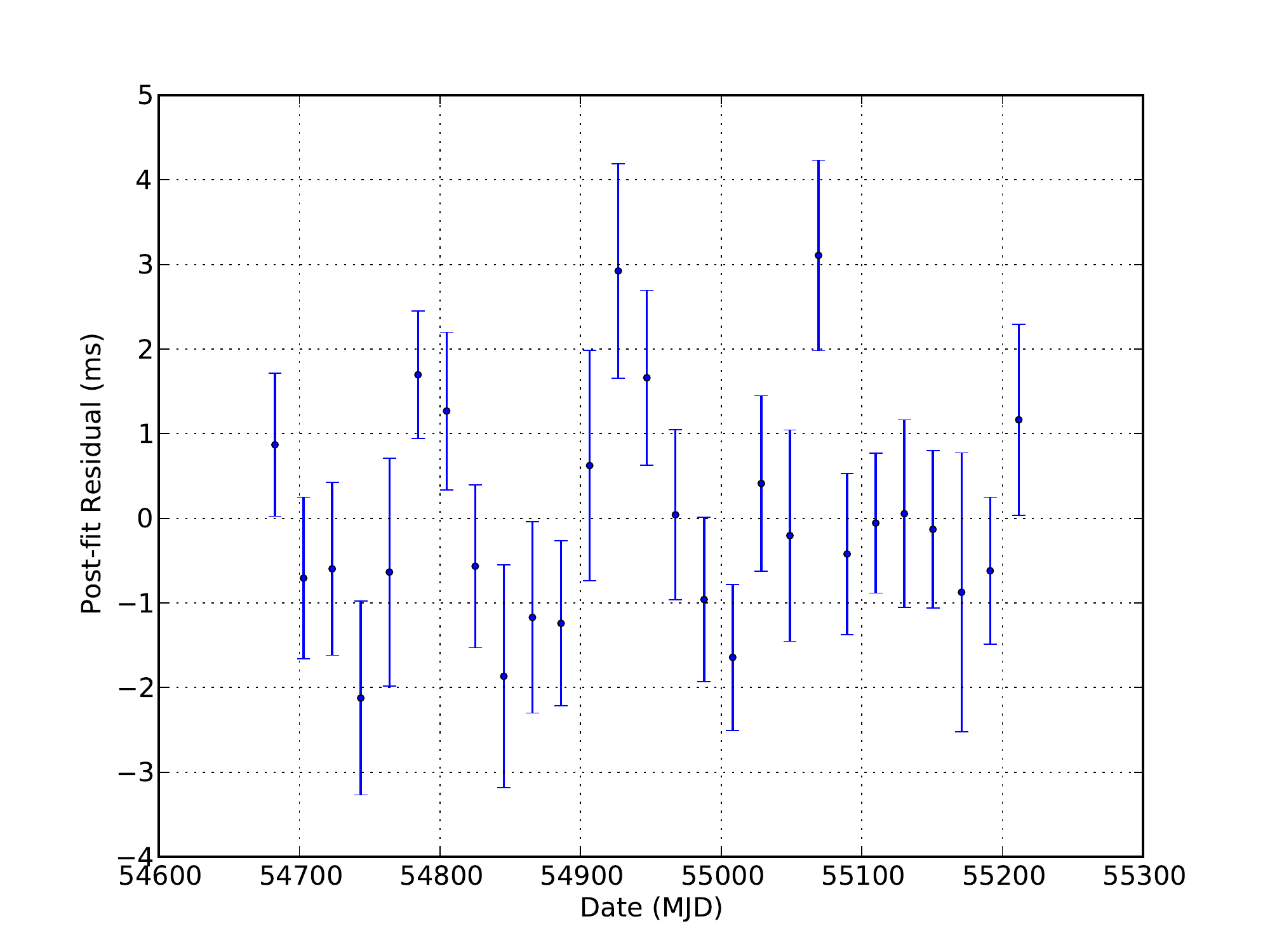}
\caption{Post-fit timing residuals for PSR J2238+5903.\label{resid:2238}}
\end{figure}

\begin{figure}
\includegraphics[width=3.0in]{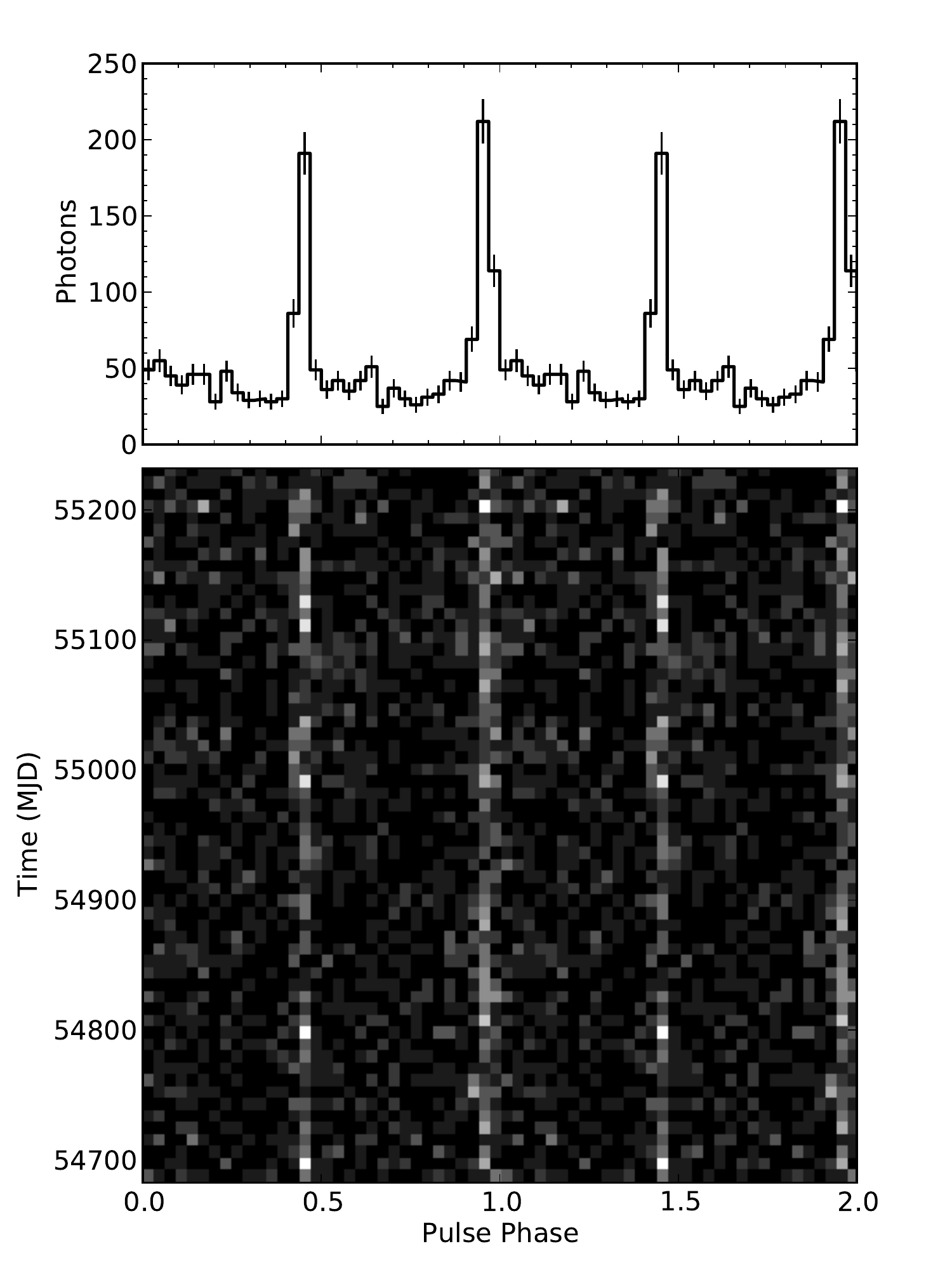}
\caption{2-D phaseogram and pulse profile of PSR J2238+5903.  Two rotations are shown on the X-axis. The photons were selected according to the ROI and $E_\mathrm{min}$ in Table~\ref{tab:2238}. The fiducial point corresponding to TZRMJD is phase 0.0.\label{phaseogram:2238}}
\end{figure}

The timing model parameters for this pulsar are displayed in Table~\ref{tab:2238} and the timing position determination, post-fit residuals, 2-D phaseogram, and folded pulse profile for this pulsar are shown in Figures \ref{pos:2238}, \ref{resid:2238}, and \ref{phaseogram:2238}, respectively.

Even though this pulsar is quite faint, it can be timed with an RMS residual of 1 ms because of its very sharp pulse profile. Consequently, we now have a very precise timing position, but there is no significant X-ray counterpart detected in the Swift image (see Figure~\ref{pos:2238}).  The pulsar is located 0.6 degree from the radio pulsar J2240+5832 detected recently in $\gamma$-rays \citep{tpc10}. The narrow pulse profile means that this pulsar can be blanked from the LAT data with a loss of only $\sim 20$\% of the exposure time.

\section{Radio Counterpart Searches}

All of these pulsars, except for PSR J1124$-$5916, were discovered in $\gamma$-ray searches and thus are $\gamma$-ray-selected pulsars, but targeted radio observations are required to determine if they are also radio quiet, or could have been discovered in radio surveys independently.  The population statistics of radio-quiet vs. radio-loud $\gamma$-ray pulsars have important implications for $\gamma$-ray emission models \citep{gsch07}. These observations are also important inputs into the population synthesis modeling of the full Galactic population of rotation-powered pulsars \citep[for example]{fk06}.

The precise positions derived from the LAT timing of these pulsars allowed us to perform deep follow up radio observations to search for pulsations from each of the new pulsars.  We used the NRAO 100-m Green Bank Telescope (GBT), the Arecibo 305-m radio telescope, and the Parkes 64-m radio telescope for these observations.  The instrument parameters used in the sensitivity calculations are shown in Table \ref{tab:radioobs}. The log of observations is shown in Table \ref{tab:radiolims} and has columns for the target name, observation code (refer to Table \ref{tab:radioobs}), observation date, observation duration ($t_\mathrm{obs}$), the R.A. and Decl. of the telescope pointing direction, the offset from the true pulsar position, an estimate of the sky temperature in that direction at the observing frequency, and our computed flux density limit ($S_\mathrm{min}$), as described below. The observations taken from the literature have $S_\mathrm{min}$ recomputed in a consistent way as well as the originally published flux limits in parentheses. In addition, the fluxes of the detected pulsars are noted with ``Det'' in parentheses.  It is notable that the flux of the detected pulsar J1907+0602 is below our nominal flux limit.  This is caused primarily by the fact that the detected pulsar has a much smaller duty cycle than the 10\% that we assume.

All observations were taken in search mode (where all data are recorded without folding at a nominal pulse period) and the data were reduced using standard pulsar analysis software, such as PRESTO \citep{rem02}.
In each case, we searched a range of dispersion measure (DM) trials out to a maximum of at least 2 times the maximum DM value predicted by the NE2001 model \citep{cl02} for that direction.
For 13 of the 16 $\gamma$-ray-selected pulsars, no radio pulsations were detected and we report upper limits in Table \ref{tab:radiolims}.  For three of the pulsars, pulsations were detected. Pulsations from PSRs J2032+4127 and J1741$-$2054 were reported by \citet{crr+09} and very faint radio pulsations from PSR J1907+06 were reported by \citet{MGROPaper}.  Here, we compile the upper limits from the literature as well as from our new observations.

We calculate upper limits in a consistent manner for all of our observations as well as those from the literature.
To compute the minimum pulsar flux that would have been detected in these observations, we use the modified radiometer equation (e.g. Lorimer \& Kramer 2005):
\begin{equation}
S_\mathrm{min} = \beta \frac{(S/N)_\mathrm{min} T_\mathrm{sys}}{G \sqrt{n_\mathrm{p} t_\mathrm{obx} \Delta f}} \sqrt{\frac{W}{P-W}}
\label{eqn:radiometer}
\end{equation}
where $\beta$ is the instrument-dependent factor due to digitization and other effects; $(S/N)_\mathrm{min} = 5$ is the threshold signal to noise for a pulsar to have been confidently detected; $T_\mathrm{sys} = T_\mathrm{rec} + T_\mathrm{sky}$, $G$ is the telescope gain, $n_\mathrm{p}$ is the number of polarizations used (2 in all cases); $t_\mathrm{obs}$ is the integration time; $\Delta f$ is the observation bandwidth; $P$ is the pulsar period; $W$ is the pulse width (for uniformity, we assume $W=0.1P$).

Because some of the observations were taken before the precise positions were known, some of the pointing directions are offset from the true direction to the pulsar. We use a simple approximation of a telescope beam response to adjust the flux sensitivity in these cases .  This factor is
\begin{equation}
f = \exp\left( \frac{-(\theta/\mathrm{HWHM})^2}{1.5} \right),
\end{equation}
where $\theta$ is the offset from the beam center and HWHM is the beam half-width at half maximum.  A computed flux limit of $S$ at the beam center is thus corrected to $S/f$ for a target offset from the pointing direction.  The resultant flux limits are compiled in Table \ref{tab:radiolims}.  \begin{deluxetable}{llrrrrrrr}
\tablewidth{0pt}
\tabletypesize{\small}
\tablecaption{Definition of radio observing codes\label{tab:radioobs}}
\tablehead{
\colhead{Obs Code} & \colhead{Telescope} & \colhead{Gain} & \colhead{Freq} & \colhead{$\Delta f$} & \colhead{$\beta$\tablenotemark{a}} & \colhead{$n_\mathrm{p}$} & \colhead{HWHM} & \colhead{$T_\mathrm{rec}$} \\
 & & (K/Jy) & (MHz) & (MHz) &   &  & (arcmin) & (K)
}
\startdata
GBT-350      & GBT     & 2.0   & 350  & 100 & 1.05 & 2  & 18.5  & 46 \\
GBT-820      & GBT     & 2.0  & 820  & 200 & 1.05 & 2  & 7.9 & 29 \\
GBT-820BCPM  & GBT     & 2.0  & 820  & 48  & 1.05 & 2  & 7.9 & 29 \\
GBT-S        & GBT     & 1.9  & 2000 & 700\tablenotemark{b} & 1.05 & 2  & 3.1 & 22 \\
AO-327       & Arecibo & 11   & 327  & 50  & 1.12 & 2  & 6.3 & 116 \\
AO-430       & Arecibo & 11   & 430  & 40  & 1.12 & 2  & 4.8 & 84 \\
AO-Lwide     & Arecibo & 10   & 1510 & 300 & 1.12 & 2  & 1.5 & 27 \\
AO-ALFA      & Arecibo & 10   & 1400 & 100 & 1.12 & 2  & 1.5 & 30 \\
Parkes-MB256 & Parkes  & 0.735& 1390 & 256 & 1.25 & 2  & 7.0 & 25 \\
Parkes-AFB   & Parkes  & 0.735& 1374 & 288 & 1.25 & 2  & 7.0 & 25 \\
Parkes-BPSR  & Parkes  & 0.735& 1352 & 340 & 1.05 & 2  & 7.0 & 25 \\
\enddata
\tablenotetext{a}{Instrument-dependent sensitivity degradation factor, see equation~\ref{eqn:radiometer}.}
\tablenotetext{b}{The instrument records 800 MHz of bandwidth, but to account for a notch filter for RFI and the lower sensitivity near the band edges, we use an effective bandwidth of 700 MHz for the sensitivity calculations.}
\end{deluxetable}

\begin{deluxetable}{llllrrrrrr}
\tabletypesize{\footnotesize}
\rotate
\tablewidth{0pt}
\tablecaption{Radio observations of $\gamma$-ray-selected pulsars\label{tab:radiolims}}
\tablehead{
\colhead{Target} & \colhead{Obs Code}  & \colhead{Date} & \colhead{$t_\mathrm{obs}$} & \colhead{R.A.\tablenotemark{a}} & \colhead{Decl.\tablenotemark{a}} & \colhead{Offset} & \colhead{$T_\mathrm{sky}$} & \colhead{$S_\mathrm{min}$} \\
 PSR ... &  & &  {(s)} & {(J2000)} & {(J2000)} & {(\arcmin)} & (K) & {($\mu$Jy)}
}
\startdata
J0007+7303 & GBT-820BCPM      & 2003 Oct 11 & 70560 & 00:07:01.6  & +73:03:08 & 0.0 & 7.7  & 12 (22\tablenotemark{b}) \\
               & GBT-S            & 2009 Aug 28 & 10000 & 00:07:01.6  & +73:03:08 & 0.0 & 0.6  &  6 \\
J0357+3205 & AO-327           & 2009 Jan 29 &  7200 & 03:57:33.1  & +32:05:03 & 4.1 & 45.7 & 43 \\
               & GBT-350          & 2010 Mar 04 &  1800 & 03:57:52.7  & +32:05:19 & 0.0 & 45.7 & 134 \\
J0633+0632 & AO-327           & 2009 Jun 27 &  3000 & 06:33:32.9  & +06:34:41 & 3.6 & 78.4 & 75 \\
               & AO-430           & 2009 Jan 30 &  4200 & 06:33:32.8  & +06:34:40 & 3.6 & 38.5 & 52 \\
               & AO-Lwide         & 2009 Jul 03 &  4200 & 06:33:44.0  & +06:32:25 & 0.1 & 1.8  & 3 \\
J1418$-$6058 & Parkes-MB256     & 2001 Feb 11 & 16900 & 14:18:41.5  & $-$60:58:11 & 0.6 & 7.8  & 32 (80\tablenotemark{c})\\
               & Parkes-AFB     & 2001 Feb 13 & 16900 & 14:18:41.5  & $-$60:58:11 & 0.6 & 7.8  & 30 (80\tablenotemark{c})\\
J1459$-$6053 & Parkes-AFB       & 2010 Feb 14 & 10200 & 14:59:30.0  & $-$60:53:21 & 0.0 & 6.0  & 38 \\
J1732$-$3131 & GBT-S            & 2009 Aug 24 & 7200  & 17:32:33.5  & $-$31:31:21 & 0.0 & 4.8  & 8 \\
               & Parkes-AFB       & 2009 Apr 14 & 16200 & 17:32:40.4  & $-$31:36:35 & 5.4 & 15.5 & 59 \\
J1741$-$2054 & Parkes-AFB       & 2000 Nov 24 & 2100  & 17:41:51.3  & $-$21:01:10 & 7.2 & 4.9  & 156 (\textbf{Det 160\tablenotemark{d}}) \\
J1809$-$2332 & Parkes-BPSR      & 2009 Apr 15 & 16200 & 18:09:50.2  & $-$23:32:23 & 0.0 & 10.6 & 26 \\
J1813$-$1246 & GBT-820          & 2009 Aug 22 & 5101  & 18:13:23.7  & $-$12:46:15 & 0.0 & 38.9 & 42 \\
               & GBT-S            & 2009 Jan 02 & 3000  & 18:13:35.9  & $-$12:48:05 & 3.6 & 3.0  & 28 \\
J1826$-$1256 & Parkes-MB256     & 2001 Feb 12 & 13234 & 18:26:04.9  & $-$12:59:48 & 3.3 & 14.0 & 49 (90\tablenotemark{c}) \\
               & Parkes-AFB     & 2001 Feb 15 & 14832 & 18:26:04.9  & $-$12:59:48 & 3.3 & 14.0 & 44 (90\tablenotemark{c}) \\
               & GBT-S        & 2010 Dec 22 & 10116 & 18:26:08.3 & $-$12:56:34 & 0.0 & 5.5 & 7 \\
               & GBT-S        & 2011 Jan 15 & 9623  & 18:26:08.3 & $-$12:56:34 & 0.0 & 5.5 & 7 \\
J1836+5925 & GBT-820BCPM      & 2002 Dec 06 & 86400 & 18:36:13.7  & +59:25:30 & 0.0 & 5.5  & 10 (17\tablenotemark{f}) \\
               & GBT-350          & 2009 Oct 24 & 7200  & 18:36:13.6  & +59:25:29 & 0.0 & 50.5 & 70 \\
J1907+0602 & AO-ALFA          & 2008 Dec 04 & 1800  & 19:07:49.5  & +06:01:52 & 1.4 & 8.8  & 22 \\
               & AO-Lwide         & 2009 Aug 21 & 3300  & 19:07:54.7  & +06:02:16 & 0.0 & 8.8  & 5 (\textbf{Det 3.4\tablenotemark{e}}) \\
J1958+2846 & AO-Lwide         & 2009 Oct 13 & 2400  & 19:58:40.3  & +28:45:54 & 0.0 & 3.2  & 5 \\
J2021+4026 & GBT-S            & 2009 Jan 04 & 3600  & 20:21:35.6  & +40:26:21 & 0.9 & 2.5  & 11 \\
               & GBT-820BCPM      & 2003 Dec 27 & 14400 & 20:21:18.1  & +40:24:35 & 3.3 & 31.9 & 51 (40\tablenotemark{g}) \\
               & GBT-820BCPM      & 2003 Dec 27 & 14400 & 20:21:21.5  & +40:23:22 & 3.9 & 31.9 & 53 (40\tablenotemark{g}) \\
J2032+4127 & GBT-S            & 2009 Jan 05 & 3600  & 20:32:13.9  & +41:22:34 & 4.8 & 2.6  & 50 (\textbf{Det 120\tablenotemark{d}}) \\
J2238+5903 & GBT-820          & 2009 Aug 22 & 4447  & 22:38:27.9  & +59:03:42 & 0.0 & 12.1 & 27 \\
               & GBT-S            & 2009 Aug 24 & 7200  & 22:38:27.9  & +59:03:42 & 0.0 & 0.9  & 7 \\
\enddata
\tablenotetext{a}{Telescope pointing direction (not necessarily source position)}
\tablenotetext{b}{\citet{hgc+04} }
\tablenotetext{c}{\citet{rrk+02}}
\tablenotetext{d}{\citet{crr+09}}
\tablenotetext{e}{\citet{MGROPaper}}
\tablenotetext{f}{\citet{hcg07}}
\tablenotetext{g}{\citet{bwa+04}}
\end{deluxetable}

\clearpage 

\begin{deluxetable}{ll}
\tablecolumns{2}
\tablewidth{0pt}
\tablecaption{PSR J0007+7303\label{tab:0007}}
\tablehead{\colhead{Parameter} & \colhead{Value}}
\startdata
Right ascension, $\alpha$ (J2000.0)\dotfill &  00:07:00.6 $\pm 0.2^s$  \\ 
Declination, $\delta$ (J2000.0)\dotfill & +73:03:07.0 $\pm 0.6\arcsec$ \\
Monte Carlo position uncertainty  & 2\arcsec \\
Pulse frequency, $\nu$ (s$^{-1}$)\dotfill & 3.165827380(3) \\ 
Frequency first derivative, $\dot{\nu}$ (s$^{-2}$)\dotfill & $-$3.6136(2)$\times$10$^{-12}$  \\ 
Frequency second derivative, $\ddot{\nu}$ (s$^{-3}$)\dotfill & $-$7(1)$\times$10$^{-23}$  \\ 
Epoch of Frequency (MJD) \dotfill & 54952 \\ 
Glitch Epoch\dotfill & 54952.652 \\ 
Glitch $\Delta \nu$ (s$^{-1}$)\dotfill & 1.759(3)$\times$10$^{-6}$ \\ 
Glitch $\Delta \dot{\nu}$ (s$^{-2}$)\dotfill & 0 \\ 
TZRMJD \dotfill &   54952.334185720257651\\
Number of photons ($n_\gamma$) \dotfill & 12790 \\
Number of TOAs \dotfill & 55 \\
RMS timing residual (ms) \dotfill & 2.2 \\
Template Profile \dotfill & KDE \\
$E_\mathrm{min}$ \dotfill & 150 MeV \\
ROI \dotfill & 1.5$^\circ$ \\
Valid range (MJD) \dotfill & 54682 -- 55222 \\
\enddata
\end{deluxetable}

\clearpage 

\begin{deluxetable}{ll}
\tablecolumns{2}
\tablewidth{0pt}
\tablecaption{PSR J0357+3205\label{tab:0357}}
\tablehead{\colhead{Parameter} & \colhead{Value}}
\startdata
Right ascension, $\alpha$ (J2000.0)\dotfill &  03:57:52.5 $\pm 0.2^s$ \\ 
Declination, $\delta$ (J2000.0)\dotfill & +32:05:25 $\pm 6\arcsec$ \\
Monte Carlo position uncertainty  & 18\arcsec \\
Pulse frequency, $\nu$ (s$^{-1}$)\dotfill & 2.251722292(3) \\ 
Frequency first derivative, $\dot{\nu}$ (s$^{-2}$)\dotfill & $-$6.61(1)$\times 10^{-14}$ \\ 
Frequency second derivative, $\ddot{\nu}$ (s$^{-3}$)\dotfill & $|\ddot{\nu}|<6\times 10^{-23}$ \\ 
Epoch of Frequency (MJD) \dotfill & 54946 \\ 
TZRMJD \dotfill &   54946.341346723796502\\
Number of photons ($n_\gamma$) \dotfill & 1335 \\
Number of TOAs \dotfill & 25 \\
RMS timing residual (ms) \dotfill & 5.3 \\
Template Profile \dotfill & 1 Gaussian \\
$E_\mathrm{min}$ \dotfill & 250 MeV \\
ROI \dotfill & 0.8$^\circ$ \\
Valid range (MJD) \dotfill & 54682 -- 55210 \\
\enddata
\end{deluxetable}

\clearpage 

\begin{deluxetable}{ll}
\tablecolumns{2}
\tablewidth{0pt}
\tablecaption{PSR J0633+0632\label{tab:0633}}
\tablehead{\colhead{Parameter} & \colhead{Value}}
\startdata
Right ascension, $\alpha$ (J2000.0)\dotfill &  06:33:44.21 $\pm 0.02^s$ \\ 
Declination, $\delta$ (J2000.0)\dotfill & +06:32:34.9 $\pm 1.6\arcsec$ \\ 
Monte Carlo position uncertainty  & 3.5\arcsec \\
Pulse frequency, $\nu$ (s$^{-1}$)\dotfill & 3.3625291588(7) \\ 
Frequency first derivative, $\dot{\nu}$ (s$^{-2}$)\dotfill & $-$8.9991(3)$\times$10$^{-13}$ \\ 
Frequency second derivative, $\ddot{\nu}$ (s$^{-3}$)\dotfill & $-$2(1)$\times$10$^{-23}$ \\ 
Epoch of Frequency (MJD) \dotfill & 54945 \\ 
TZRMJD \dotfill &  54945.385967311181439\\
Number of photons ($n_\gamma$) \dotfill & 1174 \\
Number of TOAs \dotfill & 23 \\
RMS timing residual (ms) \dotfill & 1.4 \\
Template Profile \dotfill & 2 Gaussian \\
$E_\mathrm{min}$ \dotfill & 550 MeV \\
ROI \dotfill & 0.6$^\circ$ \\
Valid range (MJD) \dotfill & 54682 -- 55208 \\
\enddata
\end{deluxetable}

\clearpage 

\begin{deluxetable}{ll}
\tablewidth{0pt}
\tablecaption{PSR J1124$-$5916\label{tab:1124}}
\tablehead{\colhead{Parameter} & \colhead{Value}}
\startdata
Right ascension, $\alpha$ (J2000.0)\dotfill &  11:24:39.0(1) \\ 
Declination, $\delta$ (J2000.0)\dotfill & $-$59:16:19(1)  \\ 
Pulse frequency, $\nu$ (s$^{-1}$)\dotfill & 7.381334652(9) \\ 
Frequency first derivative, $\dot{\nu}$ (s$^{-2}$)\dotfill & $-$4.10029(9)$\times$10$^{-11}$ \\ 
Frequency second derivative, $\ddot{\nu}$ (s$^{-3}$)\dotfill & $-$8.6(4)$\times$10$^{-22}$  \\ 
Epoch of Frequency (MJD) \dotfill & 54683.281414 \\ 
Glitch Epoch (MJD) \dotfill & 55191 \\
Glitch $\Delta\nu$ (s$^{-1}$)\dotfill & 1.18(9) $\times 10^{-7}$ \\
Glitch $\Delta\dot{\nu}$ (s$^{-2}$)\dotfill & 1.94(2) $\times 10^{-13}$ \\
TZRMJD \dotfill &  55053.0521054597626\\
TZRFREQ (MHz) \dotfill & 1371.067 \\
TZRSITE \dotfill & 7 (Parkes) \\ 
Number of photons ($n_\gamma$) \dotfill & 5030 \\
Number of TOAs \dotfill & 40 \\
RMS timing residual (ms) \dotfill & 2.8 \\
Template Profile \dotfill & 2 Gaussian \\
$E_\mathrm{min}$ \dotfill & 200 MeV \\
ROI \dotfill & 0.9$^\circ$ \\
Valid range \dotfill & 54682--55415 \\
\enddata
\end{deluxetable}

\clearpage 

\begin{deluxetable}{ll}
\tablecolumns{2}
\tablewidth{0pt}
\tablecaption{PSR J1418-6058\label{tab:1418}}
\tablehead{\colhead{Parameter} & \colhead{Value}}
\startdata
Right ascension, $\alpha$ (J2000.0)\dotfill &  14:18:42.7 $\pm 0.1^s$ \\ 
Declination, $\delta$ (J2000.0)\dotfill & $-$60:57:49 $\pm 2\arcsec$ \\ 
Monte Carlo position uncertainty  & 7\arcsec \\
Pulse frequency, $\nu$ (s$^{-1}$)\dotfill & 9.043798163(1) \\ 
Frequency first derivative, $\dot{\nu}$ (s$^{-2}$)\dotfill & $-$1.38548(8)$\times$10$^{-11}$ \\ 
Frequency second derivative, $\ddot{\nu}$ (s$^{-3}$)\dotfill & 6.4(3)$\times$10$^{-22}$ \\
Frequency third derivative, $\dddot{\nu}$ (s$^{-4}$)\dotfill & $-$8(2)$\times$10$^{-29}$ \\
Epoch of Frequency (MJD) \dotfill & 54944 \\ 
TZRMJD \dotfill &  54944.2886329214 \\
Number of photons ($n_\gamma$) \dotfill & 7283 \\
Number of TOAs \dotfill & 33 \\
RMS timing residual (ms) \dotfill & 1.9 \\
Template Profile \dotfill & KDE \\
$E_\mathrm{min}$ \dotfill & 250 MeV \\
ROI \dotfill & 0.5$^\circ$ \\
Valid range (MJD) \dotfill & 54682 -- 55205 \\
\enddata
\end{deluxetable}

\clearpage 

\begin{deluxetable}{ll}
\tablecolumns{2}
\tablewidth{0pt}
\tablecaption{PSR J1459$-$6053\label{tab:1459}}
\tablehead{\colhead{Parameter} & \colhead{Value}}
\startdata
Right ascension, $\alpha$ (J2000.0)\dotfill &  14:59:29.99 $\pm 0.06^s$ \\ 
Declination, $\delta$ (J2000.0)\dotfill & -60:53:20.7 $\pm 0.4\arcsec$ \\ 
Monte Carlo position uncertainty  & 1.3\arcsec \\
Pulse frequency, $\nu$ (s$^{-1}$)\dotfill & 9.694559498(1) \\ 
Frequency first derivative, $\dot{\nu}$ (s$^{-2}$)\dotfill & $-$2.37503(5)$\times$10$^{-12}$ \\ 
Frequency second derivative, $\ddot{\nu}$ (s$^{-3}$)\dotfill & $-$4(2)$\times$10$^{-23}$ \\ 
Epoch of Frequency (MJD) \dotfill & 54935 \\ 
TZRMJD \dotfill &  54936.19962194 \\
Number of photons ($n_\gamma$) \dotfill & 3305 \\
Number of TOAs \dotfill & 26 \\
RMS timing residual (ms) \dotfill & 1.1 \\
Template Profile \dotfill & KDE \\
$E_\mathrm{min}$ \dotfill & 350 MeV \\
ROI \dotfill & 0.7$^\circ$ \\
Valid range (MJD) \dotfill & 54682--55210 \\
\enddata
\end{deluxetable}

\clearpage 

\begin{deluxetable}{ll}
\tablecolumns{2}
\tablewidth{0pt}
\tablecaption{PSR J1732$-$3131\label{tab:1732}}
\tablehead{\colhead{Parameter} & \colhead{Value}}
\startdata
Right ascension, $\alpha$ (J2000.0)\dotfill &  17:32:33.54 $\pm 0.03^s$ \\ 
Declination, $\delta$ (J2000.0)\dotfill & $-$31:31:23 $\pm 2\arcsec$ \\ 
Monte Carlo position uncertainty  & 3\arcsec \\
Pulse frequency, $\nu$ (s$^{-1}$)\dotfill & 5.0879411200(5) \\ 
Frequency first derivative, $\dot{\nu}$ (s$^{-2}$)\dotfill & $-$7.2609(3)$\times$10$^{-13}$ \\ 
Frequency second derivative, $\ddot{\nu}$ (s$^{-3}$)\dotfill & $|\ddot{\nu}|<2\times 10^{-23}$ \\ 
Epoch of Frequency (MJD) \dotfill & 54933 \\ 
TZRMJD \dotfill &  54957.3282196892\\
Number of photons ($n_\gamma$) \dotfill & 4236 \\
Number of TOAs \dotfill & 22 \\
RMS timing residual (ms) \dotfill & 1.0 \\
Template Profile \dotfill & KDE \\
$E_\mathrm{min}$ \dotfill & 400 MeV \\
ROI \dotfill & 0.5$^\circ$ \\
Valid range (MJD) \dotfill & 54682--55207\\
\enddata
\end{deluxetable}

\clearpage 

\begin{deluxetable}{ll}
\tablecolumns{2}
\tablewidth{0pt}
\tablecaption{PSR J1741$-$2054\label{tab:1741}}
\tablehead{\colhead{Parameter} & \colhead{Value}}
\startdata
Right ascension, $\alpha$ (J2000.0)\dotfill &  17:41:57.23 $\pm 0.05^s$ \\ 
Declination, $\delta$ (J2000.0)\dotfill & -20:53:57 $\pm 19\arcsec$ \\ 
Monte Carlo position uncertainty  & 20\arcsec \\
Pulse frequency, $\nu$ (s$^{-1}$)\dotfill & 2.417209833(1) \\ 
Frequency first derivative, $\dot{\nu}$ (s$^{-2}$)\dotfill & $-$9.923(3)$\times$10$^{-14}$ \\ 
Frequency second derivative, $\ddot{\nu}$ (s$^{-3}$)\dotfill & $|\ddot{\nu}|<2\times 10^{-23}$ \\ 
Epoch of Frequency (MJD) \dotfill & 54933 \\ 
TZRMJD \dotfill &  54945.3859666189\\
Number of photons ($n_\gamma$) \dotfill & 3135 \\
Number of TOAs \dotfill & 23 \\
RMS timing residual (ms) \dotfill & 2.6 \\
Template Profile \dotfill & KDE \\
$E_\mathrm{min}$ \dotfill & 300 MeV \\
ROI \dotfill & 0.8$^\circ$ \\
Valid range (MJD) \dotfill & 54682--55208\\
\enddata
\end{deluxetable}

\clearpage 

\begin{deluxetable}{ll}
\tablecolumns{2}
\tablewidth{0pt}
\tablecaption{PSR J1809-2332\label{tab:1809}}
\tablehead{\colhead{Parameter} & \colhead{Value}}
\startdata
Right ascension, $\alpha$ (J2000.0)\dotfill &  18:09:50.31 $\pm 0.06^s$ \\ 
Declination, $\delta$ (J2000.0)\dotfill & $-$23:33:35 $\pm 51\arcsec$ \\ 
Monte Carlo position uncertainty  & 28\arcsec \\
Pulse frequency, $\nu$ (s$^{-1}$)\dotfill & 6.8125205463(3) \\ 
Frequency first derivative, $\dot{\nu}$ (s$^{-2}$)\dotfill & $-$1.59748(1)$\times$10$^{-12}$  \\ 
Frequency second derivative, $\ddot{\nu}$ (s$^{-3}$)\dotfill & $|\ddot{\nu}|<1\times 10^{-23}$ \\ 
Epoch of Frequency (MJD) \dotfill & 54935 \\ 
TZRMJD \dotfill &  54947.1551911038\\
Number of photons ($n_\gamma$) \dotfill & 10422 \\
Number of TOAs \dotfill & 27 \\
RMS timing residual (ms) \dotfill & 0.4 \\
Template Profile \dotfill & KDE \\
$E_\mathrm{min}$ \dotfill & 250 MeV \\
ROI \dotfill & 0.8$^\circ$ \\
Valid range (MJD) \dotfill & 54682--55211\\
\enddata
\end{deluxetable}

\clearpage 
\begin{deluxetable}{ll}
\tablecolumns{2}
\tablewidth{0pt}
\tablecaption{PSR J1813$-$1246\label{tab:1813}}
\tablehead{\colhead{Parameter} & \colhead{Value}}
\startdata
Right ascension, $\alpha$ (J2000.0)\dotfill &  18:13:23.77 $\pm 0.01^s$ \\ 
Declination, $\delta$ (J2000.0)\dotfill & $-$12:45:59.2 $\pm 1.5\arcsec$ \\ 
Monte Carlo position uncertainty  & 1.5\arcsec \\
Pulse frequency, $\nu$ (s$^{-1}$)\dotfill &20.802023359(3) \\ 
Frequency first derivative, $\dot{\nu}$ (s$^{-2}$)\dotfill & $-$7.60023(9)$\times$10$^{-12}$  \\ 
Frequency second derivative, $\ddot{\nu}$ (s$^{-3}$)\dotfill & $|\ddot{\nu}|<6\times 10^{-23}$ \\ 
Glitch Epoch\dotfill & 55094.1227 \\ 
Glitch $\Delta \nu$ (s$^{-1}$)\dotfill & 2.4256(9)$\times$10$^{-5}$ \\ 
Glitch $\Delta \dot{\nu}$ (s$^{-2}$)\dotfill & $-4.9(2) \times 10^{-14}$ \\ 
TZRMJD \dotfill &  54954.309848738\\
Number of photons ($n_\gamma$) \dotfill & 11611 \\
Number of TOAs \dotfill & 91 \\
RMS timing residual (ms) \dotfill & 0.7 \\
Template Profile \dotfill & KDE \\
$E_\mathrm{min}$ \dotfill & 200 MeV \\
ROI \dotfill & 1.0$^\circ$ \\
Valid range \dotfill & 54682--55226 \\
\enddata
\end{deluxetable}

\clearpage 

\begin{deluxetable}{ll}
\tablecolumns{2}
\tablewidth{0pt}
\tablecaption{PSR J1826$-$1256\label{tab:1826}}
\tablehead{\colhead{Parameter} & \colhead{Value}}
\startdata
Right ascension, $\alpha$ (J2000.0)\dotfill &  18:26:08.53 $\pm 0.01^s$ \\ 
Declination, $\delta$ (J2000.0)\dotfill & $-$12:56:33.0 $\pm 0.5\arcsec$ \\ 
Monte Carlo position uncertainty  & 17\arcsec \\
Pulse frequency, $\nu$ (s$^{-1}$)\dotfill &  9.0724588059(3) \\ 
Frequency first derivative, $\dot{\nu}$ (s$^{-2}$)\dotfill & $-$9.99654(1)$\times$10$^{-12}$  \\ 
Frequency second derivative, $\ddot{\nu}$ (s$^{-3}$)\dotfill & 1.85(5)$\times$10$^{-22}$ \\ 
Epoch of Frequency (MJD) \dotfill & 54934 \\ 
TZRMJD \dotfill &  54946.3413482956\\
Number of photons ($n_\gamma$) \dotfill & 10860 \\
Number of TOAs \dotfill & 25 \\
RMS timing residual (ms) \dotfill & 0.28 \\
Template Profile \dotfill & KDE \\
$E_\mathrm{min}$ \dotfill & 200 MeV \\
ROI \dotfill & 0.6$^\circ$ \\
Valid range \dotfill & 54682--55210 \\
\enddata
\end{deluxetable}

\clearpage 

\begin{deluxetable}{ll}
\tablecolumns{2}
\tablewidth{0pt}
\tablecaption{PSR J1836+5925\label{tab:1836}}
\tablehead{\colhead{Parameter} & \colhead{Value}}
\startdata
Right ascension, $\alpha$ (J2000.0)\dotfill &  18:36:13.69 $\pm 0.02^s$ \\ 
Declination, $\delta$ (J2000.0)\dotfill & +59:25:30.0 $\pm 0.3\arcsec$ \\ 
Monte Carlo position uncertainty  & $<3$\arcsec \\
Pulse frequency, $\nu$ (s$^{-1}$)\dotfill & 5.7715509149(4) \\ 
Frequency first derivative, $\dot{\nu}$ (s$^{-2}$)\dotfill & $-$5.007(2)$\times$10$^{-14}$ \\ 
Frequency second derivative, $\ddot{\nu}$ (s$^{-3}$)\dotfill & $-1.3$(8)$\times$10$^{-23}$ \\ 
Epoch of Frequency (MJD) \dotfill & 54935 \\ 
TZRMJD \dotfill &  54936.1996219705\\
Number of photons ($n_\gamma$) \dotfill & 14875 \\
Number of TOAs \dotfill & 26 \\
RMS timing residual (ms) \dotfill & 0.85 \\
Template Profile \dotfill & KDE \\
$E_\mathrm{min}$ \dotfill & 200 MeV \\
ROI \dotfill & 1.6$^\circ$ \\
Valid range \dotfill & 54682--55210 \\
\enddata
\end{deluxetable}

\clearpage 

\begin{deluxetable}{ll}
\tablecolumns{2}
\tablewidth{0pt}
\tablecaption{PSR J1907+0602\label{tab:1907}}
\tablehead{\colhead{Parameter} & \colhead{Value}}
\startdata
Right ascension, $\alpha$ (J2000.0)\dotfill &  19:07:54.74 $\pm 0.01^s$ \\ 
Declination, $\delta$ (J2000.0)\dotfill & +06:02:16.9 $\pm 0.3\arcsec$ \\ 
Monte Carlo position uncertainty  & 2.5\arcsec \\
Pulse frequency, $\nu$ (s$^{-1}$)\dotfill & 9.3779822336(4) \\ 
Frequency first derivative, $\dot{\nu}$ (s$^{-2}$)\dotfill & -7.63559(2)$\times$10$^{-12}$ \\ 
Frequency second derivative, $\ddot{\nu}$ (s$^{-3}$)\dotfill & 1.88(7)$\times$10$^{-22}$ \\ 
Epoch of Frequency (MJD) \dotfill & 54935 \\ 
TZRMJD \dotfill &  54947.1551911789 \\
Number of photons ($n_\gamma$) \dotfill & 10629 \\
Number of TOAs \dotfill & 27 \\
RMS timing residual (ms) \dotfill & 0.47 \\
Template Profile \dotfill & KDE \\
$E_\mathrm{min}$ \dotfill & 50 MeV \\
ROI \dotfill & 0.7$^\circ$ \\
Valid range \dotfill & 54682--55211 \\
\enddata
\end{deluxetable}

\clearpage 

\begin{deluxetable}{ll}
\tablecolumns{2}
\tablewidth{0pt}
\tablecaption{PSR J1958+2846\label{tab:1958}}
\tablehead{\colhead{Parameter} & \colhead{Value}}
\startdata
Right ascension, $\alpha$ (J2000.0)\dotfill &  19:58:40.07 $\pm 0.03^s$ \\ 
Declination, $\delta$ (J2000.0)\dotfill & +28:45:54 $\pm 1\arcsec$ \\ 
Monte Carlo position uncertainty  & 3.5\arcsec \\
Pulse frequency, $\nu$ (s$^{-1}$)\dotfill & 3.4436537099(5) \\ 
Frequency first derivative, $\dot{\nu}$ (s$^{-2}$)\dotfill & -2.5145(2)$\times$10$^{-12}$ \\ 
Frequency second derivative, $\ddot{\nu}$ (s$^{-3}$)\dotfill & 3(2)$\times$10$^{-23}$ \\ 
Epoch of Frequency (MJD) \dotfill & 54800 \\ 
TZRMJD \dotfill &  54957.3282188686 \\
Number of photons ($n_\gamma$) \dotfill & 1910 \\
Number of TOAs \dotfill & 26 \\
RMS timing residual (ms) \dotfill & 2.1 \\
Template Profile \dotfill & 2 Gaussian \\
$E_\mathrm{min}$ \dotfill & 550 MeV \\
ROI \dotfill & 0.6$^\circ$ \\
Valid range \dotfill & 54682--55210 \\
\enddata
\end{deluxetable}

\clearpage

\begin{deluxetable}{ll}
\tablecolumns{2}
\tablewidth{0pt}
\tablecaption{PSR J2021+4026\label{tab:2021}}
\tablehead{\colhead{Parameter} & \colhead{Value}}
\startdata
Right ascension, $\alpha$ (J2000.0)\dotfill &  20:21:29.99 $\pm 0.03^s$ \\ 
Declination, $\delta$ (J2000.0)\dotfill & +40:26:45.1 $\pm 0.7\arcsec$ \\ 
Monte Carlo position uncertainty  & 2.5\arcsec \\
Pulse frequency, $\nu$ (s$^{-1}$)\dotfill & 3.7690668480(6) \\ 
Frequency first derivative, $\dot{\nu}$ (s$^{-2}$)\dotfill & -7.7681(3)$\times$10$^{-13}$ \\ 
Frequency second derivative, $\ddot{\nu}$ (s$^{-3}$)\dotfill & 3.9(2)$\times$10$^{-22}$ \\ 
Epoch of Frequency (MJD) \dotfill & 54936 \\ 
TZRMJD \dotfill &  54957.3282196715\\
Number of photons ($n_\gamma$) \dotfill & 11853 \\
Number of TOAs \dotfill & 30 \\
RMS timing residual (ms) \dotfill & 2.0 \\
Template Profile \dotfill & KDE \\
$E_\mathrm{min}$ \dotfill & 400 MeV \\
ROI \dotfill & 0.7$^\circ$ \\
Valid range \dotfill & 54682--55213 \\
\enddata
\end{deluxetable}

\clearpage

\begin{deluxetable}{ll}
\tablecolumns{2}
\tablewidth{0pt}
\tablecaption{PSR J2032+4127\label{tab:2032}}
\tablehead{\colhead{Parameter} & \colhead{Value}}
\startdata
Right ascension, $\alpha$ (J2000.0)\dotfill &  20:32:13.25 $\pm 0.01^s$ \\ 
Declination, $\delta$ (J2000.0)\dotfill & +41:27:24.8 $\pm 0.3\arcsec$ \\ 
Monte Carlo position uncertainty  & 3\arcsec \\
Pulse frequency, $\nu$ (s$^{-1}$)\dotfill & 6.9809196293(4) \\ 
Frequency first derivative, $\dot{\nu}$ (s$^{-2}$)\dotfill & $-$9.9293(2)$\times$10$^{-13}$ \\ 
Frequency second derivative, $\ddot{\nu}$ (s$^{-3}$)\dotfill & $-$1.88(1)$\times$10$^{-21}$ \\ 
Epoch of Frequency (MJD) \dotfill & 54938 \\ 
TZRMJD \dotfill &  54951.224402859 \\
Number of photons ($n_\gamma$) \dotfill & 1633 \\
Number of TOAs \dotfill & 45 \\
RMS timing residual (ms) \dotfill & 0.9 \\
Template Profile \dotfill & 2 Gaussian \\
$E_\mathrm{min}$ \dotfill & 900 MeV \\
ROI \dotfill & 0.5$^\circ$ \\
Valid range \dotfill & 54682--55220 \\
\enddata
\end{deluxetable}

\clearpage

\begin{deluxetable}{ll}
\tablecolumns{2}
\tablewidth{0pt}
\tablecaption{PSR J2238+5903\label{tab:2238}}
\tablehead{\colhead{Parameter} & \colhead{Value}}
\startdata
Right ascension, $\alpha$ (J2000.0)\dotfill &  22:38:28.27 $\pm 0.04^s$ \\ 
Declination, $\delta$ (J2000.0)\dotfill & +59:03:40.8 $\pm 0.4\arcsec$ \\ 
Monte Carlo position uncertainty  & 3\arcsec \\
Pulse frequency, $\nu$ (s$^{-1}$)\dotfill & 6.1450029089(4) \\ 
Frequency first derivative, $\dot{\nu}$ (s$^{-2}$)\dotfill & $-$3.6641(2)$\times$10$^{-12}$ \\ 
Frequency second derivative, $\ddot{\nu}$ (s$^{-3}$)\dotfill & 1.1(2)$\times$10$^{-22}$ \\ 
Epoch of Frequency (MJD) \dotfill & 54800 \\ 
TZRMJD \dotfill &  54947.1551907197\\
Number of photons ($n_\gamma$) \dotfill & 1697 \\
Number of TOAs \dotfill & 27 \\
RMS timing residual ($\mu$s) \dotfill & 1171 \\
Template Profile \dotfill & 2 Gaussian \\
$E_\mathrm{min}$ \dotfill & 250 MeV \\
ROI \dotfill & 0.5$^\circ$ \\
Valid range \dotfill & 54682--55211 \\
\enddata
\end{deluxetable}

\clearpage

\section{Discussion}

We have presented precise phase-coherent timing models using 18 months of \textit{Fermi} LAT data for 17 radio-quiet or radio-faint $\gamma$-ray pulsars.  This provides precise spin parameters for all of the pulsars and position determinations of order arcsecond accuracy.

In most cases the new position determinations served to confirm previously proposed X-ray counterparts. However, in one case (PSR J1958+2846) the previously proposed counterpart is strongly discrepant with the new position determination.  In one other case (PSR J1459$-$6053), an X-ray source is apparent in a \textit{Swift} image at the pulsar position.  Lastly in 3 cases (PSRs J0633+0632, J1418$-$6058, and J2021+4026), the situation is a bit more complicated, because the observed offsets between the timing position and the X-ray counterpart position may be accounted for by the effects of timing noise on the model fits.  These were covered on a case-by-case basis.

In three of the 17 pulsars (PSRs J0007+7303, J1124$-$5916, and J1813$-$1246), we have detected a glitch.  This is not unexpected for a population of mostly young pulsars with characteristic ages of $10^4$ -- $10^5$ years. These three glitches observed in the seventeen radio quiet and radio faint pulsars are typical of the eight glitches observed in eighteen months of the \textit{Fermi} $\gamma$-ray pulsars. All pulsars observed to glitch with the LAT pulsars have spin down energies $\dot{E} > 4.5\times10^{35}$ erg s$^{-1}$. In fact, most (6/8) of the glitching pulsars are above $\dot{E} > 1 \times 10^{36}$ erg s$^{-1}$. All of the LAT glitching pulsars have characteristic ages between 1--100 kyr. A more detailed analysis of timing across glitches in $\gamma$-ray pulsars is in preparation (Dormody et al. 2010, in preparation).

\begin{figure}[htb]
\includegraphics[width=3.0in]{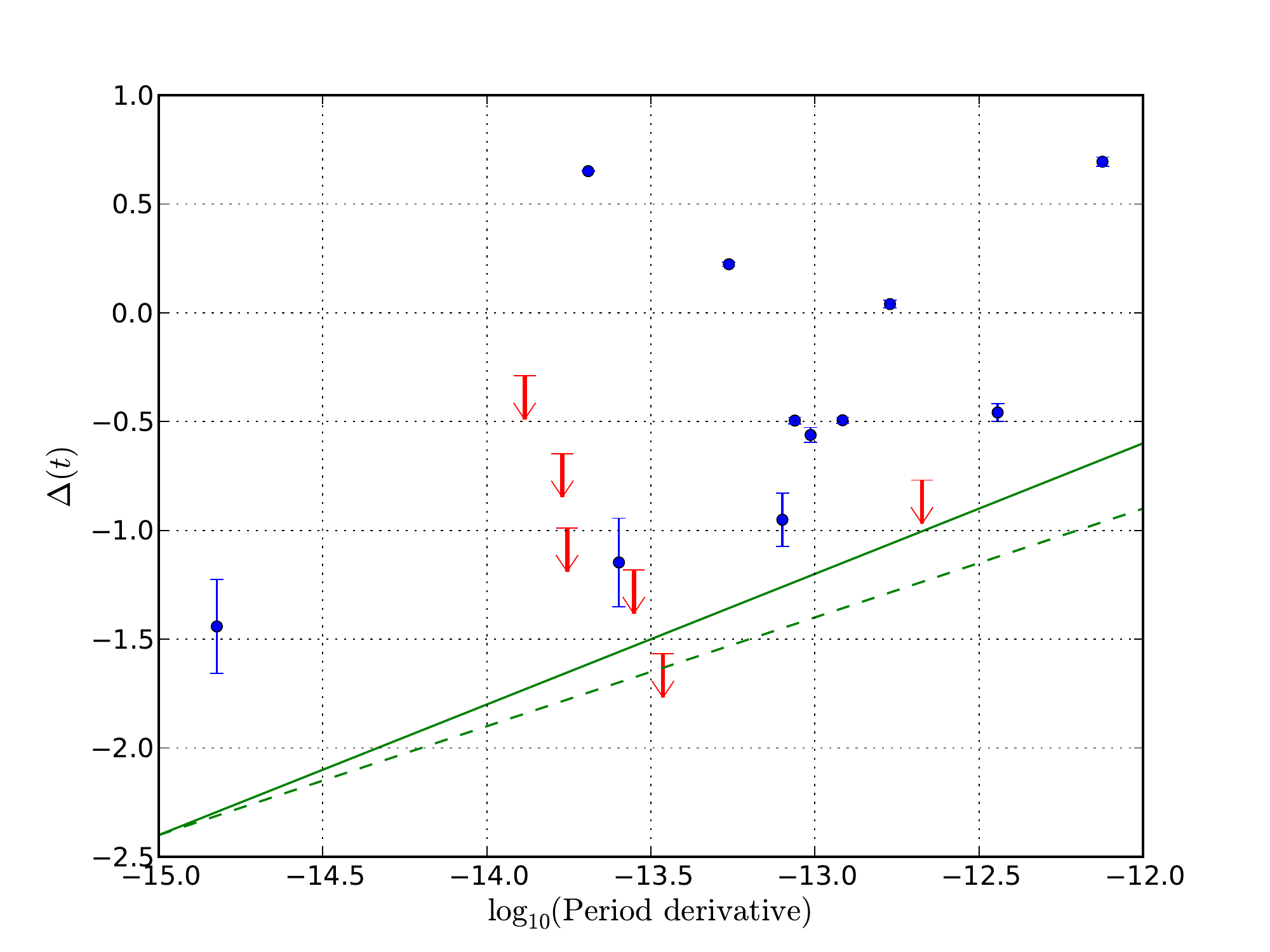}
\caption{$\Delta$ parameter characterizing timing noise vs. $\dot{\nu}$ for these pulsars.  The red arrows represent 2-$\sigma$ upper limits. The solid green line is the relation for $\Delta_8$ found by \citet{antt94}, while the dashed line is the relation found by \citet{hlk10}. Note that our data span is about half of the $10^8$ s used in the definition of $\Delta_8$.\label{fig:f2}}
\end{figure}

With 18 months of timing data, we also have measurements of $\ddot{\nu}$ for most of the pulsars.  The measured $\ddot{\nu}$s are dominated by timing noise rather than the secular spin down behavior of the pulsars.  Previously, \citet{antt94} have defined a pulsar stability parameter 
\begin{equation}
\Delta(t) = \log_{10} \left(\frac{1}{6\nu}|\ddot{\nu}|t^3\right),
\end{equation}
where $t$ is the observation duration and they define $\Delta_8 = \Delta(10^8 \mathrm{s})$.  They find a correlation of this stability parameter with pulsar period derivative ($\dot{P}$), with the form
\begin{equation}
\Delta_8 = 6.6 + 0.6 \log_{10} \dot{P}.
\end{equation}
This relationship has been re-fit using a larger sample of pulsars by \citet{hlk10}, who obtain the following parameters:
\begin{equation}
\Delta_8 = 5.1 + 0.5 \log_{10} \dot{P}.
\end{equation}
We do not have $10^8$ s of data, so we compute our $\Delta$ parameter at $t=10^{7.6}$ s. As seen in Figure \ref{fig:f2}, we see a similar correlation with period derivative, albeit with a large amount of scatter.  The two pulsars that stand out farthest from the relation as having very large $\ddot{\nu}$ for their period derivatives are PSRs J2021+4026 and J2032+4127.

These pulsars will continue to be timed regularly throughout the LAT mission.  Using the fake TOA simulation capability of \textsc{Tempo2}, we have evaluated the possibility of measuring further astrometric parameters for these pulsars.  We find that in a 10 year mission, we are unlikely to be able to detect parallax or proper motion for any of these sources.  The most nearby pulsar is $\sim 400$ pc distant, and a parallax signal at that distance is 3 $\mu$s, for a pulsar at an ecliptic latitude of 0.  Unfortunately, the nearby pulsars are also those with the longest periods and the largest RMS timing residuals.  We evaluated the possibility of detecting proper motion for a large transverse velocity of 1000 km s$^{-1}$, and again found that none of the pulsars look like promising candidates for proper motion measurements within the \textit{Fermi} mission. 

\begin{figure}
\includegraphics[width=3.5in]{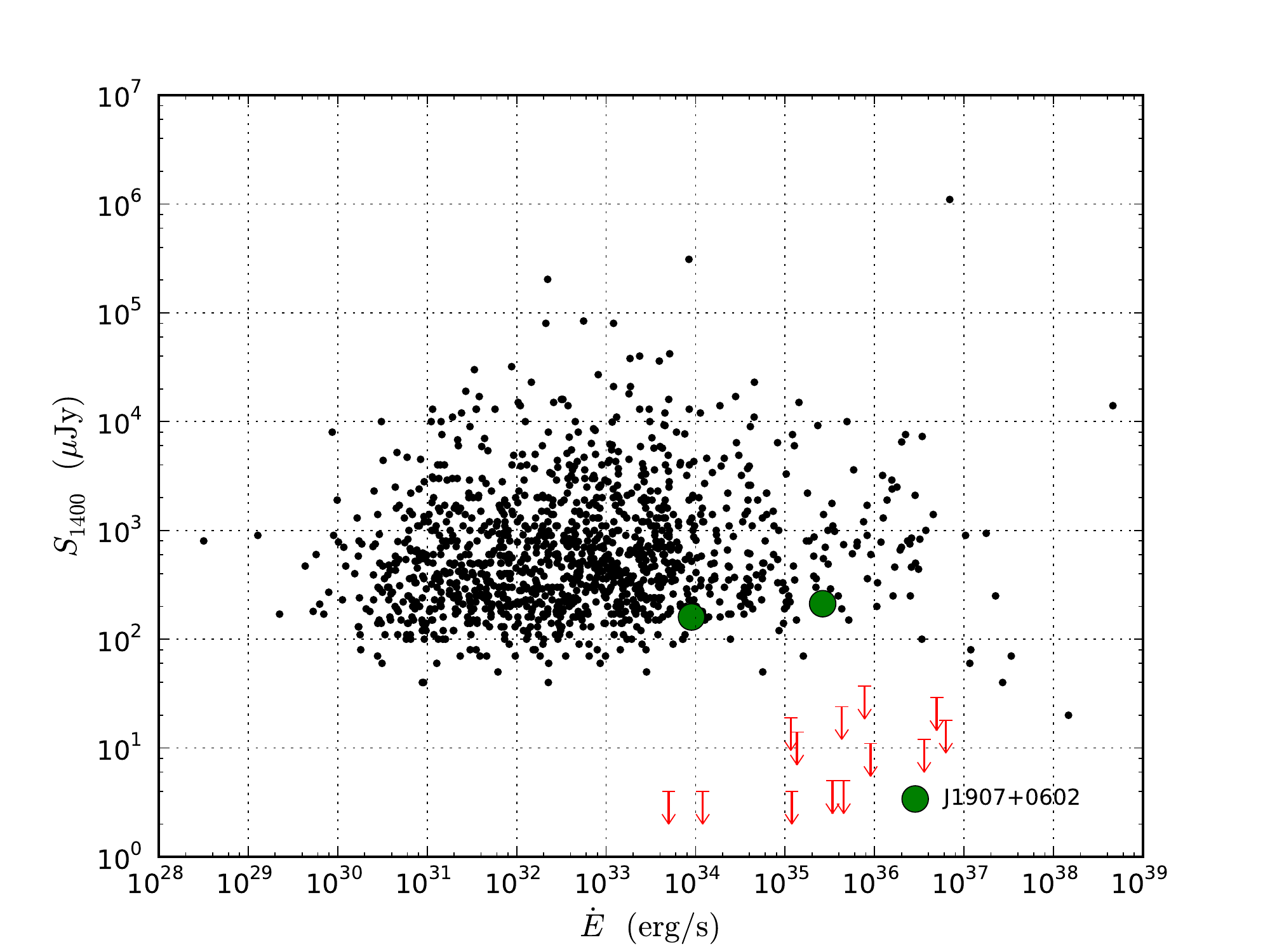} 
\caption{Summary of radio pulsation searches of $\gamma$-ray selected pulsars. The red arrows denote upper limits for the non-detections reported in Table~\ref{tab:radiolims}. The large green dots are the radio pulsation detections that have been previously reported \citep{crr+09,MGROPaper}. The black dots are the 1400 MHz flux densities of the non-millisecond pulsars in the ATNF catalog, for comparison. The detections and upper limits that were made at different observing frequencies were scaled to 1400 MHz using a typical pulsar spectral index of $1.6$. \label{fig:radiolims} } 
\end{figure}

We also made deep searches for radio pulsations from the $\gamma$-ray selected pulsars.
We compare these flux limits with the measured fluxes of the population of pulsars in the ATNF pulsar catalog \citep{mhth05} in Figure \ref{fig:radiolims}. To make the fluxes comparable, we have scaled them all to the equivalent 1400 MHz flux density using a typical pulsar spectral index of 1.6.  The upper limits we have obtained are comparable to some of the faintest known radio pulsars, but the discovery of 3.5 $\mu$Jy pulsations from PSR J1907+0602 \citep{MGROPaper} raises the possibility that some of these could yet be detected in even deeper radio searches.

The radio upper limits for 8 new $\gamma$-ray selected pulsar discovered with \textit{Fermi} are presented in \citet{BSP2}.  When combined with the results presented here, we now have deep upper limits on all known $\gamma$-ray selected pulsars. A discussion of the radio upper limits on PSR J1836+5925 was also presented by \citet{J1836}.

\acknowledgments

The authors gratefully acknowledge Masaharu Hirayama and the rest of the developers of the LAT Science Tools used for pulsar timing. This work makes use of APLpy by Eli Bressert and Thomas Robitaille.

The \textit{Fermi} LAT Collaboration acknowledges generous ongoing support
from a number of agencies and institutes that have supported both the
development and the operation of the LAT as well as scientific data analysis.
These include the National Aeronautics and Space Administration and the
Department of Energy in the United States, the Commissariat \`a l'Energie Atomique
and the Centre National de la Recherche Scientifique / Institut National de Physique
Nucl\'eaire et de Physique des Particules in France, the Agenzia Spaziale Italiana
and the Istituto Nazionale di Fisica Nucleare in Italy, the Ministry of Education,
Culture, Sports, Science and Technology (MEXT), High Energy Accelerator Research
Organization (KEK) and Japan Aerospace Exploration Agency (JAXA) in Japan, and
the K.~A.~Wallenberg Foundation, the Swedish Research Council and the
Swedish National Space Board in Sweden.

Additional support for science analysis during the operations phase is gratefully
acknowledged from the Istituto Nazionale di Astrofisica in Italy and the Centre National d'\'Etudes Spatiales in France.

The Arecibo Observatory is part of the National Astronomy and Ionosphere Center, which is operated by Cornell University under a cooperative agreement with the National Science Foundation.  The National Radio Astronomy Observatory is a facility of the National Science Foundation operated under cooperative agreement by Associated Universities, Inc. The Parkes Observatory is part of the Australia Telescope which is funded by the Commonwealth of Australia for operation as a National Facility managed by CSIRO.

{\it Facilities:} \facility{Fermi LAT}, \facility{GBT}, \facility{Arecibo}, \facility{Parkes}

\clearpage



\bibliography{journapj,psrtiming}


\appendix

\section{Calculation of Position Offsets}
\label{app:pos}


The light travel time delay $\tau$ (i.e.~Roemer delay) across the
solar system from a pulsar at ecliptic coordinates $\lambda$ (longitude) and
$\beta$ (latitude) is:
\begin{equation}
  \tau \simeq 500\,{\rm s}\ \cos(\beta)\cos\left(\theta(t) + \lambda\right),
\end{equation}
where $\theta(t)$ is the orbital phase of the Earth with respect to
the vernal equinox.  This is an approximate time delay since we are
assuming that the Earth's orbit is circular.

If a pulsar is being timed with incorrect ecliptic coordinates such
that there exist position offsets $\Delta\lambda$ and $\Delta\beta$,
there will be a differential time delay $\Delta\tau$ present in the
timing residuals:
\begin{equation}
  \Delta\tau \simeq 500\,{\rm s}\ \left[\cos(\beta+\Delta\beta) \cos\left(\theta(t) + \lambda + \Delta\lambda\right) - \cos(\beta)\cos(\theta(t) + \lambda)\right].
\end{equation}

If the positional offsets are small, such that we can use $\sin x\sim
x$, $\cos x\sim 1$, and $\Delta\beta\,\Delta\lambda\sim 0$, we can use
trigonometric identities to get:
\begin{equation}
  \Delta\tau \simeq -500\,{\rm s}\ \left[\Delta\lambda\cos(\beta)\sin(\theta(t) + \lambda) + \Delta\beta\sin(\beta)\cos\left(\theta(t) + \lambda\right)\right].
\end{equation}

Comparing the trigonometric identity $A\sin\left(\theta(t)+\phi\right) =
A\cos\phi\sin\theta(t) + A\sin\phi\cos\theta(t)$ to the equation for
$\Delta\tau$, we see that:
\begin{eqnarray}
  A\cos\phi &=& -500\,{\rm s}\ \Delta\lambda\cos\beta\\
  A\sin\phi &=& -500\,{\rm s}\ \Delta\beta\sin\beta,
\end{eqnarray}
and therefore:
\begin{eqnarray}
  \Delta\lambda &=& -\frac{A\cos\phi}{500\,{\rm s}\ \cos\beta}\\
  \Delta\beta &=& -\frac{A\sin\phi}{500\,{\rm s}\ \sin\beta}, 
\end{eqnarray}
The sinusoid amplitude $A$ and phase $\phi$ come from fits to TOA residuals. One way of doing this is by adding a binary model to a \textsc{Tempo2} fit with parameters A1 ($A$$=$A1) and T0 (with
PB$=$365.2424, the solar year; OM$=$0; and E$=$0 all held fixed in the fit). Then,
\begin{equation}
  \phi = 2\pi\frac{{\rm fmod}({\rm T0} - 51623.31250, 365.2424)}{365.2424} + \lambda + \alpha.
\end{equation}
where \texttt{fmod} is the floating point remainder function, 51623.31250 is the MJD of the vernal equinox in 2000 and $\alpha$ is a
correction for Earth's non-circular orbit that can be measured for a
particular point in the sky by fitting for T0 at several values of
simulated position offset in the ecliptic longitude direction only.
In that case, $\Delta\beta = 0$, and $\phi = 0$,
allowing us to solve for $\alpha$.

\end{document}